\documentclass[prd, 
superscriptaddress,preprintnumbers,tightenlines,nofootinbib, eqsecnum]{revtex4-2}

\usepackage{amsmath}
\usepackage{amsfonts}
\usepackage{amssymb}
\usepackage{bm}
\usepackage{hyperref}
\usepackage{mathrsfs}
\usepackage{graphicx}
\usepackage{tabularx}
\usepackage{empheq}
\usepackage{ulem}
\usepackage[usenames]{color}
\usepackage{ytableau}
\usepackage{comment}
\usepackage{etoolbox}
        
\allowdisplaybreaks

\DeclareSymbolFontAlphabet{\mathrsfs}{rsfs}
\DeclareMathAlphabet{\mathcal}{OMS}{cmsy}{m}{n}

\newcommand{\nn}{\nonumber}
\newcommand\calO{\mathcal{O}}
\newcommand{\dd}{\mathrm{d}}
\newcommand{\di}{\mathrm{i}} 
\newcommand{\de}{\mathrm{e}}

\newcommand{\calF}{\mathcal{F}}

\newcommand{\calW}{\mathcal{W}}

\newcommand{\fid}{\dot{\phi}}
\newcommand{\rd}{\dot{r}}

\newcommand{\Li}{\mathrm{Li}}

\newcommand{\hatc}{\hat{c}}
\newcommand{\hatl}{\hat{\ell}}
\newcommand{\hatn}{\hat{n}}
\newcommand{\hatS}{\hat{S}}
\newcommand{\hatW}{\hat{W}}
\newcommand{\hata}{\hat{a}}
\newcommand{\hate}{\hat{e}}
\newcommand{\hatK}{\hat{K}}
\newcommand{\hatf}{\hat{f}}
\newcommand{\hatg}{\hat{g}}
\newcommand{\hatA}{\hat{A}}
\newcommand{\hatV}{\hat{V}}
\newcommand{\hatchi}{\hat{\chi}}
\newcommand{\hatalpha}{\hat{\alpha}}
\newcommand{\hatbeta}{\hat{\beta}}

\definecolor{darkgreen}{rgb}{0,0.5,0}

\hypersetup{
 unicode=false,          
 pdftoolbar=true,        
 pdfmenubar=true,        
 pdffitwindow=false,     
 pdfstartview={FitH},    
 pdftitle={My title},    
 pdfauthor={Author},     
 pdfsubject={Subject},   
 pdfcreator={Creator},   
 pdfproducer={Producer}, 
 pdfkeywords={keyword1} {key2} {key3}, 
 pdfnewwindow=true,      
 colorlinks=true,       
 linkcolor=red,          
 citecolor=cyan,        
 filecolor=magenta,      
 urlcolor=darkgreen,           
 linktocpage=true
}

\makeatletter
\g@addto@macro\bfseries{\boldmath}
\patchcmd{\open@column@one}{\global\divide\count\footins\tw@}{}{}%
  {\PackageWarning{main}{revtex footnote patch failed}}
\makeatother

\begin{document}
	
\title{Next-to-next-to-leading gravitational radiation-reaction corrections to \\quasi-elliptic and quasi-hyperbolic compact binary orbits}

\author{Donato \textsc{Bini}}\email{donato.bini@gmail.com}  
\affiliation{Istituto per le Applicazioni del Calcolo ``M. Picone'' CNR, I-00185 Rome, Italy}

\author{Luc \textsc{Blanchet}}\email{luc.blanchet@iap.fr}
\affiliation{$\mathcal{G}\mathbb{R}\varepsilon{\mathbb{C}}\calO$, 
	Institut d'Astrophysique de Paris, UMR 7095, CNRS, Sorbonne Universit{\'e},
	98\textsuperscript{bis} boulevard Arago, 75014 Paris, France}

\author{Guillaume \textsc{Faye}}\email{faye@iap.fr}
\affiliation{$\mathcal{G}\mathbb{R}\varepsilon{\mathbb{C}}\calO$, 
	Institut d'Astrophysique de Paris, UMR 7095, CNRS, Sorbonne Universit{\'e},
	98\textsuperscript{bis} boulevard Arago, 75014 Paris, France}
    
\author{Andrea \textsc{Geralico}}\email{andrea.geralico@gmail.com}
\affiliation{Istituto per le Applicazioni del Calcolo ``M. Picone'' CNR, I-00185 Rome, Italy}

\author{Quentin \textsc{Henry}}\email{quentin.henry@uib.es}
\affiliation{Departament de F\'isica, Universitat de les Illes Balears, IAC3--IEEC, Crta. Valldemossa km 7.5, E-07122 Palma, Spain}

\author{Sara \textsc{Rufrano Aliberti}}\email{s.rufranoaliberti@ssmeridionale.it} 
\affiliation{Scuola Superiore Meridionale, Largo San Marcellino 10, I-80138, Napoli, Italy} 
\affiliation{INFN, Sezione di Napoli, Complesso Universitario di Monte S. Angelo, \\Via Cinthia Edificio 6, I-80126, Napoli, Italy}

\author{Emeric \textsc{Seraille}}\email[Corresponding author: ]{emeric.seraille@phys.ens.fr}
\affiliation{Laboratoire de Physique de l’Ecole Normale Sup{\'e}rieure, ENS, CNRS, Universit{\'e} \\ PSL, Sorbonne Universit{\'e}, Universit{\'e} Paris Cit{\'e}, F-75005 Paris, France}
\affiliation{$\mathcal{G}\mathbb{R}\varepsilon{\mathbb{C}}\calO$, 
	Institut d'Astrophysique de Paris, UMR 7095, CNRS, Sorbonne Universit{\'e},
	98\textsuperscript{bis} boulevard Arago, 75014 Paris, France}

\date{\today}

\begin{abstract}
We apply the Lagrange method of variation of constants to compute the radiation-reaction (RR) corrections to the quasi-elliptic and quasi-hyperbolic motion of compact binary systems at 4.5PN order, \textit{i.e.}, at next-to-next-to-leading order in the RR. We rely on the recent determination of the 4.5PN RR acceleration, in a general frame and in the center-of-mass frame, in harmonic coordinates. The non-local effects, due to the tails at 4PN order and to the gravitational-wave recoil at 4.5PN order, are carefully evaluated. Although the calculation is implemented differently in the quasi-elliptic and quasi-hyperbolic cases (different perturbation equations and choices of orbital parameters), we verify by analytic continuation that the results for the local effects are equivalent in the two cases.
\end{abstract}


\maketitle

\section{Introduction}\label{sec:introduction}

The most famous (and so far the only detected) sources of gravitational waves are the radiation-reaction (RR) dominated inspiral and eventual merger of two compact objects, neutron stars or black holes. For such systems, a few up to hundreds or thousands of orbital cycles can be measured (depending on the masses and detector's properties), requiring an extremely accurate modelling by relativistic post-Newtonian corrections in the orbital phasing and waveform up to very high order~\cite{3mn}. The first empirical evidence for gravitational RR came from the orbital decay of the Hulse--Taylor binary pulsar~\cite{HulseTaylor}. It confirmed the lowest-order, quadrupolar RR acceleration predicted by general relativity for a system of two point masses and, indirectly, the associated far-zone flux, given by the Einstein quadrupole formula~\cite{TW82,D83}.  

Being dissipative by nature, the gravitational RR force first appears at the so-called
second-and-a-half post-Newtonian (2.5PN) order, that is, as a formal correction proportional
to $c^{-5}$ in the limit where the speed of light $c\to\infty$. It has been computed in a large
number of important papers~\cite{CE70, BuTh70, Bu71, Miller74, Ehl80, WalkW80, K80a, K80b, PapaL81, DD81a, D82, BD84}, both for general hydrodynamical systems and for compact binaries, in various
gauges. In particular Burke and Thorne~\cite{BuTh70, Bu71} introduced a useful gauge in which the 2.5PN RR is especially simple, and Damour and Deruelle~\cite{DD81a, D82} did the computation in harmonic coordinates, which is nowadays considered to be more adapted for comparisons with alternative approaches like the Effective Field Theory (EFT).

The extension to the next, 3.5PN order, proportional to $c^{-7}$, is significantly more challenging and was obtained in a subsequent series of papers~\cite{IW93, IW95, B97, JaraS97, PW02, KFS03, NB05, itoh3, BF19}. Two approaches have been used: the ``flux-balance'' method which applies to compact binaries and assumes the standard fluxes at infinity to deduce the local RR from the balance laws~\cite{IW93, IW95}, and the ``first-principle'' approach which works out the RR directly in the local equations of motion (and checks afterwards that the balance laws are satisfied)~\cite{B97, JaraS97, PW02, KFS03, NB05, itoh3, BF19}.

Remarkably, the next RR correction does not arise at 4.5PN but instead at the 4PN order, proportional to $c^{-8}$, due to a RR correction that is induced by gravitational wave tails (non-linear interaction between the quadrupole and the total mass of the source) re-entering back into the system and affecting the equations of motion~\cite{BD88,BD92}. The 4PN RR force is accompanied by a conservative contribution which plays an important role in the conservative equations of motion at 4PN order~\cite{FStail,DJS14,GLPR16,BBFM17}.

The RR problem at 4.5PN order (2PN beyond the dominant 2.5PN RR) was first tackled for compact
binaries using the flux-balance approach, restricted to the center-of-mass frame~\cite{GII97}. A first-principle derivation, not relying on balance equations, was later attempted within EFT~\cite{LPY23}. The problem of the 4.5PN RR has been reconsidered recently and completely solved in two papers, one~\cite{BFT25} using the so-called extended Burke-Thorne coordinate system, and the other~\cite{BFST26} in the more standard harmonic coordinate system. Both calculations are valid for general orbits and in a general frame, which can be specialized to the Center-of-Mass (CM) frame. It was found in particular that the RR force acquires in the CM frame a non-local (``semi-hereditary'') contribution at the 4.5PN order, which involves the gravitational recoil of the source as well as the flux of CM position. Due to this effect, it becomes impossible (in any coordinate system) to express the RR in the CM frame at the 4.5PN order as a local or ``instantaneous'' functional of the source's parameters. 

In this respect, Refs.~\cite{BFT25, BFST26} are in conflict with the flux-balance approach extended to 4.5PN order in~\cite{GII97}, which assumed as a basic hypothesis that the RR force in the CM frame is a local functional of the source parameters, thereby missing the physical consequence of the binary’s GW recoil. It was proved~\cite{BFT25, BFST26} that the equations of motion including the 4.5PN RR force in a general frame imply the full set of flux-balance laws, for energy, angular momentum, linear momentum and CM position, with the standard expressions for the fluxes. Furthermore, when reduced to the CM frame, including the above mentioned non-local contribution of the recoil, they imply the flux-balance laws for energy and angular momentum, with the standard fluxes in the CM, and with the non-local contribution of the recoil contributing to the Schott terms. It was finally shown that the local part of the RR acceleration and of the balance laws for energy and angular momentum in the CM frame were consistent with the flux-balance approach in~\cite{GII97}, in the sense that there is a unique set of gauge parameters corresponding to the choices made in~\cite{BFT25, BFST26} of extended Burke-Thorne and harmonic coordinates. Last but not least, the 4.5PN calculation in harmonic coordinates~\cite{BFST26} was shown to match perfectly the recent RR post-Minkowskian EFT calculation~\cite{BDG24}, when it is truncated at 4.5PN order and keeping only terms up to order $G^2$ (see tables IV and V of Ref.~\cite{BDG24}).

With the 4.5PN RR acceleration completed, we can study its effect on the orbital motion. The RR, which is dissipative, acts as a perturbation of the conservative dynamics, whose
solution can be parametrized in a quasi-Keplerian (QK) form in the CM frame, using polar
coordinates in the orbital plane and a set of orbital elements. We shall consider the perturbation of the RR force at the 2PN \textit{relative} order (\textit{i.e.}, up to 4.5PN order), hence we shall need the QK parametrization at the 2PN order. The latter is known to 3PN order~\cite{DD85, DS88, SW93, MGS04, CGHL18} (see also~\cite{CTG22}) for both the quasi-elliptic and quasi-hyperbolic cases, which can in fact be retrieved from each other by analytic continuation (AC). The expressions of the orbital elements as functions of the conserved energy and angular momentum of the orbit generally depend on the coordinate system, chosen here to be the harmonic coordinate system, consistent with the RR force in~\cite{BFST26}. The QK parametrization is now known including aligned-spin effects~\cite{THS13,HK23} up to 3.5PN order, adiabatic tidal effects to relative 2PN~\cite{HH25} and magnetic dipole interaction to relative 1PN~\cite{HHLL23b}. 

Motivated by the need for high-accuracy GW templates for quasi-eccentric binaries and by the comparison with EFT results in the quasi-hyperbolic case, we shall determine the corrections due to RR in the phasing of GWs, by the Lagrange method of variation of constants~\cite{BrouwerClemence}. In the case of eccentric orbits, this will include not only the secular adiabatic corrections varying on the long RR time scale, but also fast oscillations at the orbital period, and modulations at the intermediate precession time scale. This method was pioneered in this context by Damour, Gopakumar and Iyer~\cite{DGI04} and developed in further papers~\cite{KG06, BDG23, BGR25}, see in particular the hyperbolic version of the method which was applied to RR terms at 2.5PN~\cite{BDG23} and 3.5PN~\cite{BGR25} orders.

In the present paper, we extend the previous results by including the complete RR terms up to 4.5PN order in harmonic coordinates for both elliptic-like and hyperbolic-like orbits. We shall pay particular attention to the non-local terms, at the 4PN order due to the RR part of the tail, and at the 4.5PN order due to the contribution of the source's GW recoil. In particular, the case of the recoil is known from classic PN calculations~\cite{BoR61,Papa71,Th80} applied to compact binaries~\cite{Fit83,Wi92,BQW05}. Here we shall adopt a multi-scale approach for the computation of the recoil in the elliptic case, to obtain an expression which is purely oscillating and is therefore expected to be accurate over a long period of time, much longer than the orbital time scale and also the precession time scale, although still smaller than the RR time scale. This multi-scale approach goes beyond  the standard PN calculation of the recoil (\textit{e.g.}~\cite{Fit83}) but is nevertheless fully consistent with it in the appropriate limit.

The plan of the paper is as follows. Sec.~\ref{sec:recap} summarizes the known results for the RR acceleration of compact binaries up to 4.5PN order. In Sec.~\ref{sec:Lagrange} we investigate the QK parametrization at 2PN order for the elliptic-like and hyperbolic-like cases, and implement the Lagrange method of variation of constants in the two cases. Our results are presented in Sec.~\ref{sec:reslocal} for local contributions, where the elliptic and hyperbolic results are shown to be equivalent by analytic continuation, and in Sec.~\ref{sec:resnonlocal} for non-local contributions, at 4PN for the tail term and at 4.5PN for the recoil. The main text ends with a short conclusion in Sec.~\ref{sec:conclusion}. In App.~\ref{app:QK} we present the details for the QK parametrization at 3PN order for elliptic and hyperbolic orbits, and show their equivalence by analytic continuation. App.~\ref{app:formulas} is devoted to explicit expressions of definite and indefinite technical integrals in the elliptic-like case, and App.~\ref{app:IijFourier} to formulas for the Fourier coefficients of the quadrupole moment. The final expressions of the full results are extremely lengthy; we therefore relegate them, as well as most of the important formulas in the main text, to the Supplemental Material~\cite{ancillary}.

\section{Recapitulation of the 4.5PN radiation reaction acceleration}
\label{sec:recap}

The total relative acceleration ${\bm A} \equiv {\bm A}_1-{\bm A}_2$ between the two point masses (without spins) in an asymptotically flat inertial frame is split into a conservative part (admitting the usual ten Poincar\'e invariants), and a dissipative RR part due to the GW emission, say ${\bm A} = {\bm A}^{\rm cons} + {\bm A}^{\rm RR}$. To 2PN order the conservative piece reads
\begin{equation}\label{eq:Acons}
	{\bm A}^{\rm cons} = {\bm A}^{\rm cons}_\text{N} + {\bm A}^{\rm cons}_\text{1PN} + {\bm A}^{\rm cons}_\text{2PN} + \calO\left(\frac{1}{c^6}\right)\,,
\end{equation}
where ${\bm A}^{\rm cons}_\text{N} = - \frac{G m}{r^2}\bm{n}$ is the Newtonian acceleration and the 1PN and 2PN corrections are well known (see, \textit{e.g.}~\cite{BlanchetLR_2024}) and will be considered here in harmonic coordinates and in the CM frame. On the other hand the RR piece starts at the 2.5PN order and is given, up to 2PN relative order (\textit{i.e.}, up to the 4.5PN total level), by
\begin{equation}\label{eq:Arr}
	{\bm A}^{\rm RR} = {\bm A}^{\rm RR}_\text{2.5PN} + {\bm A}^{\rm RR}_\text{3.5PN} + {\bm A}^{\rm RR}_\text{4PN\,tail} + {\bm A}^{\rm RR}_\text{4.5PN} + \calO\left(\frac{1}{c^{10}}\right)\,.
\end{equation}
The leading 2.5PN and next-to-leading 3.5PN corrections are also well known (in various coordinate systems) and will also be considered here in harmonic coordinates and in the CM frame, where they are given for general binary orbits by~\cite{DD81a, D82, D83, PW02, NB05}
\begin{subequations}\label{eq:accCM23p5PN}
\begin{align}\label{eq:accCM2p5PN}
{\bm A}^{\rm RR}_\text{2.5PN} &= \frac{G^2 m^2 \nu }{r^3c^5}\Bigg\{\bm{n} \left(\frac{136}{15} \frac{G m}{r}\dot{r} +\frac{24}{5}\dot{r} v^2\right)-\bm{v} \left(\frac{24}{5}\frac{G m}{r}+\frac{8}{5} v^2\right)\Bigg\} \,, \\
\label{eq:accCM3p5PN}
{\bm A}^{\rm RR}_\text{3.5PN} &= \frac{G^2 m^2 \nu }{r^3c^7} \Bigg\{\bm{n} \Bigg[\frac{G^2 m^2}{r^2} \dot{r} \left(-\frac{3956}{35}-\frac{184}{5} \nu \right)+\frac{G m}{r}  \dot{r}^3 \left(-\frac{294}{5}-\frac{376}{5} \nu \right) \nn\\
&\qquad\qquad\qquad+\frac{G m}{r} \dot{r} v^2\left(-\frac{692}{35}+\frac{724}{15} \nu \right) -112\, \dot{r}^5 +
\dot{r}^3 v^2 (114+12 \nu ) + \dot{r} v^4 \left( -\frac{366}{35}-12 \nu\right) \Bigg]\nn\\
&\qquad\qquad~ +\bm{v} \Bigg[\frac{G^2 m^2}{r^2}\left(\frac{1060}{21}+\frac{104}{5} \nu \right) + \frac{G m}{r} \dot{r}^2 \left(\frac{82}{3}+\frac{848}{15} \nu\right) + \frac{G m}{r} v^2 \left(-\frac{164}{21}-\frac{148}{5} \nu \right) \nn\\
&\qquad\qquad\qquad +120\,
\dot{r}^4+ \dot{r}^2 v^2 \left(-\frac{678}{5}-\frac{12}{5} \nu \right)+  v^4\left(\frac{626}{35}+\frac{12}{5} \nu\right)\Bigg]\Bigg\} \,,
\end{align}
\end{subequations}
see for instance Eqs.~(5.8) and (5.9) in Ref.~\cite{NB05}. The relative distance is $r=\vert\bm{x}\vert\equiv\vert\bm{y}_1-\bm{y}_2\vert$, we pose $\bm{n}=\bm{x}/r$, $\bm{v}=\dd\bm{x}/\dd t$ and $\dot{r}=\dd r/\dd t=\bm{n}\cdot\bm{v}$, and mass parameters are $m=m_1+m_2$, the symmetric mass ratio $\nu=m_1m_2/m^2$ and, for future use, the mass difference $\delta=(m_1-m_2)/m$. The next-order RR term actually arises at 4PN order (corresponding to 1.5PN order beyond leading) and is entirely due to the gravitational quadrupole tail~\cite{BD88}. It is made of the non-local (in time) integral
\begin{equation}\label{eq:tail}
	A^{i\,{\rm RR}}_{\text{4PN\,tail}} = -\frac{4G^2M}{5 c^8} \,x^j
	\!\int_0^{+\infty} \!\dd\tau
	\ln\left(\frac{\tau}{\tau_0}\right)\left[I_{ij}^{(7)}(t-\tau) +
	I_{ij}^{(7)}(t+\tau)\right]\,,
\end{equation}
where $M$ is the total Arnowitt–Deser–Misner (ADM) mass of the system, $I_{ij}=m\nu x^{\langle i}x^{j\rangle}$ denotes the binary's Newtonian mass-type quadrupole moment (with $x^{\langle i}x^{j\rangle}=x^{i}x^{j}-\frac{1}{3}\delta^{ij}r^2$), with $I_{ij}{}^{\!\!(n)}$ denoting the time derivatives which are to be computed here using the Newtonian law of motion, and where $\tau_0$ is an arbitrary time scale, which is actually easily checked to cancel out between the two terms of Eq.~\eqref{eq:tail}.\footnote{The conservative part of the 4PN tail term is given by~\cite{BD88,FStail,DJS14,GLPR16,BBFM17}
\begin{equation}\label{eq:tailcons}
	A^{i\,{\rm cons}}_{\text{4PN\,tail}} = -\frac{4G^2M}{5 c^8} \,x^j
		\!\int_0^{+\infty} \!\dd\tau
		\ln\left(\frac{c\tau}{2r}\right)\left[I_{ij}^{(7)}(t-\tau) -
		I_{ij}^{(7)}(t+\tau)\right] = -\frac{4G^2M}{5 c^8} \,x^j \!\mathop{\text{Pf}}_{2r/c}\int_{-\infty}^{+\infty}\frac{\dd t'}{\vert t - t'\vert} I_{ij}^{(6)}(t')\,,
\end{equation}
where Pf refers to the Hadamard partie finie. Assuming a matter system invariant under time reversal, in the sense that $I_{ij}(-t) = I_{ij}(t)$, the integral~\eqref{eq:tailcons} is also invariant, while by contrast Eq.~\eqref{eq:tail} changes sign, hence it is associated with dissipative RR effects. In this paper we do not need to consider the conservative tail integral~\eqref{eq:tailcons} which falls into the remainder term of Eq.~\eqref{eq:Acons}. 
}

At the 4.5PN order, the structure of the RR acceleration is more complicated than those at 2.5 and 3.5PN due to the contribution of RR in the definition of the CM frame. As shown in~\cite{BFT25}, the novel feature is a supplementary non-local term at order 4.5PN, of a nature different from the tail term~\eqref{eq:tail}, coming from the contribution of radiation (in addition to the usual matter contribution) in the definition of the frame of the CM. Hence we pose
\begin{equation}\label{45PNsplit}
	{\bm A}^{\rm RR}_\text{4.5PN} = {\bm A}^{\rm RR}_\text{4.5PN\,mat} + {\bm A}^{\rm RR}_\text{4.5PN\,rad}\,,
\end{equation}
where the matter part (which is local) neglects the radiation in the definition of the CM, while the radiation part takes into account the (integrated) radiation fluxes at infinity. The complete matter part of the 4.5PN RR acceleration in harmonic coordinates reads
\begin{align}\label{eq:accCM4pPNmat}
{\bm A}^{\rm RR}_\text{4.5PN\,mat} &= \frac{G^2 m^2 \nu}{r^3c^9}  \Bigg\{\bm{n} \Bigg[\frac{G^3 m^3}{r^3}  \dot{r} \left(\frac{336922}{945}+\frac{20644}{35} \nu-\frac{3632}{105} \nu ^2\right)  +\frac{G^2 m^2}{r^2}  \dot{r}^3 \left(-\frac{83177}{945}+\frac{41524}{135} \nu -\frac{30076}{105} \nu ^2\right)  \nn\\
&\qquad\qquad\qquad +\frac{G^2 m^2}{r^2}\dot{r} v^2 \left(-\frac{129769}{315}-\frac{58468}{315} \nu +\frac{4636}{105} \nu^2\right) +\frac{G m}{r}  \dot{r}^5  \left(\frac{261883}{105}+\frac{5316}{5} \nu -\frac{1776}{5}\nu ^2\right)  \nn\\
&\qquad\qquad\qquad  +\frac{G m}{r}   \dot{r}^3 v^2 \left(-\frac{22133}{6}-\frac{47643}{35} \nu +\frac{16028}{35} \nu^2\right)  +\frac{G m}{r}  \dot{r} v^4 \left(\frac{14307}{14}+\frac{46337}{105} \nu -\frac{2659}{21} \nu^2\right)  \nn\\
&\qquad\qquad\qquad   + \dot{r}^7(-180-504 \nu)
+ \dot{r}^5 v^2 \left(\frac{329}{2}+1106 \nu+21 \nu ^2\right) + \dot{r}^3 v^4 \left(\frac{88}{7}-\frac{4362}{7} \nu-66 \nu ^2\right) \nn\\
&\qquad\qquad\qquad  + \dot{r} v^6 \left(-\frac{1643}{210}+\frac{1248}{35} \nu +45 \nu^2\right)
\Bigg] \nn\\
&\qquad\qquad +\bm{v} \Bigg[\frac{G^3 m^3}{r^3} \left(-\frac{499286}{2835}-\frac{1376}{5} \nu+\frac{272}{35}\nu^2\right)+\frac{G^2 m^2}{r^2}\dot{r}^2\left(\frac{85991}{315}-\frac{97228}{315} \nu+\frac{4604}{21} \nu^2\right) \nn\\
&\qquad\qquad\qquad +\frac{G^2 m^2}{r^2}  v^2 \left(\frac{47459}{315}+\frac{66632}{315} \nu -\frac{1796}{35} \nu ^2\right) +
\frac{G m}{r} \dot{r}^4  \left(-\frac{86323}{105}-\frac{10646}{35} \nu+\frac{1192}{7} \nu^2\right)\nn\\
&\qquad\qquad\qquad 
+ \frac{G m}{r} \dot{r}^2 v^2 \left(\frac{224351}{210}+\frac{7177}{35} \nu-\frac{2816}{15} \nu ^2\right) 
+ \frac{G m}{r} v^4 \left(-\frac{10747}{70}-\frac{539}{15} \nu+\frac{1769}{35} \nu ^2\right) \nn\\
&\qquad\qquad\qquad  +  \dot{r}^6(350+420 \nu )+ \dot{r}^4 v^2 \left(-\frac{1007}{2} -984 \nu -3 \nu ^2\right)+ \dot{r}^2 v^4\left(\frac{5778}{35}+\frac{4350}{7} \nu+\frac{54}{5} \nu^2\right)  \nn\\
&\qquad\qquad\qquad + v^6\left(-\frac{4873}{630} -\frac{298}{5} \nu-\frac{39}{5} \nu ^2\right) \Bigg]\Bigg\} \,.
\end{align}
The radiation part of the RR acceleration, defined in Eq.~(6.12) of~\cite{BFT25}, is given by\footnote{In the Supplemental Material~\cite{ancillary}, the radiation part of the acceleration is further split into a local and a non-local piece,
\begin{align*}
& A^{i\,{\rm RR}}_\text{4.5PN\,rad} = A^{i\,{\rm RR}}_\text{4.5PN\,rad\,loc} + A^{i\,{\rm RR}}_\text{4.5PN\,rad\,non-loc}\,,\\
\text{where}\quad & A^{i\,{\rm RR}}_\text{4.5PN\,rad\,loc} = \frac{G\delta}{r^2 c^2}\left(2n^iv^j+n^jv^i\right)\calF_{\bm{G}}^j\,,\qquad
A^{i\,{\rm RR}}_\text{4.5PN\,rad\,non-loc} = \frac{G\delta}{r^2c^2}\left(2n^iv^j+n^jv^i\right)\Pi^j\,,
\end{align*}
the latter containing the non-local recoil integral~\eqref{recoil}. This split is used for all
results except the scattering angle, which is given in the alternative split~\eqref{45PNsplitGII}.}
\begin{equation}\label{eq:45PNnonloc}
	A^{i\,{\rm RR}}_\text{4.5PN\,rad} = \frac{G \delta}{r^2c^2}\left(2 n^iv^j+n^jv^i\right)\Bigl[\Pi^j + \calF_{\bm{G}}^j\Bigr]\,,
\end{equation}
where $\calF_{\bm{P}}^i$ and $\calF_{\bm{G}}^i$ respectively denote the fluxes of linear momentum $\bm{P}$ and center-of-mass position $\bm{G}$, both arising at the 3.5PN order and given in terms of source multipole moments, \textit{viz.} the mass and current type quadrupoles $I_{ij}$ and $J_{ij}$, and mass octupole $I_{ijk}$, by
\begin{subequations}\label{eq:balancePG}
\begin{align}
	\calF_{\bm{P}}^i &= \frac{G}{c^7} \left[ \frac{2}{63}
	I_{ijk}^{(4)}I_{jk}^{(3)} + \frac{16}{45}
	\varepsilon_{ijk} I_{jl}^{(3)} J_{kl}^{(3)}
	\right] + \calO\left(\frac{1}{c^{9}}\right)\,,\\
	\calF_{\bm{G}}^i &= \frac{2}{21} \frac{G}{c^7}
	\left[ I_{ijk}^{(3)} I_{jk}^{(3)}\right] + \calO\left(\frac{1}{c^{9}}\right)\,.
\end{align}
\end{subequations}
Inserting the expressions of Newtonian source moments in the CM frame we explicitly get
\begin{subequations}\label{eq:balancePGexpl}
\begin{align}
	\label{eq:fluxPexpl}
\bm{\calF}_{\bm{P}} &= \frac{G^3 m^4 \delta \nu^2}{c^7 r^4} \Bigg\{ \dot{r} \bm{n} \Bigg(\frac{32 G m}{35 r} -  \frac{24}{7} \dot{r}^2 + \frac{88}{21} v^2\Bigg) + \bm{v}\Bigg(- \frac{64 G m}{105 r} + \frac{304}{105} \dot{r}^2 -  \frac{80}{21} v^2\Bigg)  \Bigg\} + \calO\left(\frac{1}{c^{9}}\right)\,,\\
 	\label{eq:fluxGexpl}
\bm{\calF}_{\bm{G}} &= \frac{G^2 m^3 \delta \nu^2}{c^7 r^2} \Bigg\{\bm{n} \Bigg(\frac{272G m}{105r} \dot{r}^2 -  \frac{64G m}{15r} v^2 + \frac{48}{35} \dot{r}^2 v^2 -  \frac{32}{35} v^{4}\Bigg) +  \dot{r} \bm{v} \Bigg(\frac{16 G m}{21 r} -  \frac{24}{7} \dot{r}^2 + \frac{24}{7} v^2\Bigg) \Bigg\} + \calO\left(\frac{1}{c^{9}}\right) \,.
\end{align}
\end{subequations}
Alternatively, we can introduce another decomposition of the RR acceleration 
\begin{equation}\label{45PNsplitGII}
	{\bm A}^{\rm RR}_\text{4.5PN} = {\bm A}^{\rm RR\,GII}_\text{4.5PN} + {\delta \bm A}^{\rm RR}_\text{4.5PN}\,,
\end{equation}
given by Eqs.~(7.8a) and~(7.10) in~\cite{BFT25} where
\begin{equation}\label{eq:45PNnonlocGII}
	\delta A^{i\,{\rm RR}}_\text{4.5PN} = \frac{G \delta}{r^2c^2}\left(2 n^iv^j+n^jv^i\right)\Bigl[\Pi^j + \calF_{\bm{G}}^j\Bigr] - \frac{\delta}{m c^2} v^i v^j \biggl[\calF_{\bm{P}}^j + \frac{\dd{\calF}_{\bm{G}}^j}{\dd t}\biggr]\,.
\end{equation}
The GII part is exactly given by the end result of Gopakumar, Iyer and Iyer (GII)~\cite{GII97}, when their gauge parameters take the unique values corresponding to harmonic coordinates.\footnote{See Eqs.~(6.20) in~\cite{BFST26}, and see also the gauge parameters corresponding to Burke-Thorne coordinates in Eqs.~(7.12) in~\cite{BFT25}.} 
The non-local contribution (\textit{stricto sensu}) in the 4.5PN acceleration is the integral of the flux of linear momentum $\calF_{\bm{P}}$ or total GW recoil of the source which is therefore defined by
\begin{equation}\label{recoil}
	\Pi^i(t) = \int_{-\infty}^t \!\dd \tau\, \calF_{\bm{P}}^i(\tau) \,.
\end{equation}
A careful treatment of this term is necessary in our calculation, both for elliptic and hyperbolic orbits.

\section{The Lagrange method of variation of constants}
\label{sec:Lagrange}

To obtain RR corrections to orbital parameters (and hence to the orbit itself) we use Lagrange's method of variation of constants~\cite{DGI04, KG06, BDG23, BGR25}. To this aim we start from the quasi-Keplerian (QK) representation of the \textit{conservative} orbit in harmonic coordinates and in the CM frame. The constant parameters in the QK parametrization of the conservative orbit are then varied to account for the presence of the RR corrections. For convenience and consistency checks we perform two separate calculations: (i) in the quasi-elliptic case, using a particular set of \textit{six} parameters, including two parameters which are actually constant in the present case of a fixed orbital plane; (ii) in the quasi-hyperbolic case, with a different set of \textit{four} ``dynamical'' parameters. The two calculations will be shown to be equivalent \textit{via} analytic continuation (AC).

\subsection{Quasi-elliptic orbit}
\label{sec:elliptic}

To 2PN order the QK representation of the conservative, quasi-elliptic bound orbit, which represents an explicit solution of the 2PN conservative dynamics~\eqref{eq:Acons}, is derived in Refs.~\cite{DD85,DS88,MGS04}. Using polar coordinates $\{r, \phi\}$, where $\phi$ is the orbital phase, we have
\begin{subequations}\label{eq:QK2PNelliptic}
\begin{align}
	r &= a_r\bigl( 1 -e_r \cos u\bigr)\,, \label{eq:QK2PNelliptica}\\
	\ell &= u - e_t \sin u + f_t \sin V_\phi +g_t\bigl( V_\phi - u\bigr)\,, \\
	\phi &= \omega + K\Bigl[ V_\phi + f_\phi \sin 2V_\phi + g_\phi \sin 3V_\phi\Bigr]\,. \label{eq:phielliptic}
\end{align}
\end{subequations}
The terms with $f_t$, $g_t$, $f_\phi$ and $g_\phi$ arise only at the 2PN order. The motion is parametrized by the eccentric anomaly $u$, while the mean anomaly $\ell$, in the case of the conservative motion, is a linear function of time. We pose $\dot{\ell}=n$ with $n$ the mean motion or radial frequency ($n=2\pi/P$ where $P$ is the orbital period). The mean anomaly $\ell$ is defined up to a constant we shall denote $c_\ell$. The argument of the periastron (or angular distance from ascending node to periastron) is $\omega$. The auxiliary quantity $V_\phi$ is given by two equivalent forms,
\begin{subequations}\label{eq:defV}
\begin{align}	
	V_\phi &= 2 \arctan\left[\sqrt{\frac{1+e_\phi}{1-e_\phi}} \tan\left(\frac{u}{2}\right)\right] + 2\pi\left(\left\lfloor\frac{u-\pi}{2\pi}\right\rfloor+1 \right) \label{defVa} \\[-0.1cm] 
	&
	= u + 2 \arctan\left[\frac{\beta_\phi\sin u}{1-\beta_\phi\cos u}\right]\,, \label{defVb}\\[-0.2cm]
	\frac{\dd V_\phi}{\dd u} &= \frac{\sqrt{1 - e_\phi^2}}{1-e_\phi \cos u}\,,
\end{align}
\end{subequations}
where $\lfloor\cdots\rfloor$ in the first form~\eqref{defVa} is the floor function (\textit{i.e.}, $\lfloor x\rfloor$ is the greatest integer less than or equal to $x$), and we pose $\beta_\phi\equiv [1-(1-e_\phi^2)^{1/2}]/e_\phi$ in the second form~\eqref{defVb}. Although neither $V_\phi$ nor $u$ is periodic, the combination $V_\phi-u$ is a periodic function of the eccentric anomaly $u$ with period $2\pi$. 

The point of the QK solution of the equations of motion is that all the constant parameters $n$, $K$, $a_r$, $e_t$, $e_r$, $e_\phi$, $f_t$, $g_t$, $f_\phi$ and $g_\phi$ are uniquely specified, in a given coordinate system, by the (specific) conserved energy and angular momentum of the orbit, defined by
\begin{align}\label{eq:EJ}
	\tilde{E} \equiv \frac{E-m c^2}{m\nu}\,,\qquad\qquad \tilde{J} \equiv \frac{J}{m\nu}\,,
\end{align}
where $E$ is the relativistic energy and $J$ the norm of the angular momentum. Note that the eccentric anomaly $u$ should be considered as a function of the conserved energy and angular momentum, and of time through $\ell$. It is convenient to introduce some definition of the semimajor axis $a$ and eccentricity $e$ by the Newtonian formulas
\begin{align}\label{eq:Newtonianae}
	a \equiv - \frac{G m}{2 \tilde{E}}\,,\qquad e \equiv \sqrt{ 1 + \frac{2\tilde{E} \tilde{J}^2}{G^2m^2}}\,. 
\end{align}
The parameters $n$ and $K$ are gauge invariant (they take the same expression in terms of $a$ and $e$ for the solution of the equations of motion in any coordinate system) and given by 
\begin{subequations}\label{parametersnK}
	\begin{align}
	n(a,e) &= \sqrt{\frac{G m}{a^3}}\left\{ 1 + \frac{1}{8} (-15+\nu) \frac{G m}{a c^2} + \frac{1}{128}\Bigl[555+30\nu+11\nu^2+\frac{192}{\sqrt{1-e^2}}(-5+2\nu)\Bigr] \left(\frac{G m}{a c^2}\right)^2 \right\}\,,\label{parametersn}\\
	K(a,e) &= 1 + \frac{3 G m}{a c^2(1-e^2)}\left\{ 1 - \frac{1}{4}\Bigl[ 5-2\nu - \frac{35-10\nu}{1-e^2} \Bigr]\frac{G m}{a c^2} \right\}\,.
	\end{align}
\end{subequations}
The orbital precession is $\Delta\phi = 2\pi k$, with $k=K-1$ denoting the relativistic precession. The other parameters $a_r$, $e_t$, $e_r$, $e_\phi$, $f_t$, $g_t$, $f_\phi$ and $g_\phi$ are gauge dependent and we use their expressions in terms of $a$ and $e$ valid in harmonic coordinates. In Appendix~\ref{app:QK} we provide all the parameters up to 3PN order in modified harmonic coordinates.

\begin{figure}[t]
\centering
\includegraphics[width=0.4\textwidth]{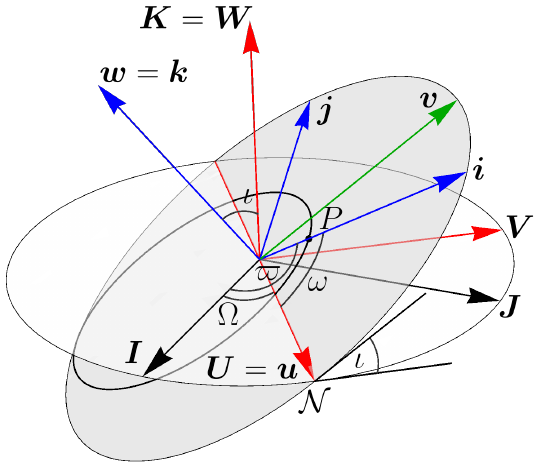}
\caption{Illustration of our definitions for osculating angles. The orthonormal basis $\{\bm{I}, \bm{J}, \bm{K}\}$ defines an inertial frame, with $\bm{I}$ and $\bm{J}$ in the ``plane of the sky'' and $\bm{K}$ being in the direction of an observer. The orientation of the orbital plane is given by the inclination angle $\iota$ and the longitude $\Omega$ of the ascending node $\mathcal{N}$ with respect to the observer. We define also the orthonormal basis $\{\bm{U}, \bm{V}, \bm{W}\}$, where $\bm{U}$ is in the direction of the node, $\{\bm{u}, \bm{v}, \bm{w}\}$ where $\bm{w}$ is the oriented normal to the orbital plane, and $\{\bm{i}, \bm{j}, \bm{k}\}$ where $\bm{i}$ is in the direction of the periastron $P$. The angular distance between the periastron and the node is $\omega$, and it is customary to define the longitude of the periastron as $\varpi\equiv\omega+\Omega$, although $\Omega$ and $\omega$ span different planes (see, \textit{e.g.},~\cite{BrouwerClemence}).}
\label{fig:angles}
\end{figure}

Thus the 2PN conservative dynamics~\eqref{eq:QK2PNelliptic} depends on four constants $a$, $e$, $c_\ell$ and $\omega$, to which we find it convenient to add two parameters specifying the orientation of the orbital plane, even though the orbital plane remains fixed in our case. Following the notation of celestial mechanics~\cite{BrouwerClemence} we add the inclination of the orbit $\iota$ and the longitude of the ascending node $\Omega$. Furthermore, still following standard notation, we introduce the longitude of the periastron $\varpi\equiv\omega+\Omega$ and the parameter $\bar{\varepsilon}$ defined by $\bar{\varepsilon} \equiv \varpi+c_\ell$.
The 2PN conservative solution is thus described by the set of constant parameters ($\alpha=1, \cdots, 6$)
\begin{align}\label{eq:calpha}
	c_\alpha = \bigl(a, e, \bar{\varepsilon}, \varpi, \iota, \Omega\bigr)\,.
\end{align}
Recall that in our notation $a$ and $e$ are defined in terms of the conserved energy and angular momentum~\eqref{eq:EJ} by the Newtonian formulas~\eqref{eq:Newtonianae}. Our conventions for the various celestial-mechanics angles are shown in Fig.~\ref{fig:angles}.

By the method of variation of constants, the effect of the RR perturbing acceleration~\eqref{eq:Arr} is to make the six parameters~\eqref{eq:calpha} functions of time, while the functional form of the solution~\eqref{eq:QK2PNelliptic} is maintained. We need also to maintain the functional form of $\dot{r}$, $\dot{\phi}$ (as well as $\dd\ell/\dd u$) using formulas straightforwardly deduced from~\eqref{eq:QK2PNelliptic}. Furthermore, we need to promote two angles $\ell$ and $\Lambda$ to be specific functions of time~\cite{DGI04}. We rewrite the phase angle~\eqref{eq:phielliptic} in the perturbation case as
\begin{subequations}\label{eq:phaseangle}
\begin{align}
	\label{eq:phaseanglea}
	\phi &= \omega + \Lambda + \calW\,,\\
	\text{with}\quad \calW(\ell) &\equiv K\Bigl[ V_\phi - \ell + f_\phi \sin 2V_\phi + g_\phi \sin 3V_\phi\Bigr]\,.
\end{align}
\end{subequations}
The function $\calW$ is a periodic function of time, $\calW(t+P)=\calW(t)$ where $P$ is the radial period, which is equivalent to saying that $\calW(\ell+2\pi)=\calW(\ell)$. Next, in the perturbed case, the mean anomaly $\ell$ is taken as
\begin{align}\label{eq:ell}
	\ell(t) = \int_{t_0}^t n\bigl[a(t'),e(t')\bigr]\,\dd t' + c_\ell(t)\,,
\end{align}
where $t_0$ is an irrelevant constant, and where in our notation the ``integration constant'' parameter reads $c_\ell(t)=\bar\varepsilon(t)-\varpi(t)$. As a result, $\ell$ can be viewed as a function of time and of the orbital elements, $\ell(t)=F_\ell\bigl(t,\bar\varepsilon(t),\varpi(t)\bigr)$ with $F_\ell(t,\bar\varepsilon,\varpi)\equiv\int_{t_0}^t n\bigl[a(t'),e(t')\bigr]\mathrm{d}t'+\bar\varepsilon-\varpi$. Next, the angle $\Lambda$ in~\eqref{eq:phaseanglea} is defined as the integral
\begin{subequations}\label{eq:Lambda}
	\begin{align}
       \Lambda(\ell) = \int_{\ell_0}^\ell K\bigl[a(\ell'),e(\ell')\bigr]\dd\ell'\,,
\end{align}
where $\ell_0=\ell(t_0)$, which is a short-hand for the more detailed expression
\begin{align}
	\Lambda(t) = \int_{t_0}^t K\bigl[a(t'),e(t')\bigr]\biggl(n\bigl[a(t'),e(t')\bigr]+\frac{\dd c_\ell}{\dd t'}\biggr)\dd t'\,.
\end{align}
\end{subequations}
As we see from~\eqref{eq:phaseanglea} the argument of the periastron, which reads in our notation $\omega(t) = \varpi(t) - \Omega(t)$, naturally plays the role of ``integration constant'' for the angle $\Lambda$. Likewise, through Eq.~\eqref{eq:Lambda}, $\Lambda$ is regarded as a function of $t$, $\bar\varepsilon$ and $\varpi$. The partial derivatives of these two angles are
\begin{subequations}\label{eq:derivativeslandlambda}
\begin{align}
  & \frac{\partial \ell}{\partial t} =n \, , & &  \frac{\partial \ell}{\partial \overline{\varepsilon}} = -\frac{\partial \ell}{\partial \varpi} = 1\, , & &  \frac{\partial \ell}{\partial a} = \frac{\partial \ell}{\partial e} = \frac{\partial \ell}{\partial \iota} =  \frac{\partial \ell}{\partial \Omega}  = 0\,,\\ & \frac{\partial \Lambda}{\partial t} = K \frac{\partial \ell}{\partial t}\,, & &  \frac{\partial \Lambda}{\partial c_\alpha} = K \frac{\partial \ell}{\partial c_\alpha}\,. & &
\end{align}
\end{subequations}
The perturbation equations for the parameters $c_\alpha(t)$ defined by~\eqref{eq:calpha} are
\begin{align}\label{eq:perteqs0}
	\sum_{\beta=1}^6 \bigl[c_\alpha, c_\beta\bigr]\frac{\dd c_\beta}{\dd t} = A_\alpha^\text{RR}\,,
\end{align}
and dictate the slow evolution of the parameters (with respect to a typical orbital scale) through RR effects. The Lagrange brackets are given by\footnote{For reference the Lagrange brackets corresponding to the set of parameters~\eqref{eq:calpha} are given, when restricted to Newtonian order, by~\cite{BrouwerClemence}
\begin{align*}
	&\bigl[a,\bar{\varepsilon}\bigr] = -\frac{a n}{2}\,,\qquad \bigl[a,\varpi\bigr] = \frac{a n}{2}\left(1-\sqrt{1-e^2}\right)\,,\qquad\bigl[a,\Omega\bigr] = \frac{a n}{2}\sqrt{1-e^2}\bigl(1-\cos\iota\bigr)\,,\\
	&\bigl[e,\varpi\bigr] = \frac{e a^2 n}{\sqrt{1-e^2}}\,,\qquad\bigl[e,\Omega\bigr] = -\frac{e a^2 n}{\sqrt{1-e^2}}\bigl(1-\cos\iota\bigr)\,,\qquad\bigl[\iota,\Omega\bigr] = a^2 n \sqrt{1-e^2} \sin\iota \,,
\end{align*}
all the other brackets being zero.}
\begin{align}\label{Lagrange}
	\bigl[c_\alpha, c_\beta\bigr] \equiv \frac{\partial \bm{x}}{\partial c_\alpha}\cdot\frac{\partial \bm{v}}{\partial c_\beta} - \frac{\partial \bm{x}}{\partial c_\beta}\cdot\frac{\partial \bm{v}}{\partial c_\alpha}\,.
\end{align}
The right-hand side of~\eqref{eq:perteqs0} is defined from the RR acceleration~\eqref{eq:Arr} as 
\begin{align}
	A_\alpha^\text{RR} \equiv \frac{\partial \bm{x}}{\partial c_\alpha}\cdot\bm{A}^\text{RR}\,,
\end{align}
where the dot indicates the usual Euclidean scalar product, and is composed of small 2.5PN, 3.5PN and 4.5PN terms. Here we remind that $\bm{x}$ and $\bm{v}$ are the relative position and velocity of the binary in the fixed global inertial frame with respect to which the parameters $c_\alpha$ are defined (as we said, including the parameters $\Omega$ and $\iota$ specifying the orientation of the orbital plane). By inverting the equations~\eqref{eq:perteqs0} we readily obtain
\begin{align}\label{eq:perteqs}
	\frac{\dd c_\alpha}{\dd t} = \mathcal{A}_\alpha^\text{RR} \equiv \sum_{\beta} \bigl\{c_\alpha, c_\beta\bigr\} A_\beta^\text{RR}\,,
\end{align}
where the Poisson brackets, which are the inverse of the Lagrange brackets (\textit{i.e.}, $\sum_\beta\{c_\alpha, c_\beta\}[c_\beta, c_\gamma]=\delta_{\alpha\gamma}$) read 
\begin{align}
	\bigl\{c_\alpha, c_\beta\bigr\}
	= \frac{\partial c_\alpha}{\partial \bm{v}}\cdot\frac{\partial c_\beta}{\partial \bm{x}} - \frac{\partial c_\beta}{\partial \bm{v}}\cdot\frac{\partial c_\alpha}{\partial \bm{x}}\,. 
\end{align}
The next step is to solve the perturbation equations to obtain the explicit corrections to the orbit using the post-adiabatic approximation up to relative 2PN order, following Ref.~\cite{DGI04}. The right-hand side of the equations~\eqref{eq:perteqs}, which is a small (at least 2.5PN) quantity is a periodic function of time $t$ through the mean anomaly $\ell(t)$, and is also an independent functional of the slowly varying parameters $c_\beta(t)$, hence
\begin{align}\label{eq:perteqs2}
	\frac{\dd c_\alpha}{\dd t} = \mathcal{A}_\alpha^\text{RR}\bigl[t,c_\beta(t)\bigr]\,. 
\end{align}
The function of the \emph{independent} variables $t$ and $c_\beta$,  $\mathcal{A}^\text{RR}_\alpha[t,c_\beta]$, is an algebraic combination of the $A^\text{RR}_\beta$ and of the Lagrange brackets~\eqref{Lagrange}, all of which are scalar products of the form $(\partial\bm{a}/\partial c_\alpha)\cdot\bm{b}$ or $(\partial\bm{a}/\partial c_\alpha)\cdot(\partial\bm{b}/\partial c_\beta)$, where $\bm{a}(\bm{x},\bm{v})$ and $\bm{b}(\bm{x},\bm{v})$ are functions of the position and velocity. Let us now prove that these quantities, and hence $\mathcal{A}^\text{RR}_\alpha$, are periodic in $t$ with period $P$ (up to higher-order corrections), a property which is crucial in our approach. For this purpose, it is sufficient to show that, in the appropriate comoving basis $(\bm{n},\bm{\lambda},\bm{\ell})$, with $\bm{n}=\bm{x}/r=\bm{n}(\phi)$, $\bm{\lambda}=\partial \bm{n}/\partial\phi$ and $\bm{\ell}$ normal to the orbital plane ($\bm{\ell}=\bm{w}$ in the notation of Fig.~\ref{fig:angles}), the components of $\bm{a}$ and of $\partial\bm{a}/\partial c_\alpha$ are $2\pi$-periodic in $\ell$ at fixed orbital elements. Since $\bm{x}=r\,\bm{n}$ and $\bm{v}=\dot{r}\,\bm{n}+r\,\dot{\phi}\,\bm{\lambda}$, the components of $\bm{a}$ depend only on $r$, $\dot{r}$ and $\dot{\phi}$. Those are $2\pi$-periodic in $\ell$ because, when $u\to u+2\pi$, Eq.~\eqref{defVb} gives $V_\phi\to V_\phi+2\pi$, hence $\ell\to\ell+2\pi$ by Eqs.~\eqref{eq:QK2PNelliptic}, while $r$, $\dot{r}$ and $\dot{\phi}$ remain unchanged. By contrast, note that the vector $\bm{a}$ itself is not periodic, since it depends on $\cos \phi$ and $\sin \phi$ through $\bm{n}$ and $\bm{\lambda}$, whereas the shift $\ell\to\ell+2\pi$ translates to $\phi\to\phi+2\pi K$ for the phase, due to the periastron advance. However, $\partial\phi/\partial c_\alpha$ is $2\pi$-periodic in $\ell$, because, by Eqs.~\eqref{eq:derivativeslandlambda}, the
derivative of the non-periodic piece $\Lambda$ with respect to $\bar\varepsilon$ or $\varpi$ reduces to $\pm K$, which does not depend explicitly on $\ell$. The components of $\partial\bm{a}/\partial c_\alpha$, which involve only $r$,
$\dot{r}$, $\dot{\phi}$, their derivatives with respect to $c_\alpha$, and $\partial\phi/\partial c_\alpha$,	are therefore $2\pi$-periodic, so that $\mathcal{A}^\text{RR}_\alpha[t,c_\beta]$ is $P$-periodic in $t$.

The method adopted here exploits this periodicity. It consists of splitting the solution into a part evolving slowly on the RR time scale and a part which essentially oscillates at the orbital period. Thus we pose 
\begin{equation}\label{eq:decomposition}
c_\alpha(t)=\overline{c}_\alpha(t)+\delta^\text{RR} c_\alpha(t)\,,
\end{equation}
where $\overline{c}_\alpha(t)$ is obtained from the equations when the RR force is averaged over an orbit, and where $\delta^\text{RR} c_\alpha(t)$ represents the small RR-driven corrections (containing 2.5PN, 3.5PN and 4.5PN terms) which oscillate at the orbital period. By definition the averaged evolution (on an orbital time scale) is called the adiabatic approximation and obeys the simpler differential equation 
\begin{align}\label{eq:adiabapprox}
	\frac{\dd \overline{c}_\alpha}{\dd t} = \overline{\mathcal{A}}{}_\alpha^\text{RR}\bigl[\overline{c}_\beta(t)\bigr]\,, 
\end{align}
where the right-hand side is the average of the RR force over the short orbital time scale, keeping fixed the values of the parameters, that are themselves solutions of Eq.~\eqref{eq:adiabapprox}:
\begin{align}\label{eq:average}
	\overline{\mathcal{A}}{}_\alpha^\text{RR}\bigl[\overline{c}_\beta\bigr] \equiv \frac{1}{P} \int_0^P \dd t \,\mathcal{A}_\alpha^\text{RR}\bigl[t,\overline{c}_\beta\bigr]\,,
\end{align}
where $P=2\pi/n$ is the radial period. Plugging~\eqref{eq:decomposition} into~\eqref{eq:perteqs2} we note that in the RR force which is a small, at least 2.5PN quantity, we can replace the parameters $c_\beta$ by their adiabatic approximation $\overline{c}_\beta$, since the error is at least of order of the square of RR effect, \textit{i.e.} 5PN which is negligible here. Hence the equation for the fast RR oscillations:
\begin{align}\label{eq:deltarrc}
	\frac{\dd \delta^\text{RR} c_\alpha}{\dd t} = \delta \mathcal{A}_\alpha^\text{RR}\bigl[t,\overline{c}_\beta(t)\bigr]\,,
\end{align}
where we denote $\delta \mathcal{A}_\alpha^\text{RR}[t,\overline{c}_\beta]=\mathcal{A}_\alpha^\text{RR}[t,\overline{c}_\beta]-\overline{\mathcal{A}}{}_\alpha^\text{RR}[\overline{c}_\beta]$, which has zero average over the orbital time scale. We solve Eq.~\eqref{eq:deltarrc} by requiring that the solution has zero average,
\begin{align}\label{eq:deltarrcaverage}
	\frac{1}{P} \int_0^P \dd t \,\delta^\text{RR} c_\alpha[t,\overline{c}_\beta] = 0\,.
\end{align}
Finally, we shall obtain the full solution as the sum of the adiabatic approximation $\overline{c}_\alpha$ which represents a slow, secular drift on the RR time scale much longer than the orbital period, and fast oscillations $\delta^\text{RR} c_\alpha$ that are at least of order 2.5PN and are zero on average over one orbit.  

Once the orbital elements are known, we obtain the angles $\ell$ and $\Lambda$ by inserting the expressions found for $c_\alpha=\overline{c}_\alpha + \delta^\text{RR}c_\alpha$ into Eqs.~\eqref{eq:ell}--\eqref{eq:Lambda}, then expanding linearly in the $\delta^\text{RR} c_\alpha$'s, since quadratic corrections can be safely neglected. This yields the natural splittings $\ell=\overline{\ell} + \delta^\text{RR} \ell$ and $\Lambda= \overline{\Lambda} + \delta^\text{RR} \Lambda$, where the secular parts read
\begin{subequations}
\begin{align}
  \overline{\ell}(t) &= \int_{t_0}^t n\bigl[\overline{a}(t'),\overline{e}(t')\bigr]\,\dd t' + \overline{c}_\ell(t) \,, & \overline{\Lambda}(t) &=  \int_{t_0}^t K \bigl[\overline{a}(t'),\overline{e}(t')\bigr] \left(n \bigl[\overline{a}(t'),\overline{e}(t')\bigr] + \frac{\dd \overline{c}_\ell(t')}{\dd t'}\right) \,\dd t' \,,
\end{align}
\end{subequations}
while the RR oscillatory parts are given by
\begin{subequations}
\label{eq:deltalandLambda}
\begin{align}
\delta^\text{RR}\ell(t) &= \int_{t_0}^t \delta^\text{RR} n(t') \, \dd t'+ \delta^\text{RR} c_\ell(t) \,, \\ \delta^\text{RR} \Lambda(t) &=  \int_{t_0}^t \left[K[\overline{a}(t'),\overline{e}(t')] \left( \delta^\text{RR} n(t') + \frac{\dd \delta^\text{RR} c_\ell(t')}{\dd t'} \right) +  \delta^\text{RR} K(t')\, \left( n[\overline{a}(t'),\overline{e}(t')] + \frac{\dd \overline{c}_\ell(t')}{\dd t'} \right) \right]\, \dd t' \,,
\end{align}
\end{subequations}
modulo higher order terms, the integration constants being fixed by the zero-average condition \eqref{eq:deltarrcaverage}. The functions $\delta^\text{RR} n(t)$ and $\delta^\text{RR} K(t)$ entering the above expressions are defined by linearity as
\begin{align}
\delta^\text{RR} n &= \frac{\partial n(\overline{a}, \overline{e})}{\partial \overline{a}}\,\delta^\text{RR} a + \frac{\partial n(\overline{a}, \overline{e})}{\partial \overline{e}} \,\delta^\text{RR} e \,, &
\delta^\text{RR} K &= \frac{\partial K (\overline{a}, \overline{e})}{\partial \overline{a}}\, \delta^\text{RR} a + \frac{\partial K(\overline{a}, \overline{e})}{\partial \overline{e}} \,\delta^\text{RR} e \,.
\end{align}

\subsection{Quasi-hyperbolic orbit}
\label{sec:hyperbolicic}

For the quasi-hyperbolic computation it is convenient to use some different conventions for the parameters, adopting the ones in the previous 3.5PN calculation in~\cite{BGR25}. Furthermore the presentation of the perturbation equations will closely follow that of Ref.~\cite{DGI04}, instead of the more ``celestial-mechanics'' approach for the quasi-elliptic case we followed in the previous subsection. Nevertheless we shall be able to prove that the results of the two calculations in the elliptic and hyperbolic cases are equivalent through analytic continuation. The functional form for the hyperbolic-like QK solution to the unperturbed (conservative) equations of motion at 2PN order reads
\begin{subequations}\label{sys0}
	\begin{align}
		r &= \hatS(\hatl,\hatc_1,\hatc_2)\,,\\
		\dot r &= \hatn(\hatc_1,\hatc_2) \frac{\partial \hatS(\hatl,\hatc_1,\hatc_2)}{\partial \hatl}
		\,,\\
		\phi &= \hatc_\phi + \hatW(\hatl,\hatc_1,\hatc_2) \,,\\ 
		\dot \phi &= \hatn(\hatc_1,\hatc_2) \frac{\partial \hatW(\hatl,\hatc_1,\hatc_2)}{\partial \hatl}
		\,,
	\end{align}
\end{subequations}
where we systematically endow the parameters in the hyperbolic-like case with an overhat to distinguish them from the parameters (and different conventions) used in the elliptic-like case. Here again, in the conservative case, $\hatl$ is a linear function of time and we denote $\hatl = \hatn(\hatc_1,\hatc_2)(t-t_0)+\hatc_\ell$, where $\hatc_\ell$ is an integration constant and $t_0$ is an arbitrary reference time value. Therefore, as we see from Eqs.~\eqref{sys0}, the unperturbed solution depends on the four integration constants: $\hatc_1$, $\hatc_2$, $\hatc_\ell$ and $\hatc_\phi$. These correspond to the four constants $a$, $e$, $\bar{\varepsilon}$ and $\varpi$ in the elliptic case, see Eq.~\eqref{eq:calpha}. The two other constants $\iota$ and $\Omega$ of the elliptic case simply play a spectator role.

As in the previous subsection, to obtain next-to-next-to-leading RR corrections to the orbit it is enough to consider the 2PN-accurate QK parametrization of the conservative, hyperbolic-like motion. Paralleling Eqs.~\eqref{eq:QK2PNelliptic} in the elliptic case, this is written in standard form as
\begin{subequations}
\begin{align} 
	\hatS &= \hata_r \bigl(\hate_r \cosh v-1\bigr)
	\,,\\
	\hatW &= \hatK\Bigl[\hatV_\phi+\hatf_\phi \sin 2\hatV_\phi+\hatg_\phi \sin 3\hatV_\phi\Bigr]
	\,,\\
	\hatl &= \hate_t \sinh v-v + \hatf_t \sin \hatV_\phi + \hatg_t \hatV_\phi\,,
\end{align}
\end{subequations}
where $v$ is the eccentric anomaly, and 
\begin{equation}\label{Vdef}
\hatV_\phi = 2 \arctan\Biggl[\sqrt{\frac{\hate_\phi+1}{\hate_\phi-1}}\tanh \left(\frac{v}{2}\right) \Biggr]\,.
\end{equation}
All the parameters $\hatn$, $\hatK$, $\hata_r$, $\hate_t$, $\hate_r$, $\hate_\phi$, $\hatf_t$, $\hatg_t$, $\hatf_\phi$ and $\hatg_\phi$ (with $\hatf_t$, $\hatg_t$, $\hatf_\phi$, $\hatg_\phi$ being purely 2PN) can be expressed as functions of the two integrals of motion $\hatc_1$ and $\hatc_2$. The auxiliary variable $v$ used to parametrize the orbit should be considered as a function of $\hatl$, $\hatc_1$ and $\hatc_2$, just as $\hatS$ and $\hatW$. The constants $\hatc_1$ and $\hatc_2$ may be chosen as the specific energy and angular momentum of the system, Eqs.~\eqref{eq:EJ}. Similarly to~\eqref{eq:Newtonianae}, we may also define the parameters  
\begin{align}\label{eq:Newtonianaetilde}
 	\hata \equiv \frac{G m}{2 \tilde{E}}\,,\qquad \hate \equiv \sqrt{ 1 + \frac{2\tilde{E} \tilde{J}^2}{G^2m^2}}\,.
\end{align}
For instance, we have, for the gauge invariants $\hatn$ and $\hatK$: 
\begin{subequations}\label{parametersnKhyper}
\begin{align}
 \hatn(\hata) &= \sqrt{\frac{G m}{\hata^3}}\left\{ 1 + \frac{1}{8} (15-\nu) \frac{G m}{\hata c^2} + \frac{1}{128}\Bigl[555+30\nu+11\nu^2\Bigr] \left(\frac{G m}{\hata c^2}\right)^2 \right\}\,,\\
\hatK(\hata,\hate) &= 1 + \frac{3 G m}{\hata c^2(\hate^2-1)}\left\{ 1 + \frac{1}{4}\Bigl[ 5-2\nu + \frac{35-10\nu}{\hate^2-1} \Bigr]\frac{G m}{\hata c^2} \right\}\,.
\end{align}
\end{subequations}
Notice that while in the elliptic case $n$ depends on both the energy $E$ and angular momentum $J$, with the dependence on $J$ starting at 2PN order, see Eq.~\eqref{parametersn}, in the hyperbolic case $\hat{n}$ depends only on the energy $E$. We relegate the explicit expressions of all the parameters to Appendix~\ref{app:QK}.

The perturbed solution is then assumed to have the same functional form as the background solution~\eqref{sys0} --- where the explicit time dependence for the various RR quantities is entirely expressed via the variable $\hatl$ ---, allowing then the four constants $\hatc_1$, $\hatc_2$, $\hatc_\ell$ and $\hatc_\phi$ to be functions of time, \textit{i.e.}, $\hatc_1(t)$, $\hatc_2(t)$, $\hatc_\ell(t)$ and $\hatc_\phi(t)$, and the variable $\hatl(t)$ to be given by
\begin{equation}\label{ldef}
\hatl(t)=\int_{t_0}^t \dd t' \, \hatn\bigl[\hatc_1(t'),\hatc_2(t')\bigr] + \hatc_\ell(t)\,, 
\end{equation}
which is equivalent to~\eqref{eq:ell} in the elliptic case. The evolving constants $\hatc_1(t)$, $\hatc_2(t)$, $\hatc_\ell(t)$ and $\hatc_\phi(t)$ then satisfy the following set of equations
\begin{subequations}\label{LagrangeHyp}
\begin{align}
\frac{\dd \hatc_1}{\dd t} &=  \frac{\partial{\hatc_1({\mathbf x},{\mathbf v})}}{\partial {\mathbf v}} \cdot {\bm A}^{\rm RR}
\,,\\
\frac{\dd \hatc_2}{\dd t} &=  \frac{\partial{\hatc_2({\mathbf x},{\mathbf v})}}{\partial {\mathbf v}} \cdot {\bm A}^{\rm RR}
\,,\\
\frac{\dd \hatc_\ell}{\dd t} &= -\left(\frac{\partial \hatS}{\partial \hatl}\right)^{-1}\left[\frac{\partial \hatS}{\partial  \hatc_1}\frac{\dd \hatc_1}{\dd t}+\frac{\partial \hatS}{\partial \hatc_2}\frac{\dd \hatc_2}{\dd t}\right]  
\,,\\
\frac{\dd \hatc_\phi}{\dd t} &= -\frac{\partial \hatW}{\partial \hatl}\frac{\dd \hatc_\ell}{\dd t}-\frac{\partial \hatW}{\partial \hatc_1}\frac{\dd \hatc_1}{\dd t}-\frac{\partial \hatW}{\partial \hatc_2}\frac{\dd \hatc_2}{\dd t}\, ,
\end{align}
\end{subequations}
where ${\bm A}^{\rm RR}$ denotes the (relative) RR acceleration~\eqref{eq:Arr}.
Notice that the quantities on the right-hand side of Eqs.~\eqref{LagrangeHyp} appear as multiplied by RR quantities and hence all take their unperturbed values (which for the purpose of the present paper means expansion up to the fractional 2PN order). The equations~\eqref{LagrangeHyp} can be put also in a form analogous to~\eqref{eq:perteqs} by means of Poisson brackets. Note also that in perturbations, the variations of the parameters $\hat{c}_\ell$ and $\hat{c}_\phi$ are related through the analytic continuation [see Eq.~\eqref{eq:AnalyticContinuationQKparameters}] to the corresponding variations of the parameters $c_\ell$ and $\omega$ in the quasi-elliptic case, by\footnote{Similarly, $\dd\hata_r/\dd t = - \dd a_r/\dd t$ and $\dd \hate_r/\dd t = \dd e_r/\dd t$. Strictly speaking the correspondence by analytic continuation holds only for the local parts of the acceleration and breaks down for the non-local contributions.}
\begin{subequations}\label{eq:relationcellcphi}
\begin{align}
	\frac{\dd\hatc_\ell}{\dd t} &=  \di \biggl[\bigl(1+g_t\bigr) \frac{\dd c_\ell}{\dd t} + \ell \frac{\dd g_t}{\dd t} \biggr] \,,\\
	\frac{\dd\hatc_\phi}{\dd t} &= \frac{\dd\omega}{\dd t} - \ell \frac{\dd K}{\dd t}\,.
\end{align}
\end{subequations}

Denoting by $\mathcal{F}_E$ and $\mathcal{F}_J$ the radiated energy flux and the norm of the
angular momentum flux, defined at future null infinity, the following standard balance laws hold
\begin{subequations}\label{balance_1}
	\begin{align}
		\frac{\dd E^{\rm sys}}{\dd t}+\frac{\dd E^{\rm Schott}}{\dd t} &= -\mathcal{F}_E\,,\\
		\frac{\dd J^{\rm sys}}{\dd t}+\frac{\dd J^{\rm Schott}}{\dd t} &= -\mathcal{F}_J\,,
	\end{align}
\end{subequations}
where $E^{\rm sys}$ and $J^{\rm sys}$ are the energy and angular momentum associated with the conservative part of the acceleration~\eqref{eq:Acons}, and which are conserved in the absence of radiation. This implies that the (opposite of the) far-zone fluxes are equal to the sum of the time
derivatives of the energy and angular momentum Schott terms, which represent contributions
of the gravitational field (at orders 2.5PN, 3.5PN, 4PN and 4.5PN), and of the mechanical
(\textit{i.e.}, matter-system) fluxes, given by
\begin{subequations}\label{EJsysdef}
\begin{align}
	\frac{\dd E^{\rm sys}}{\dd t} &\equiv \frac{\partial E^{\rm sys}}{\partial {\bm v}}\cdot{\bm A}^{\rm RR}\,,\\
	\frac{\dd J^{\rm sys}}{\dd t} &\equiv \frac{\partial J^{\rm sys}}{\partial {\bm v}}\cdot{\bm A}^{\rm RR}{}\,,
\end{align}
\end{subequations}
where ${\bm  A}^{\rm RR}$ is the relative RR acceleration. Integrating the fluxes over the whole scattering process then leads to the total (gauge-invariant) radiated energy and angular momentum by the system,
\begin{subequations}
	\begin{align}
E^{\rm rad} &= \int_{-\infty}^{+\infty} \dd t \,\mathcal{F}_E(t)\,,\\
J^{\rm rad} &= \int_{-\infty}^{+\infty} \dd t \,\mathcal{F}_J(t)\,.
\end{align}
\end{subequations}
We notice that the inclusion of Schott terms in the flux balance equation is crucial in the hyperbolic case, due to the non-local part of the radiation-reaction acceleration. The 4.5PN radiation Schott terms for the energy and angular momentum are given in Eqs.~(6.15) and of Ref.~\cite{BFST26}. At this stage, the Schott term associated with the RR part of the 4PN tail term is also important~\cite{T25}.

Let us go back to the set of equations~\eqref{LagrangeHyp} for the perturbed constants. Following previous works~\cite{DGI04, KG06, BDG23, BGR25}, we choose here $\hatc_1$ and $\hatc_2$ to be identified with the following two orbital parameters: $\hatc_1=\hata_r$ and $\hatc_2=\hate_r$. These quantities can be expressed in terms of the energy and angular momentum, or equivalently the parameters~\eqref{eq:Newtonianaetilde}, as 
\begin{subequations}
	\begin{align}
		\hata_r(\hata,\hate) &= \hata \left\{ 1 + \frac{1}{4} (7-\nu)\frac{G m}{\hata c^2} + \Bigl[\frac{1}{16}(1+\nu^2)+\frac{4-7\nu}{\hate^2-1}\Bigr]\left(\frac{G m}{\hata c^2}\right)^2 \right\}\,,\\
		\hate_r^2(\hata,\hate) &= \hate^2 + \frac{1}{4} \Bigl[-24+4\nu+5(\hate^2-1)(-3+\nu)\Bigr]\frac{G m}{\hata c^2} \nn\\
		&\quad + \frac{1}{8}\biggl[60+148\nu+2\nu^2-\frac{128-224\nu}{\hate^2-1}
		+(\hate^2-1)(80-45\nu+4\nu^2)\biggr]\left(\frac{G m}{\hata c^2}\right)^2\,. 
	\end{align}
\end{subequations}
We provide the complete expressions of all the parameters for both elliptic-like and hyperbolic-like cases in terms of energy and angular momentum up to 3PN order in Appendix~\ref{app:QK}. If needed, the energy $E$ and angular momentum $J$ can be expressed as functions of the original variables $\bm{x}$ and $\bm{v}$, and can be found up to 4PN order in Sec.~III of~\cite{BBFM17}. Finally in the quasi-hyperbolic case we choose to work with the four parameters ($A=1, \cdots, 4$)
\begin{align}\label{eq:ctildeA}
	\hatc_A = \bigl(\hata_r, \hate_r, \hatc_\ell, \hatc_\phi\bigr)\,.
\end{align}
The next step, as in the elliptic case, is to decompose the four varying constants $\hatc_A(t)$ as
\begin{equation}
\hatc_A(t) = \hatc^{\,\text{cons}}_A + \delta^{\rm RR}\hatc_A(t)\,,
\end{equation}
with the boundary conditions fixed so that $\delta^{\rm RR}\hatc_A(t)$ vanishes in the limit $t\to-\infty$. The system of equations~\eqref{LagrangeHyp} then yields an evolution system for the RR corrections $\delta^{\rm RR}\hatc_A(t)$, whose right-hand side is given (at our approximation) by explicit functions of time involving the unperturbed (constant) values corresponding to the conserved dynamics, say $\hatc^{\,\text{cons}}_A$. It is convenient to integrate perturbed quantities with respect to the  eccentric anomaly $v$  as follows 
\begin{equation}
\delta^{\rm RR}\hatc_A(v) = \int_{-\infty}^{v} \frac{\dd \hatc_A}{\dd t} \frac{\dd t}{\dd v}\dd v\,.
\end{equation}
The total variation between $-\infty$ and $+\infty$ is then simply given (because of our choice of boundary conditions) by the value at $v=+\infty$, which will be denoted by
\begin{equation}
	\bigl[\delta^{\rm RR}\hatc_A\bigr] \equiv \int_{-\infty}^{+\infty} \frac{\dd \hatc_A}{\dd t} \frac{\dd t}{\dd v}\dd v\,.
\end{equation}
Finally, the correction to the orbit follows from the decomposition $r(t)=r^{\rm cons}(t)+\delta^{\rm RR} r(t)$, $\phi(t)=\phi^{\rm cons}(t)+\delta^{\rm RR} \phi(t)$, with the conservative part given by the conservative dynamics~\eqref{eq:Acons}, and the RR correction explicitly given by
\begin{subequations}\label{rrorb}
\begin{align}
\delta^{\rm RR} r(t) &= \delta^{\rm RR}\hatS(\hatl, \hata_r, \hate_r)\,,\\
\delta^{\rm RR} \phi(t) &= \delta^{\rm RR}\hatc_\phi(t) + \delta^{\rm RR}\hatW(\hatl, \hata_r, \hate_r)\,,
\end{align}
\end{subequations}
where we employ the notation
\begin{subequations}\label{DX_notation}
\begin{align}
	\delta^{\rm RR} X(\hatl, \hata_r, \hate_r) &\equiv \frac{\partial X}{\partial \hatl} \delta^{\rm RR} \hatl(t)+\frac{\partial X}{\partial \hata_r} \delta^{\rm RR} \hata_r (t)
	+\frac{\partial X}{\partial \hate_r} \delta^{\rm RR} \hate_r(t)\,,\\
	\text{with}\quad\delta^{\rm RR} \hatl(t) &= \int_{t_0}^t \dd t' \,\delta^{\rm RR}\hatn(t') + \delta^{\rm RR} \hatc_\ell(t) \equiv \delta^{\rm RR}_0 \hatl(t)+\delta^{\rm RR} \hatc_\ell(t)\,,
\end{align}
\end{subequations}
the second definition following from Eq.~\eqref{ldef}. The total change of the azimuthal angle $[\delta^{\rm RR}\phi]$ gives the RR contribution to the (relative) scattering angle, \textit{i.e.},
\begin{equation}
\chi^{\rm RR} = \bigl[\delta^{\rm RR}\phi\bigr]\,.
\end{equation}
Following the decomposition~\eqref{eq:Arr} we write
\begin{equation}\label{chirr}
	\chi^{\rm RR} =  \chi^{\rm RR}_\text{2.5PN} +  \chi^{\rm RR}_\text{3.5PN} +  \chi^{\rm RR}_\text{4PN\,tail} + \chi^{\rm RR}_\text{4.5PN}+ \calO\left(\frac{1}{c^{10}}\right)\,,
\end{equation}
where we remind that the 2.5PN and 3.5PN terms are local, the 4PN tail term is non-local, while the 4.5PN term is composed of both local and non-local terms, and we may write
\begin{equation}\label{45PNchi}
	\chi^{\rm RR}_{\rm 4.5PN} = \chi^{\rm RR\,GII}_{\rm 4.5PN} + \delta\chi^{\rm RR}_{\rm 4.5PN}\,.
\end{equation}
With our definition in Eq.~\eqref{45PNsplitGII} the local part of $\chi^{\rm RR}_{\rm 4.5PN}$  corresponds exactly to the 4.5PN contribution to the acceleration (${\bm A}^{\rm RR\,GII}_\text{4.5PN}$) obtained in~\cite{GII97}, and where the GII gauge parameters are replaced by values corresponding to harmonic coordinates as computed in~\cite{BFST26}, and the non-local part follows from the supplementary acceleration~\eqref{eq:45PNnonlocGII} (${\delta \bm A}^{\rm RR}_\text{4.5PN}$) physically due to the linear momentum and CM radiation losses at that order. Note that in fact the only non-local term in~\eqref{eq:45PNnonlocGII} is the recoil integral given by Eq.~\eqref{recoil}. We will find that the (local) 2.5PN, 3.5PN, (non-local) 4PN tail and (local $+$ non-local) 4.5PN contributions agree with previous results, see Eqs.~(C1) and (C8)--(C11) of Ref.~\cite{BDG23}, which we quote below for the sake of completeness. The RR corrections to the scattering angle up to 4.5PN order are given by 
\begin{subequations}
\begin{align}
\chi^{\rm RR}_{\rm 2.5PN}(\hata_r,\hate_r)&=\left( \frac{Gm}{c^2\hata_r}\right)^{5/2}\frac{2\nu}{15(\hate_r^2-1)^{5/2}}\left[\frac{72\hate_r^4+1069\hate_r^2+134}{3\hate_r^2}+\frac{121\hate_r ^2+304}{\sqrt{\hate_r^2-1}} {\rm arccos}\left(-\frac{1}{\hate_r}\right) \right]\,,\\
\chi^{\rm RR}_{\rm 3.5PN}(\hata_r,\hate_r)&= \left( \frac{Gm}{c^2\hata_r}\right)^{7/2}\frac{\nu}{(\hate_r^2-1)^{7/2}}\left[C_2^{3.5\rm PN}{\rm arccos}^2\left(-\frac{1}{\hate_r}\right)+\frac{C_1^{3.5\rm PN}}{\sqrt{\hate_r^2-1}}{\rm arccos}\left(-\frac{1}{\hate_r}\right)+C_0^{3.5\rm PN}\right]\,,\\
\chi^{\rm RR}_{\rm 4.5PN}(\hata_r,\hate_r)&= \left( \frac{Gm}{c^2\hata_r}\right)^{9/2} \frac{\nu}{(\hate_r^2-1)^{9/2}}\left[C_2^{4.5\rm PN}{\rm arccos}^2\left(-\frac{1}{\hate_r}\right)+\frac{C_1^{4.5\rm PN}}{\sqrt{\hate_r^2-1}}{\rm arccos}\left(-\frac{1}{\hate_r}\right)+C_0^{4.5\rm PN}\right]\,,
\end{align}
\end{subequations}
\noindent where the 3.5PN and 4.5PN coefficients read respectively as
\begin{subequations}
\begin{align}
C_2^{3.5\rm PN}&= \frac{336}{5}\hate_r^2+\frac{384}{5}\,,\\
C_1^{3.5\rm PN}&= \left(\frac{2783}{420}+\frac{47}{15}\nu\right) \hate_r^4+\left(-\frac{260}{3}\nu-\frac{1507}{7}\right) \hate_r^2-\frac{1832}{15}\nu-\frac{14594}{105}\,,\\
C_0^{3.5\rm PN}&= \left(\frac{8}{5}\nu+\frac{288}{35}\right) \hate_r^4+\left(-\frac{1253}{45}\nu-\frac{1396049}{6300}\right) \hate_r^2-\frac{7498}{45}\nu-\frac{71683}{450}+\left(-\frac{64}{5}\nu+\frac{39394}{1575}\right)\frac{1}{\hate_r^2}\,,
\end{align}
\end{subequations}
\noindent and
\begin{subequations}
\begin{align}
C_2^{4.5\rm PN}&= \left(-\frac{1716}{35}+\frac{94}{5}\nu\right) \hate_r^4+\left(-\frac{10008}{35}-\frac{2624}{5}\nu\right) \hate_r^2-480\nu+\frac{16904}{35}\,,\\
C_1^{4.5\rm PN}&= \left(\frac{9}{20}\nu^2+\frac{7783}{840}\nu+\frac{82489}{1680}\right) \hate_r^6+\left(\frac{49}{3}\nu^2+\frac{48821}{84}\nu-\frac{417001}{3780}\right) \hate_r^4+\left(\frac{514}{5}\nu^2+\frac{427622}{105}\nu-\frac{1607}{63}\right) \hate_r^2\nn\\
&+88\nu^2+\frac{19066}{15}\nu-\frac{19882}{27}\,,\\
C_0^{4.5\rm PN}&=\left(-\frac{2}{5}\nu^2+\frac{242}{35}\nu+\frac{808}{45}\right) \hate_r^6+\left(\frac{1367}{180}\nu^2+\frac{72587}{2520}\nu+\frac{28987039}{176400}\right) \hate_r^4+\left(\frac{365}{6}\nu^2+\frac{72257}{18}\nu-\frac{147017953}{793800}\right) \hate_r^2\nn\\
&+\frac{5956}{45}\nu^2-\frac{98228321}{99225}+\frac{1299217}{630}\nu+\left(\frac{36}{5}\nu^2-\frac{56108}{315}\nu+\frac{16847071}{99225}\right)\frac{1}{\hate_r^2}\,,
\end{align}
\end{subequations}
while the 4PN tail contribution is given in a large-$\hate_r$ expanded form as follows
\begin{equation}
\label{chi_rr_tail}
\chi_{4\rm PN\, tail}^{\rm RR}(\hata_r,\hate_r)=\left(\frac{Gm}{c^2 \hata_r}\right)^4  \nu  \left[\frac{7168}{45\hate_r^5} +\frac{573\pi^3}{20\hate_r^6}+\left(\frac{4096}{9}+\frac{153856}{675}\pi^2\right) \frac{1}{\hate_r^7}+\mathcal{O}\left(\frac{1}{\hate_r^8}\right)\right]\,.
\end{equation}
Notice that up to the 4.5PN level the RR scattering angle~\eqref{45PNchi} was already computed in Ref.~\cite{BDG23} through the linear-response formula~\cite{BD12}, involving the radiative losses of energy and angular momentum. Time-asymmetric RR effects (such as the RR-squared term) 
instead start contributing at the (absolute) 5PN order, beyond our present 4.5PN goal. Here we find that the direct computation using the 4.5PN RR acceleration~\cite{BFST26} recovers the expected result from the linear-response formula, Eq.~(13.4) in~\cite{BDG23}, with $\chi^{\rm RR}_{\rm 4.5PN} = \chi^{\rm RR\,GII}_{\rm 4.5PN}$.
In fact, we will show  below that there is no contribution to the scattering angle coming from the supplementary term in~\eqref{45PNchi}: 
\begin{equation}\label{45PNchinonloc0}
	\delta \chi^{\rm RR}_\text{4.5PN} = 0\,.
\end{equation}

\section{Results for the local part} 
\label{sec:reslocal} 

After employing the method of Section~\ref{sec:Lagrange}, the solutions are straightforwardly obtained using the explicit indefinite integrals listed in Appendix~\ref{app:formulas} for the elliptic-like case.
For simplicity, we only present the results of the oscillatory parts of the six orbital elements in the quasi-elliptic case at leading order,
\begin{subequations}\label{OscilatoryPartEllipticLocal}
	\begin{align}
		\delta^\text{RR} a &= - \Bigl(\frac{G m}{a c^2}\Bigr)^{5/2} \frac{2 a e \nu \sin u}{45 \bigl(1 -  e^2\bigr)^{3}} \left[  \frac{602 + 673e^2}{\chi} + \frac{(1-e^2)(314 + 111 e^2)}{\chi^2} + \frac{2(1-e^2)^2(49 + 36 e^2)}{\chi^3} + \frac{210(1-e^2)^3}{\chi^4} \right] \nn\\
         & -\frac{2 a \nu(96+292 e^2 + 37 e^4)}{15 (1-e^2)^{7/2}} \Bigl(\frac{G m}{a c^2}\Bigr)^{5/2} (2 w + e \sin u) + \calO\left(\frac{1}{c^7}\right)\,,\\
		\delta^\text{RR} e &= - \Bigl(\frac{G m}{a c^2}\Bigr)^{5/2} \frac{\nu \sin u}{45 \bigl(1 -  e^2\bigr)^{2}} \left[  \frac{134 + 1069 e^2 + 72 e^4}{\chi} + \frac{134 + 157 e^2 -291 e^4}{\chi^2} + \frac{2(1-e^2)^2(49 + 36 e^2)}{\chi^3} \right. \nn \\
        & \left. + \frac{210(1-e^2)^3}{\chi^4} \right]  -\frac{e \nu(304+121 e^2)}{15} \Bigl(\frac{G m}{a c^2(1-e^2)}\Bigr)^{5/2} (2 w + e \sin u) + \calO\left(\frac{1}{c^7}\right)\,,\\
		\delta^\text{RR} \bar{\varepsilon} &= \Bigl(\frac{G m}{a c^2}\Bigr)^{5/2} \frac{\beta \nu}{45 e\sqrt{1-e^2}} \left[  -\frac{288\big(1-e^2 + \sqrt{1-e^2}\big)}{\chi} + \frac{12\big(43 - 43e^2 + 40 \sqrt{1-e^2}\big)}{\chi^2} - \frac{8(14 - 23 e^2 + 9 e^4)}{\chi^3} \right. \nn \\
        & \left. + \frac{210(1-e^2)^2}{\chi^4} -\frac{192 + 96 e^2 - 288 e^4}{1-e^2} -\frac{326 - 51 e^2}{\sqrt{1-e^2}}\right] + \calO\left(\frac{1}{c^7}\right)\,,\\
		\delta^\text{RR} \varpi &= \Bigl(\frac{G m}{a c^2}\Bigr)^{5/2} \frac{\nu \sqrt{1-e^2}}{45 e^2} \left[   \frac{36}{\chi^2} - \frac{8(14 - 9 e^2)}{\chi^3} + \frac{210(1-e^2)}{\chi^4} -\frac{134 + 141 e^2 }{(1-e^2)^{3/2}}\right] + \calO\left(\frac{1}{c^7}\right) \,,\\
        \delta^\text{RR} \Omega &= \delta^\text{RR} \iota= 0\,,
	\end{align}
\end{subequations}
where $\chi\equiv1-e\cos u$, $w\equiv\frac{1}{2}(V-u)$ and $\beta$ are given by~\eqref{eq:defV2}. Next, using the above results, we have integrated as well the oscillatory parts of $\ell$ and $\Lambda$, see Eqs.~(\ref{eq:deltalandLambda}). At leading order, they both coincide: $\delta^\text{RR} \ell = \delta^\text{RR} \Lambda$, thus we only display the first
\begin{align}
		\delta^\text{RR} \ell &= -\left( \frac{G m}{a c^2} \right)^{5/2} \frac{\nu}{45}\Bigg\{ \frac{1}{e^2}\left[ \frac{6e^2(97-12e^2)}{(1-e^2)\chi} + \frac{36 - 201 e^2}{\chi^2} - \frac{8(14- 23 e^2 + 9 e^4)}{\chi^3} + \frac{210(1-e^2)^2}{\chi^4}\right] \nn \\
        & + \frac{9 (96 + 292e^2 +37 e^4)}{(1-e^2)^{7/2}}\left[ \chi -\frac{\chi^2}{2} + 2\Re[\mathrm{Li}_2(\beta\de^{\di u})] + 2 e w \sin u \right]  - \frac{3(602 + 673e^2)}{(1-e^2)^3}\left[ \chi + \ln\left( \frac{2\chi}{1+\sqrt{1-e^2}}\right) \right] \nn \\
        & +\frac{3(866 + 1621 e^2 + 39 e^4 + 24 e^6)}{(1-e^2)^3} - \frac{134 + 1514 e^2 + 2201 e^4 - 24 e^6}{e^2(1-e^2)^{5/2}}\Biggr\}+ \calO\left(\frac{1}{c^7}\right)\,,
\end{align}
and the dilogarithm function $\mathrm{Li}_2(z)$ is defined by~\eqref{eq:Li2}. The full results to NNLO for the evolution equations and their solutions are given in the Supplemental Material~\cite{ancillary}. Notice that besides the usual functions $\chi^{-n}$, $\sin u \,\chi^{-n}$ and $w \,\chi^{-n}$, the results contain as well non-trivial functions of the form $w^2 \chi^{-n}$, $\ln \chi\,\chi^{-n}$ and $\Re[\mathrm{Li}_2(\beta\de^{\di u})]\chi^{-n}$. 
A similar structure and formulas hold also for hyperbolic-like orbits. Note that once the solution is obtained, say for the hyperbolic case ($A=1, \cdots, 4$)
\begin{equation}
	\delta^{\rm RR}\hatc_A = \delta^{\rm RR}\hatc_{A\,\text{2.5PN}} + \delta^{\rm RR}\hatc_{A\,\text{3.5PN}} + \delta^{\rm RR}\hatc_{A\,\text{4.5PN}} \,, 
\end{equation}
(with all terms given in the Supplemental Material~\cite{ancillary}) the perturbed orbit $\delta^{\rm RR}r(t)$ and $\delta^{\rm RR}\phi(t)$ can then be easily computed by Eqs.~\eqref{rrorb}--\eqref{DX_notation}, with $\delta^{\rm RR}\hatl(t)$ given in~\cite{ancillary}. To present our results in the hyperbolic case we use the convenient notation $\hatchi_r(v) \equiv \hate_r \cosh v -1$ ($v$ being the eccentric anomaly in the hyperbolic case) together with 
\begin{subequations}
	\begin{align}
		\hatalpha_r &\equiv\sqrt{\frac{\hate_r+1}{\hate_r-1}}\,,\\
		\hatV_r(v) &\equiv 2 \arctan\left[\hatalpha_r \tanh\left(\frac{v}{2}\right)\right]\,,\\
		\hatA_r(v) &\equiv \frac{\hatV_r(v)}{2} + \arctan\hatalpha_r\,,
	\end{align}
\end{subequations}
where the definition of the function $\hatV_r$ parallels that in Eq.~\eqref{Vdef}.
We also use 
\begin{subequations}
	\begin{align}
		{\mathfrak L}(v) &= - \frac{\di}{2}\Bigl[{\rm Li}_2(z) - {\rm Li}_2(z^*)\Bigr]\,,\\
		\text{with}\quad z &\equiv  \de^{v} \hatbeta_r 
		= -\text{exp}\bigl[v+2\di\,{\rm arctan}(\hatalpha_r)\bigr]\,,
	\end{align}
\end{subequations}
where $\hat\beta_r=\frac{1-\di \sqrt{\hate_r^2-1}}{\hate_r}$ is the analogue of $\beta_r = \frac{1-\sqrt{1-e_r^2}}{e_r}$ in the elliptic case. Here $z^*$ denotes the complex conjugate and ${\rm Li}_2(z)$ is the dilogarithm function. The quantity ${\mathfrak L}(v)$ is real-valued, and can be written in terms of the Clausen function of the second order (see Ref.~\cite{BGR25}).

We find it convenient to distinguish in the 4.5PN RR acceleration ${\bm A}^{\rm RR}_\text{4.5PN} = {\bm A}^{\rm RR\,GII}_\text{4.5PN} + {\delta \bm A}^{\rm RR}_\text{4.5PN}$, Eq.~\eqref{45PNsplitGII}, a local and a non-local part, ${\bm A}^{\rm RR}_\text{4.5PN} = {\bm A}^{\rm RR\, loc}_\text{4.5PN} + {\bm A}^{\rm RR \, nonloc}_\text{4.5PN}$, with
\begin{subequations}
\begin{align}
{\bm A}^{\rm RR\, loc}_\text{4.5PN} &= {\bm A}^{\rm RR\,GII}_\text{4.5PN}  + {\delta \bm A}^{\rm RR\,\text{no}\,\Pi}_\text{4.5PN}\,,\\
\label{Arrnonloc1}
{\bm A}^{i\,\rm RR\, nonloc}_\text{4.5PN} &\equiv  {\delta \bm A}^{i\, \rm RR\, \Pi}_\text{4.5PN} =  \frac{G \sqrt{1-4\nu}}{r^2c^2}\left(2 n^iv^j+n^jv^i\right) \Pi^j\,,
\end{align}
\end{subequations}
where $\delta A^{i\,{\rm RR}}_\text{4.5PN}=\delta A^{i\,{\rm RR}\,\Pi}_\text{4.5PN}+\delta A^{i\,{\rm RR\, no}\,\Pi}_\text{4.5PN}$ is defined in Eq. \eqref{eq:45PNnonlocGII}. We will adopt hereafter the same decomposition for all relevant quantities. For example, we denote
\begin{equation}
\delta \chi^{\rm RR}_\text{4.5PN} = \chi^{{\rm RR}\, \Pi}_\text{4.5PN} +  \chi^{{\rm RR\, no}\, \Pi }_\text{4.5PN}\,.
\end{equation}
As anticipated above, the total change of the azimuthal angle over the whole scattering process due to the GII part of the RR acceleration  yields the expected value  $\chi^{\rm RR}_{\rm 4.5PN}$ of the (relative) radiation-reacted scattering angle, whereas the remaining (no $\Pi$) part gives
\begin{align}
    \chi^{\rm RR \,  no \,\Pi}_{\rm 4.5PN}
    &=-\left(\frac{G m}{c^2 \hata_r}\right)^{9/2}\frac{\nu^2(1-4\nu)}{\hate_r^2(\hate_r^2-1)^{9/2}}\left[
	\frac{\hate_r^2}{\sqrt{\hate_r^2-1}}\left(\frac{74 \hate_r^6}{15}+\frac{2552 \hate_r^4}{21}+\frac{8784 \hate_r^2}{35}+\frac{2432}{35}\right){\rm arccos}\left(-\frac{1}{\hate_r}\right)\right.\nn\\
	&\left.
	+\frac{196654 \hate_r^6}{3675}+\frac{563188 \hate_r^4}{2205}+\frac{1492784 \hate_r^2}{11025}+\frac{28544}{11025}
	\right]
	\,.
\end{align} 
Integrating the kinematical part of the fluxes, Eq. \eqref{EJsysdef}, gives 
\begin{subequations}
	\begin{align}
	 E_{\rm sys}^{\rm loc} &\equiv  E_{\rm sys}^{\rm GII}+\delta   E_{\rm sys}^{{\rm 4.5PN\,no}\,\Pi}
	\,,\\
	 J_{\rm sys}^{\rm loc} &\equiv  J_{\rm sys}^{\rm GII}+\delta   J_{\rm sys}^{{\rm 4.5PN\,no}\,\Pi}
	\,,
	\end{align}
\end{subequations}
with $E_{\rm sys}^{\rm GII}=-E_{\rm rad}$, $J_{\rm sys}^{\rm GII}=-J_{\rm rad}$, and
\begin{subequations}\label{deltaEJsysloc}
	\begin{align}
	\delta  E_{\rm sys}^{{\rm 4.5PN\,no}\,\Pi} &= m c^2\left(\frac{Gm}{c^2\hata_r}\right)^{11/2}\frac{\nu^3(1-4\nu)}{(\hate_r^2-1)^{11/2}}\left[
	\left(\frac{37 \hate_r^8}{15}+\frac{8207 \hate_r^6}{105}+\frac{700 \hate_r^4}{3}+\frac{4344 \hate_r^2}{35}+\frac{928}{105}\right){\rm arccos}\left(-\frac{1}{\hate_r}\right)\right.\nn\\
	&\left.
	+\sqrt{\hate_r^2-1}\left(\frac{334469 \hate_r^6}{11025}+\frac{2233799 \hate_r^4}{11025}+\frac{418786 \hate_r^2}{2205}+\frac{29448}{1225}\right)
	\right]
	\,,\\
	\delta  J_{\rm sys}^{{\rm 4.5PN\,no}\,\Pi} &= \frac{Gm^2}{c} \left(\frac{Gm}{c^2\hata_r}\right)^4\frac{\nu^3(1-4\nu)}{(\hate_r^2-1)^4}\left[
	\left(\frac{37 \hate_r^6}{15}+\frac{1604 \hate_r^4}{35}+\frac{952 \hate_r^2}{15}+\frac{928}{105}\right){\rm arccos}\left(-\frac{1}{\hate_r}\right)\right.\nn\\
	&\left.
	+\sqrt{\hate_r^2-1}\left(\frac{36499 \hate_r^4}{1575}+\frac{119662 \hate_r^2}{1575}+\frac{33784}{1575}\right)
	\right]
	\,.
\end{align}
\end{subequations}
The fractionally 2PN-accurate instantaneous radiated energy $  E_{\rm rad}=  E_{\rm rad}^{\rm 2PN,\,inst}$ and angular momentum $  J_{\rm rad}=  J_{\rm rad}^{\rm 2PN,\,inst}$ can be found  in Ref.~\cite{BDG21}, Eqs.~(C7)--(C9) and Eqs.~(E4)--(E6), respectively.

\section{Results for the Non-local terms} 
\label{sec:resnonlocal} 

\subsection{The non-local term associated with the source's recoil at 4.5PN}

As we have seen in Eq.~\eqref{eq:45PNnonloc} the RR acceleration at the 4.5PN order in the center-of-mass frame contains the recoil of the source due to GW emission, given by the non-local (or semi-hereditary\footnote{Following the terminology of Ref.~\cite{BD92}, we refer to integrals given by time anti-derivatives of local terms as ``semi-hereditary''; applying multiple time derivatives to a semi-hereditary term reduces it to be local.}) integral of the total flux of linear momentum $\calF_{\bm{P}}^i$, \textit{i.e.} 
\begin{equation}\label{Piintegral}
	\Pi^i(t) =  \int_{- \infty}^t \dd \tau \calF_{\bm{P}}^i(\tau) \,.
\end{equation}
The flux of linear momentum takes the general form in the CM frame $\bm{\calF}_{\bm{P}} = a\,\bm{n} + b\,\bm{v}$, where the coefficients $a$ and $b$ are functions of the phase space variables $r$, $\dot{r}$ and $\dot{\phi}$ (but not of the phase $\phi$ itself), where $r$ is the radial separation, $\dot{r}=\bm{n}\cdot\bm{v}$ and the orbital frequency $\dot{\phi}$ enters through $v^2=\dot{r}^2+r^2\dot{\phi}^2$. The coefficients $a$ and $b$ can be read off from Eq.~\eqref{eq:fluxPexpl} at the leading 3.5PN order and from Eq.~\eqref{eq:alpha45PN} at the next-to-leading 4.5PN order. 

\subsubsection{Quasi-elliptic case}
\label{sec:4p5PNnonlocalelliptic}

We start by presenting the standard PN integration of the balance equation $\dd\Pi^i/\dd t={\cal F}_{\bm{P}}^i$ for the recoil, following the derivation of Ref.~\cite{Fit83}. At the leading order the flux is 3.5PN, and given at that order by Eq.~\eqref{eq:fluxPexpl}, so that one can consider a Newtonian elliptic orbit. This is only a lowest-order approximation, since starting at 2.5PN the orbit undergoes a decay during the inspiral due to RR, so that one should allow the orbital elements to evolve to compute the flux to get a more physically meaningful description of the process. However, as we shall see later, this is a higher-order effect which can be neglected here.

Expressing the recoil integral~\eqref{Piintegral} using the Newtonian QK representation (and therefore, neglecting the orbital precession) and using the formulas in App.~\ref{app:formulas} to integrate over the eccentric anomaly $u$, the result in the inertial basis $(\bm{i}, \bm{j})$ (with $\bm{i}$ pointing in the constant direction of the periastron) is~\cite{Fit83} 
\begin{align}\label{eq:Pifitchett}
	\bm{\Pi} &= \frac{G^4 m^5 \delta \nu^2}{a^4 c^7 \chi^6} \Biggl\{ \frac{4}{315 e}\biggl(-45 - 45 e^4 - 15 e^2 \biggl [-6 + \bigl(-2 + \chi \bigr) \chi \biggr] + \chi \biggl [-30 + \chi \bigl(15 + 4 \chi + 56 \chi^4\bigr)\biggr]\biggr) \bm{i}\nn\\
	& - \frac{1}{1260 \bigl(1 -  e^2\bigr)^4 }\Biggl(42 e \bigl(312 + 456 e^2 + 37 e^4\bigr) \chi^6 \Bigl(u+2\arctan\Bigl[\frac{\beta\sin u}{1-\beta \cos u}\Bigr]\Bigr)\nn\\
	& + 2 \sqrt{1-e^2} \Biggl[360 \bigl(1 -  e^2\bigr)^5 + 600 \bigl(1 -  e^2\bigr)^4 \chi + 6 \bigl(1 -  e^2\bigr)^3 \bigl(80 + 81 e^2\bigr) \chi^2 + 14 \bigl(1 -  e^2\bigr)^2 \bigl(32 + 129 e^2\bigr) \chi^3 \nn\\
	& + 7 \chi^4 \Biggl(64 \bigl(1 + \chi \bigr) + e^2 \biggl [566 - 3 e^2 \bigl(173 + 37 e^2 - 283 \chi \bigr) + 1502 \chi \biggr]\Biggr)\Biggr] \sin u\Biggr)\bm{j} \Biggr\} + \bm{\Pi}_0\,,
\end{align}
where $\beta \equiv \frac{1-\sqrt{1-e^2}}{e}$ and $\chi = 1 - e \cos u$. Although there is an apparent divergence when $e\to 0$ in~\eqref{eq:Pifitchett}, the circular orbit limit is finite and given by
\begin{align}\label{eq:Picirc}
	\bm{\Pi}_\text{circ} = -\frac{464}{105} \frac{G^4 m^5 \delta \nu^2}{a^4 c^7}\,\bm{n}\,,
\end{align}
since, as we shall see in Sec.~\ref{sec:secular} below, the constant $\bm{\Pi}_0$ vanishes in this case. The circular orbit limit can also be retrieved directly from $\bm{\calF}_{\bm{P}} = a\,\bm{n} + b\,\bm{v}$, noticing that in this case $a$ and $b$ are constant, and that $\dot{\bm{n}}=\bm{v}/r$ and $\dot{\bm{v}}=-r \dot{\phi}^2\bm{n}$ which permits the direct integration, with result $\bm{\Pi}_\text{circ} = r b_\text{circ}\bm{n}$.

The first problem with the formula~\eqref{eq:Pifitchett} is that, since the integration has been performed using the \textit{indefinite} integrals in App.~\ref{app:formulas}, the result is obtained up to an uncontrolled constant vector denoted $\bm{\Pi}_0$. To solve this problem we show that the contribution to the equations of motion of the constant vector $\bm{\Pi}_0$ is actually pure gauge. Indeed from Eq.~\eqref{eq:45PNnonloc} we know that this constant will affect the RR CM acceleration through the term 
\begin{equation}\label{eq:45PNPI0}
	A^{i\,{\rm RR}}_\text{4.5PN\,$\Pi_0$} = \frac{G \delta}{r^2c^2}\left(2 n^iv^j+n^jv^i\right)\Pi_0^j \,,
\end{equation}
but~\eqref{eq:45PNPI0} takes the form of a gauge transformation (or shift) of the acceleration, namely it reads
\begin{align}\label{eq:45PNnonloc0}
	A^{i\,{\rm RR}}_\text{4.5PN\,$\Pi_0$}  = \frac{\dd^2 \xi^{i}}{\dd t^2} - \xi^j \frac{\partial A^i}{\partial x^j} + \cdots \,,
\end{align}
where to the lowest order $\bm{A}\equiv \bm{A}^\text{cons}_\text{N}$ denotes the Newtonian conservative acceleration~\eqref{eq:Acons} and higher order corrections are indicated by dots. The gauge vector in~\eqref{eq:45PNnonloc0} is given by
\begin{align}\label{eq:xi}
	\xi^{i} = - \frac{\delta}{c^2 m}\,v^i x^j \Pi_0^j + \cdots \,.
\end{align}
Thus we conclude that the arbitrary constant $\bm{\Pi}_0$ is harmless as it can be gauged away and will never appear in the Poincaré invariant quantities. However it should take a definite value in each coordinate system.

The second problem, pointed out in Ref.~\cite{Fit83}, is that the integrated component in the direction $\bm{i}$ (the $x$-component) of the total recoil integral~\eqref{eq:Pifitchett} oscillates with the orbital period, whereas the component along $\bm{j}$ ($y$-component) contains both periodic contributions and a secular contribution (proportional to $u$) that therefore accumulates with each revolution. The presence of the secular term at 3.5PN order, although formally correct, results from neglecting the orbital precession and implies that the result is valid over a time scale long with respect to the orbital period but short with respect to the precession period. Another drawback is that, in order to perform the computation including the secular term of Eq.~\eqref{eq:Pifitchett}, one must reconsider the post-adiabatic formalism of Sec.~\ref{sec:elliptic} because it assumes the functions in Eq.~\eqref{eq:deltarrc} to be periodic. We shall see in Sec.~\ref{sec:secular} how to remedy this behaviour and obtain an expression of the recoil integral which will be valid over a much longer period, long with respect to the precession period but of course still small with respect to the RR period. The treatment of the secular term is reported in Sec.~\ref{sec:secular} below.

\subsubsection{Quasi-hyperbolic case}

Let us consider the 4.5PN non-local (\textit{i.e.}, $\Pi$) contribution to the RR acceleration, Eq. \eqref{Arrnonloc1}, \textit{i.e.}
\begin{equation}\label{ArrPi}
	\delta{\mathbf A}^{i\, {\rm RR}\,\Pi}_{\rm 4.5PN}=\frac{G \sqrt{1-4\nu}}{r^3 c^2}\left(2 x^iv^j+x^jv^i\right)\hat{\Pi}^j\,,
\end{equation}
with $\hat{\Pi}^i=\hat{\Pi}^i(t)$ denoting the integrated flux of the linear momentum components given by the non-local-in-time integral (6.1a) of Ref. \cite{BFST26}. 
In terms of the auxiliary variable $v$ we find
\begin{subequations}\label{eq:hyperbolicPIresult}
	\begin{align}
		\hat\Pi^x(v)&=-\frac{\sqrt{1-4\nu}}{c^7}\left(\frac{Gm}{\hata_r}\right)^4\frac{m\nu^2}{\hate_r}\left[\frac{16}{315\hatchi_r^3}+\frac{4(\hate_r^2-1)}{21\hatchi_r^4}+\frac{8(\hate_r^2-1)}{21\hatchi_r^5}+\frac{4(\hate_r^2-1)^2}{7\hatchi_r^6}\right]
		\,,\\
		\hat\Pi^y(v)&=- \frac{\sqrt{1-4\nu}}{c^7}\left(\frac{Gm}{\hata_r}\right)^4\frac{m\nu^2}{(\hate_r^2-1)^{7/2}}\left\{
		\left(\frac{37 \hate_r^4}{15}+\frac{152 \hate_r^2}{5}+\frac{104}{5}
		\right)\frac{\hate_r\hat{A}_r(v)}{\sqrt{\hate_r^2-1}} \right.\nn\\
		&+\left(\frac{283 \hate_r^4}{30}+\frac{751 \hate_r^2}{45}+\frac{32}{45}
		\right)\frac1{\hate_r}\left(1+\frac{\hate_r\sinh(v)}{\hatchi_r}\right)+\frac{\sinh(v)}{\hatchi_r}\left[
		\left(\frac{37 \hate_r^4}{30}+7 \hate_r^2+\frac{32}{45}\right)\frac{\hate_r^2-1}{\hatchi_r}
		\right.\nn\\
		& \left.\left.
        +\left(\frac{43 \hate_r^2}{15}+\frac{32}{45}\right)\frac{(\hate_r^2-1)^2}{\hatchi_r^2}
		+\left(\frac{27 \hate_r^2}{35}+\frac{16}{21}\right)\frac{(\hate_r^2-1)^3}{\hatchi_r^3}
		+\frac{20}{21}\frac{(\hate_r^2-1)^4}{\hatchi_r^4}
        +\frac{4}{7}\frac{(\hate_r^2-1)^5}{\hatchi_r^5}
		\right]
		\right\}
		\,.
	\end{align}
\end{subequations}
The result for the recoil integral~\eqref{eq:hyperbolicPIresult} in the hyperbolic case agrees through the analytic continuation with the result~\eqref{eq:Pifitchett} in the elliptic case modulo a constant.

For the contribution to the equations of motion we find
\begin{subequations}
	\begin{align}
		\delta{\hat{\mathbf A}}^{{\rm RR}\,x}_{\rm 4.5PN\,\Pi}&=
        \left(\frac{Gm}{\hata_r^2} \right) \sqrt{\frac{Gm}{c^2 \hat a_r}} \frac{1}{mc}
         \frac{\sqrt{1-4\nu}}{\hatchi_r^4}\left[
		3\sinh(v)(\cosh(v)-\hate_r)\hat\Pi^x
		-\sqrt{\hate_r^2-1}(3\cosh(v)^2-2\hate_r\cosh(v)-1)\hat\Pi^y
		\right]
		\,,\\
		\delta{\hat{\mathbf A}}^{{\rm RR}\,y}_{\rm 4.5PN\,\Pi}&=
        \left(\frac{Gm}{\hata_r^2} \right) \sqrt{\frac{Gm}{c^2 \hat a_r}} \frac{1}{mc}
       \frac{\sqrt{1-4\nu}}{\hatchi_r^4}\sqrt{\hate_r^2-1}\left[
		-(3\cosh(v)^2-\hate_r\cosh(v)-2)\hat\Pi^x
		+3\sqrt{\hate_r^2-1}\sinh(v)\cosh(v)\hat\Pi^y
		\right]
		\,.\nn\\
	\end{align}
\end{subequations}
The (negative) total variation of $\hat\Pi^i(v)$ along the orbit gives the components of the radiated linear momentum $\hat P_{\rm rad}^i=-[\hat \Pi^i]$, with $\hat P_{\rm rad}^x=0$ and
\begin{equation}
	\hat P_{\rm rad}^y =mc \left(\frac{Gm}{c^2 \hata_r}\right)^4\frac{\nu^2\sqrt{1-4\nu}}{(\hate_r^2-1)^4} \left[
	\left(\frac{104}{5}+\frac{152}{5}\hate_r^2+\frac{37}{15}\hate_r^4\right)\hate_r\arccos\left(-\frac1{\hate_r}\right)
	+\frac{(\hate_r^2-1)^{1/2}}{\hate_r}\left(\frac{64}{45}+\frac{1502}{45}\hate_r^2+\frac{283}{15}\hate_r^4\right)
	\right]\,,
\end{equation}
in agreement with Eqs. (G3)--(G5) of Ref.~\cite{BDG21}.

The solutions for the RR corrections $\delta^{\rm RR}\hat c_A^{\Pi}(t)$ are listed in  \cite{ancillary}. The contribution to the relative radiation-reacted scattering angle 
$ \chi^{{\rm RR}\, \Pi}_\text{4.5PN} =[\delta^{\rm RR} \phi]^{\rm 4.5PN}_{\Pi} $ turns out to be 
\begin{equation}
	[\delta^{\rm RR} \phi]^{\rm 4.5PN}_{\Pi} = [\delta^{\rm RR} \hat c_\phi^{\Pi}]-\frac{[\delta^{\rm RR} \hat e_r^{\Pi}]}{\hat e_r\sqrt{\hat e_r^2-1}} =-[\delta^{\rm RR} \phi]^{\rm 4.5PN}_{{\rm no}\,\Pi}=-\chi^{{\rm RR\, no}\, \Pi}_\text{4.5PN}\,,
\end{equation}
as mentioned above.

Integrating the kinematical part of the fluxes gives
\begin{subequations}
\label{deltaEJsysPI}
	\begin{align}
		\delta   E_{\rm sys}^{{\rm 4.5PN}\,\Pi}&=mc^2 \left(\frac{Gm}{c^2 \hata_r}\right)^{11/2}\frac{\nu^3(1-4\nu)}{(\hate_r^2-1)^{11/2}}\left[
		\left(-\frac{5533 \hate_r^6}{105}-\frac{4063 \hate_r^4}{15}-\frac{4736 \hate_r^2}{35}+\frac{1256}{105}\right){\rm arccos}\left(-\frac{1}{\hate_r}\right)\right.\nn\\
		&\left.
		+\frac{\sqrt{\hate_r^2-1}}{\hate_r^2}\left(-\frac{126464 \hate_r^8}{11025}-\frac{2281819 \hate_r^6}{11025}-\frac{104249 \hate_r^4}{441}+\frac{23866 \hate_r^2}{3675}+\frac{64}{45}\right)
		\right]
		\,,\\
		\delta   J_{\rm sys}^{{\rm 4.5PN}\,\Pi}&= \frac{Gm^2}{c} \left(\frac{Gm}{c^2\hata_r}\right)^4 \frac{\nu^3(1-4\nu)}{(\hate_r^2-1)^4}\left[
		\left(-\frac{1879 \hate_r^4}{105}-\frac{1096 \hate_r^2}{15}-\frac{3112}{105}\right){\rm arccos}\left(-\frac{1}{\hate_r}\right)\right.\nn\\
		&\left.
		+\frac{\sqrt{\hate_r^2-1}}{\hate_r^2}\left(-\frac{6784 \hate_r^6}{1575}-\frac{32269 \hate_r^4}{525}-\frac{9346 \hate_r^2}{175}-\frac{64}{45}\right)
		\right]
		\,.
	\end{align}
\end{subequations}
At the 4.5PN level the Schott terms associated with the energy and angular momentum have a non-local (\textit{i.e.}, $\Pi$) part given in Eqs.~(6.15a) and~(6.15b) of Ref.~\cite{BFST26}, respectively. 
Evaluating these terms at $t\to+\infty$ then gives 
\begin{subequations}
\label{EJschottPI}
	\begin{align}
		  E_{\rm Schott}^{{\rm 4.5PN}\,\Pi}&=-\frac{\sqrt{\hate_r^2-1}}{\hate_r}c\left(\frac{Gm}{c^2 \hata_r}\right)^{3/2} \nu \sqrt{1-4\nu} \hat P_{\rm rad}^y
		\,,\\
		 J_{\rm Schott}^{{\rm 4.5PN}\,\Pi}&=-\frac{\hate_r^2-1}{\hate_r}\frac{Gm}{c^2} \nu \sqrt{1-4\nu} \hat P_{\rm rad}^y
		\,.
	\end{align}
\end{subequations}
Note that the following relation holds 
\begin{equation}
	\frac{  E_{\rm Schott}^{{\rm 4.5PN}\,\Pi}}{  J_{\rm Schott}^{{\rm 4.5PN}\,\Pi}}=\frac{c^3}{Gm}\left(\frac{Gm}{c^2 \hata_r }\right)^{3/2}\frac1{\sqrt{\hate_r^2-1}}
	=\frac{c^3}{Gm} \frac{p_\infty^2}{j} \,,
\end{equation}
where at the leading-order $j=\sqrt{\frac{c^2 \hat{a}_r (\hat{e}_r^2-1)}{G m}}$ denotes the dimensionless angular momentum and  $p_\infty= \sqrt{\frac{G m}{\hat{a}_r c^2}}$.

It is easy to see that the contribution to the energy and angular momentum due to the Schott terms, Eq. \eqref{EJschottPI}, exactly cancels that associated with the (local + non-local) 4.5PN RR acceleration $\delta A_{\rm RR}^i$, Eqs. \eqref{deltaEJsysloc} and \eqref{deltaEJsysPI}, \textit{i.e.}, 
\begin{subequations}
	\begin{align}
		\delta   E_{\rm sys}^{{\rm 4.5PN\,no}\,\Pi}+\delta   E_{\rm sys}^{{\rm 4.5PN}\,\Pi}&=- E_{\rm Schott}^{{\rm 4.5PN}\,\Pi}
		\,,\\
		\delta   J_{\rm sys}^{{\rm 4.5PN\,no}\,\Pi}+\delta   J_{\rm sys}^{{\rm 4.5PN}\,\Pi}&=-  J_{\rm Schott}^{{\rm 4.5PN}\,\Pi}
		\,.
	\end{align}
\end{subequations}

\subsection{Multi-scale treatment of the secular term in the recoil for the elliptic case}
\label{sec:secular} 

As we shall now detail, a major difference arises with the non-local time integration in~\eqref{Piintegral}, in that it induces a ``PN order reduction'' in the bound elliptic case, with respect to the computation of local terms. The explicit expression of $\bm{\Pi}$ for elliptic orbits contains a 2.5PN contribution, even if the flux $\bm{\calF}_{\bm{P}}$ is 3.5PN before integration. Although this effect may be surprising at first glance, it is perfectly consistent in the appropriate limit with the previous result~\eqref{eq:Pifitchett}, and will allow us to compute the gauge constant vector $\bm{\Pi}_0$ in~\eqref{eq:Pifitchett} corresponding to harmonic coordinates. 

To proceed, it is convenient to introduce the moving orthonormal basis $\{\bm{n},\bm{\lambda}\}$ in the orbital plane, such that the velocity reads $\bm{v} = \dot{r} \bm{n} + r \dot{\phi} \bm{\lambda}$, and we further define the complex basis $\{\bm{m},\bm{m}^*\}$ such that $\bm{m} = \frac{1}{\sqrt{2}}(\bm{n}+ \di \bm{\lambda})$ together with its complex conjugate $\bm{m}^*$ (we have $\bm{m}^2={\bm{m}^*}^2=0$ and $\bm{m}\cdot\bm{m}^*=1$). With this notation the flux becomes
\begin{subequations}
	\begin{align}
		\bm{\calF}_{\bm{P}} &= \alpha \,\bm{m} + \text{c.c.}\,,\\
		\text{where}\quad\alpha &= \frac{1}{\sqrt{2}}\left( a + b \dot{r} - \di \,b \,r \dot{\phi}\right)\,.
	\end{align}
\end{subequations}
Anticipating the PN order reduction, we need the explicit expression of $\alpha$ at 4.5PN,
\begin{align}\label{eq:alpha45PN}
\alpha(r,\rd,\dot\phi) &= \frac{4\sqrt{2} G^3m^4\delta\nu^2}{105 c^7r^4}\Biggl\{
5r^2\rd\fid^2 - 2 \rd^3 + 4\frac{Gm}{r}\bigl(\rd + 2\mathrm{i} r\fid\bigr)
+ 2\mathrm{i}\bigl(6r\rd^2\fid + 25r^3\fid^3\bigr)  \nn \\
&\qquad + \frac{1}{18c^2}\biggl[\frac{G^2m^2}{r^2}\Bigl(2(74\nu-129) \rd - 8\mathrm{i} (189+17\nu) r\fid\Bigr) \nn \\
&\qquad\qquad + \frac{Gm}{r}\Bigl(4(53\nu-197) \rd^3 + 6(235\nu-832) r^2\rd\fid^2
+ 6\mathrm{i} (248\nu-11) r \rd^2\fid + 9\mathrm{i} (162\nu-907) r^3\fid^3\Bigr) \nn \\
&\qquad\qquad + 12(13-4\nu) \rd^5 + 6(70\nu-97) r^2 \rd^3\fid^2 + 9(259-174\nu) r^4\rd\fid^4 \nn \\
&\qquad\qquad + \mathrm{i}\Bigl(12(35\nu-83) r \rd^4\fid + 18(83\nu-345) r^3 \rd^2\fid^3
+ 111(25-28\nu) r^5\fid^5\Bigr)\biggr]\Biggr\} + \mathcal{O}\Bigl(\frac{1}{c^{11}}\Bigr)\,.
\end{align}
To perform the integration~\eqref{Piintegral} we express the moving basis at the time $\tau$ in terms of the basis at the current time $t$ and the elapsed phase difference, namely\footnote{We define $\psi\equiv\phi-\omega$ where $\omega$ is constant in the adiabatic approximation, see Eq.~\eqref{eq:phielliptic}. By definition of the phase angle $\psi$ we have
\begin{align*}
	\bm{n}(\tau) &= \bm{n}(t) \cos{\big(\psi(t)-\psi(\tau)\big)} - \bm{\lambda}(t) \sin{\big(\psi(t)-\psi(\tau)\big)}\,, \\
	\bm{\lambda}(\tau) &= \bm{n}(t) \sin{\big(\psi(t)-\psi(\tau)\big)} + \bm{\lambda}(t) \cos{\big(\psi(t)-\psi(\tau)\big)}\,.
\end{align*}}
\begin{align}\label{eq:mpast}
	\bm{m}(\tau) = \bm{m}(t) \,\text{exp}\bigl[\di \bigl(\psi(t)-\psi(\tau)\bigr)\bigr]\,,
\end{align}
which allows us to write the recoil integral in the form 
\begin{align}\label{eq:Piadvanced}
	\bm{\Pi}(t) =  \bm{m}_0 \int_{-\infty}^t \dd \tau\, \alpha(\tau) \,\text{exp}\bigl[ - \di \psi(\tau)\bigr] + \text{c.c.}\,.
\end{align}
where we define the constant vector $\bm{m}_0 \equiv \bm{m}(t) \,\text{exp}[\di  \psi(t) ]$.

Next, we can express $\alpha(r,\dot{r},\dot{\phi})$ in terms of the orbital elements $a_r$, $e_\phi$ and the eccentric anomaly $u$ using the QK parametrization of the motion in~\eqref{eq:QK2PNelliptic}, and we recall that $a_r$ and $e_\phi$ are related to the Newtonian definitions~\eqref{eq:Newtonianae} by the formulas in App.~\ref{app:QK}. For the practical computation, we decompose following~\eqref{eq:phaseangle} the orbital phase as
\begin{align}\label{eq:phase_adapted_expansion}
	\psi(t) = \Lambda(t)+\calW(t) = K\, V_{\phi}(t) + \mathcal{D}(t) + \mathcal{P}(t)\,,
\end{align}
where $\mathcal{D} \equiv \Lambda - K\,\ell$ vanishes for the conservative dynamics and $\mathcal{P}\equiv K[f_\phi \sin 2V_\phi + g_\phi \sin 3V_\phi]$ is a periodic function of period $P$ that starts at the 2PN order. We shall show the following properties: (i) $\mathcal{P}$ can be neglected in the integral computation,  (ii) the orbital elements $\bar{a}_r$, $\bar{e}_{\phi}$ and $\bar{k}$ may be considered constant in those integrals, (iii) the lower bound ($\tau \rightarrow -\infty$) of the integral~\eqref{eq:Piadvanced} does not contribute to the integral value at this PN order. First, as it is periodic (and actually continuous), $\mathcal{P}$ remains bounded at all times, so the factor $\text{exp}[\di(\mathcal{P}(t)-\mathcal{P}(\tau))]$ can be expanded in the usual PN way and, then, be neglected.
Then, for simplicity, we shall use $V_\phi$ directly as the integration variable instead of $u$, as we found that this is the most convenient variable for the integration. Neglecting the 2PN periodic contribution $\mathcal{P}$ to the phase, the recoil $\bm{\Pi}$  is given by
\begin{align}
	\label{eq:alphaqaev}
	\bm{\Pi}(t) = \bm{m}_0 \sum_{q = -q_0}^{q_1} \int_{- \infty}^{t} \dd \tau\,  \frac{\dd V_\phi}{\dd\tau} \beta_q(\bar{a}_{r}(\tau),\bar{e}_{\phi}(\tau))\de^{\di (\Phi_q(\tau)+\bar{\omega}(t))}+ \text{c.c.}\,,
\end{align}
where we have defined $\Phi_q(\tau) = (q-1) V_\phi(\tau) - \bar{\omega}(\tau) - \bar{k}(\tau) V_\phi(\tau) - \mathcal{D}(\tau) = q V_\phi - \phi + \mathcal{P}$, neglecting relative 2.5PN corrections in the amplitudes and the phase, and factorized (for later convenience) the ``Jacobian'' of the transformation $\tau\longrightarrow V_\phi(\tau)$, \textit{i.e.}
\begin{align}\label{eq:jacobian}
	\frac{\dd V_\phi}{\dd \tau} = \frac{n \sqrt{1-e_\phi^2}}{(1-e_t\cos u)(1-e_\phi\cos u)} = \frac{n}{\sqrt{1-e_\phi^2}} \frac{(1 + e_\phi\cos V_\phi)^2}{1-e_t e_\phi+(e_\phi - e_t)\cos V_\phi}\,.
\end{align}
To prove (ii) and (iii), we compute $\dd \Phi_q/\mathrm{d}\tau = [(q-1)-\bar{k}] \dd V_\phi/ \dd \tau + \mathcal{O}(c^{-5})$. Because $\dd V_\phi/\dd \tau$ is always positive, it follows that, whenever $q\neq 1$, the phase $\Phi_q$, for which $\dd \Phi_q(t) / \dd t\simeq (q-1)\, \dd V_\phi(t)/\dd t $, is strictly monotonic in $t$, increasing for $q>1$, and decreasing for $q<1$. For the slowly oscillating mode $q=1$,  $\dd \Phi_1(t) / \dd t \simeq -\bar{k}(t)\dd V_\phi(t) / \dd t < 0 $, since $\bar{k}>0$, so that $\Phi_1(t)$ is also monotonic. This implies that the relation $z=\xi \,|\Phi_q(t)-\Phi_q(\tau)|$, with $\xi = \dot{\bar{n}}(t)/\bar{n}^2(t)$ being a small adiabatic parameter that compensates for the rapid oscillations of $\Phi_q(\tau)$, always constitutes an admissible change of variable. It can be checked that the amplitudes $\frac{\dd V_\phi}{\dd\tau} \beta_q(\bar{a}_{r}(z),\bar{e}_{\phi}(z))$ are actually slowly varying. Applying the same techniques as detailed in Sec.~V\,B of Ref.~\cite{BFHLT23b} to compute the integral given by Eq.~(5.8), which has the same structure as Eq.~\eqref{eq:alphaqaev}, except for the presence of an extra logarithmic factor, we conclude that properties (ii) and (iii) hold.
Finally, the  $\mathcal{D}$ contribution can also be safely dropped since its contribution to the phase difference $\psi(t)-\psi(\tau)$ vanishes in the adiabatic approximation.

To summarize, we have just shown that, for the phase, we can simply take $\psi = \phi-\omega = K\, V_{\phi}$, where $V_\phi$ is defined by~\eqref{eq:defV}. In terms of $a_r$, $e_\phi$ and $V_\phi$, the coefficients $\alpha(r,\dot{r},\dot{\phi})$ in~\eqref{eq:Piadvanced} take the form
\begin{equation}\label{eq:alphaaev}
	\alpha(a_r,e_\phi,V_\phi) =  \frac{\dd V_\phi}{\dd\tau} \sum_{q=-q_0}^{q_1} \beta_q (a_r,e_\phi) \, \de^{\di q V_\phi} \,.
\end{equation}
This factorization is always possible (modulo a small PN expansion of~\eqref{eq:jacobian} to re-express $e_t$ in terms of $e_\phi$) as the flux $\bm{\calF}_{\bm{P}}$ is made of powers of the type $(1 + e_\phi \cos V_\phi)^p$ with $p \geq 2$,\footnote{This is a consequence of the form of $\alpha(\tau)$ which is a sum of products of $1/\chi_\phi$ and $\sin u/\chi_\phi$, which are given by
\begin{equation} \label{eq:amplitude_structure}
	\frac{1}{\chi_\phi} = \frac{1 + e_\phi \cos V_\phi}{1-e_\phi^2}\,, \qquad
	\frac{\sin u}{\chi_\phi} = \frac{\sin V_\phi}{\sqrt{1-e_\phi^2}}\,,
\end{equation}
where $\chi_{\phi} \equiv 1 - e_{\phi} \cos u$, so that its expansion on the basis $\de^{\di q V_{\phi}}$ is straightforward.
} and the constant bounds $q_0$ and $q_1$ of the sum~\eqref{eq:alphaaev} depend solely on the PN order of $\alpha$. Notice that the sum in~\eqref{eq:alphaaev} is finite and that this is just a convenient expression of $\alpha$ and not an infinite Fourier decomposition. For the sake of readability, we do not provide the explicit expressions of the $\beta_q$, which are quite lengthy. 

The oscillatory structure of $\alpha$ in~\eqref{eq:alphaaev} combines with the structure of the phase $\psi=K V_\phi$,  which gives
\begin{align}\label{eq:PiSourceGen}
	\bm{\Pi} =  \bm{m}_0 \sum_{q=-q_0}^{q_1} \int_{-\infty}^{V_\phi} \dd v \, \beta_q \bigl[a_r(v),e_\phi(v)\bigr]\,\text{exp}\bigl[ \di\bigl(q-K(v)\bigr) v\bigr] + \text{c.c.}\,.
\end{align}
where we have performed the change of variable $\tau\longrightarrow v=V_\phi(\tau)$. We shall compute the integral~\eqref{eq:PiSourceGen} in the adiabatic limit, \textit{i.e.}, assuming that the parameters $a_r(\tau)$, $e_\phi(\tau)$ and $K(\tau)$ can be replaced by their adiabatic counterparts which evolve on the RR time scale, evaluated at the current time: $\overline{a}_r$, $\overline{e}_\phi$ and $\overline{K}=1+\overline{k}$. Furthermore, recalling that each term of the integral is oscillating on a much shorter time scale than the secular evolution of the orbital elements $(\overline{a}_r,\overline{e}_\phi,\overline{k})$,  those can be considered constant. As a result, we can write 
\begin{align}\label{eq:PiSourceGenbis}
	\bm{\Pi} =  \bm{m}_0 \sum_{q=-q_0}^{q_1} \beta_q \bigl(\overline{a}_r, \overline{e}_\phi\bigr) \int_{-\infty}^{V_\phi} \dd v\, \,\text{exp}\bigl[ \di\bigl(q-1-\overline{k}\bigr) v\bigr] + \text{c.c.}\,.
\end{align}
It remains to compute the integral. Since the orbital elements are slowly varying we can use in the adiabatic approximation the formula
\begin{align}\label{eq:Idef}
	\int_{-\infty}^{V_\phi} \dd v\, \,\text{exp}\bigl[ \di\bigl(q-1-\overline{k}\bigr) v\bigr] =  \frac{\text{exp}\bigl[ \di\bigl(q-1-\overline{k}\bigr) V_\phi\bigr]}{\di(q-1-\overline{k})}\,,
\end{align}
and we obtain the recoil in the form
\begin{align}\label{eq:PiSourceGen3}
	\bm{\Pi} =  \bm{m}_0 \sum_{q=-q_0}^{q_1} \beta_q  \frac{\text{exp}\bigl[ \di\bigl(q-1-\overline{k}\bigr) V_\phi\bigr]}{\di(q-1-\overline{k})} + \text{c.c.}\,.
\end{align}
The justifications of the arguments leading to the result~\eqref{eq:PiSourceGen3}, as well as generalizations involving a logarithmic kernel, are discussed in a similar context in Sec.~IV of~\cite{ABIQ04} and Sec.~V\,B of Ref.~\cite{BFHLT23b}, where the method for including post-adiabatic corrections is presented in detail in the case of quasi-circular orbits. See notably Eq.~(5.18) of~\cite{BFHLT23b} and Eq.~(4.22) of~\cite{ABIQ04}. We recall that the bound $-\infty$ of the initial integral~\eqref{eq:PiSourceGen} does not play any role in the adiabatic approximation and does not affect the result~\eqref{eq:PiSourceGen3}.

As mentioned above, the main feature of this computation concerns the slowly oscillating terms $q=1$ because they yield a $1/\overline{k} = \calO(c^2)$ factor which reduces the order of those terms in the PN expansion.\footnote{The fact that the slowly oscillating terms arise from the mixing between the oscillatory part of $\alpha$ in Eq.~\eqref{eq:alphaaev} and the phase factor $\exp(- \di \psi)$ containing the precession is crucial to the PN order reduction. Such terms cannot appear in the hyperbolic case because $\alpha$ does not contain an oscillatory part when expressed in terms of $(\hat{a}_r,\hat{e}_r, \hat{V}_\phi)$ since all exponentials are real. This means that a combination $(\hat{K}-1)\hat{V}_\phi$ cannot arise in an exponential. Thus, it is sufficient to consider the leading-order flux in the unbound hyperbolic case instead of the next-to-leading flux in the bound case.} For consistency, $\overline{k}$ must include its 2PN term (\textit{i.e.} be known at relative 1PN order) to be consistent at the 3.5PN order in $\bm{\Pi}$. On the other hand, one can take $\overline{k}=0$ for all terms with $q\neq 1$ since its contribution will be of higher PN order. Hence
\begin{align}\label{eq:PiSourceGen4}
	\bm{\Pi} &= \bm{m}_0 \biggl[ \beta_1 \frac{\de^{-\di\overline{k} V_\phi}}{-\di \overline{k}} + \sum_{q\not= 1} \beta_q \frac{\de^{\di(q-1) V_\phi}}{\di(q-1)}\biggr] + \text{c.c.}\,.
\end{align}
The method leading to~\eqref{eq:PiSourceGen4} differs from the standard PN calculation in Sec.~\ref{sec:4p5PNnonlocalelliptic} only in the computation of the $\beta_1$ term which is thus responsible for the 1PN order reduction. In standard harmonic coordinates, the full result to 3.5PN expressed back in the comoving basis $\{\bm{n},\bm{\lambda}\}$ reads, in terms of $a$ and $e$
\begin{align}\label{eq:PiEllipticRes}
	&\bm{\Pi} = \frac{G^3 m^4 \delta \nu^2}{a^3 c^5 \chi (1
		- e^2)^{3}} \Biggl\{\frac{1}{90}\bigl(312
	+ 456 e^2
	+ 37 e^4\bigr)  \biggl [\bigl(-1
	+ e^2 + \chi \bigr) \bm{n}
	+ e \sqrt{1
		-  e^2} \sin\bigl(u\bigr) \bm{\lambda} \biggr] \\
	&\quad - \frac{1}{15120 e^2(1-e^2)\chi^4} \frac{G m }{a c^2}  \Biggl[\Biggl( - 144 e^2 \bigl(1
	- e^2\bigr)^5
	- 7728 \bigl(1
	- e^2\bigr)^4 e^2 \chi  - 24 \bigl(836
	+ 291 e^2\bigr) \bigl(1
	- e^2\bigr)^3 e^2 \chi^2 \nn\\
	&\quad- 21 \bigl(1
	- e^2\bigr)^2 \biggl [37 e^6 \bigl(-15 + \nu \bigr)
	- 312 \bigl(9
	+ \nu \bigr)
	- 16 e^2 \bigl(331
	+ 9 \nu \bigr)
	+ e^4 \bigl(-4221
	+ 419 \nu \bigr)\biggr] \chi^3\nn\\
	&\quad - \bigl(1
	- e^2\bigr) \biggl [6552 \bigl(9
	+ \nu \bigr)
	+ 3 e^8 \bigl(456
	+ 1813 \nu \bigr)
	- 8 e^2 \bigl(34547+ 3465 \nu \bigr)
	+ 21 e^4 \bigl(3913
	+ 12633 \nu \bigr)
	+ 4 e^6 \bigl(15160
	+ 44919 \nu \bigr)\biggr] \chi^4\nn\\
	&\quad + \biggl [-351808
	+ 229236 e^2
	+ 59632 e^4
	- 10287 e^6
	+ 84 \bigl(-444+ 3227 e^2
	+ 2253 e^4
	+ 74 e^6\bigr) \nu \biggr] e^2 \chi^5\Biggr) \bm{n} \nn\\
	&\quad + e \sqrt{1 -  e^2} \Biggl( 8784 e^2 \bigl(1
	- e^2\bigr)^4
	+ 30912 \bigl(1
	- e^2\bigr)^3 e^2 \chi
	+ 504 \bigl(124 + 37 e^2\bigr) \bigl(1
	- e^2\bigr)^2 e^2 \chi^2 \nn\\
	&\quad + 21 \bigl(1
	- e^2\bigr) \biggl [37 e^6 \bigl(-15
	+ \nu \bigr) - 312 \bigl(9
	+ \nu \bigr)
	- 16 e^2 \bigl(113
	+ 9 \nu \bigr)
	+ e^4 \bigl(-1269
	+ 419 \nu \bigr)\biggr] \chi^3\nn\\
	&\quad + \biggl [3 e^6 \bigl(456
	+ 1813 \nu \bigr)
	+ 84 e^2 \bigl(1921
	+ 3263 \nu \bigr)
	- 8 \bigl(37949  + 3843 \nu \bigr)
	+ e^4 \bigl(67633
	+ 180453 \nu \bigr)\biggr] e^2 \chi^4\Biggr) \sin\bigl(u\bigr) \bm{\lambda} \Biggr]\Biggr\} \,,\nn
\end{align}
where for convenience we used the Newtonian definitions of the parameters $a$ and $e$ in~\eqref{eq:Newtonianae}, which, as we discussed before, should be understood as representing their constant parts only.

We observe that the expression~\eqref{eq:PiEllipticRes} is purely oscillating and there is no explicit secular term as in Eq.~\eqref{eq:Pifitchett}. Hence, this result will stay accurate over time intervals much longer than both the orbital and the precession periods. The price to pay is the presence of the leading 2.5PN term given by the first line in~\eqref{eq:PiEllipticRes}. Note that this 2.5PN term affects the evolution equations for the orbital parameters at 3.5PN and must therefore be added to the local contribution obtained in \cite{KG06} to complete the RR corrections in harmonic coordinates already at 3.5PN.
If we formally re-expand the phase factors in this 2.5PN term, assuming that the precession is a small PN term, we recover the secular term of the standard PN calculation in Eq.~\eqref{eq:Pifitchett}. What remains is a constant which can thus be identified with the constant $\bm{\Pi}_0$ in Eq.~\eqref{eq:Pifitchett} which is therefore determined in this approach. Furthermore, we have seen in Eqs.~\eqref{eq:45PNPI0}--\eqref{eq:xi} that $\bm{\Pi}_0$ is a gauge constant and therefore the procedure here will give its value in the harmonic coordinates we are using, but it could be set to zero in the appropriate coordinate system. In conclusion, the above computation is fully consistent with the standard calculation in Sec.~\ref{sec:4p5PNnonlocalelliptic}, while remaining valid over time scales much longer than the precession period and conserving the initial harmonic coordinates. 

In detail, the constant we obtain (corresponding to our harmonic coordinate system) is
\begin{align}\label{eq:Piconst}
	&\bm{\Pi}_0 = \frac{G^3 m^4 \delta \nu^2 e_r}{a_r^3 c^5 (1
		- e_r^2)^{3}} \Biggl\{
	- \frac{1}{90}\bigl(312 + 456 e_r^2 + 37 e_r^4\bigr) \\
	&+ \frac{1}{15120(1 - e_r^2)}\frac{G m}{a_r c^2}\Bigl[734144 - 44352 \nu - 336 e_r^2 \bigl(-4261 + 30 \nu \bigr) + e_r^6 \bigl(-30363 + 6216 \nu \bigr) + 8 e_r^4 \bigl(28202 + 8967 \nu \bigr)\Bigr] \Bigg\}\bm{i}\,,\nn
\end{align}
which is manifestly constant and is along the periastron direction $\bm{i}$ (see Fig.~\ref{fig:angles}). In Eq.~\eqref{eq:Piconst} we have employed the QK parameters $a_r$ and $e_r$ to emphasize that this constant term is actually zero in the case of a circular orbit where $e_r=0$.\footnote{Although the parametrization by the Newtonian-like $a$ and $e$ is convenient in our practical calculations, one should keep in mind that the limit of a circular orbit is pathological with the variable $e$ since the circular-orbit condition [see Eq.~\eqref{eq:QK2PNelliptica}] is $e_r=0$ which implies
\begin{align*}
	e_\text{circ} =\di \sqrt{\left(\frac{G m}{a c^2}\right)\frac{9+\nu}{8}}+\mathcal{O}\left(\frac{1}{c^{3}}\right)\,.
\end{align*}
}

Since the recoil integral contains a term at order 2.5PN due to the above PN order reduction, to remain consistent at the 4.5PN order in the equations of motion and the evolution equations of the orbital elements, one must compute the contribution of $\bm{\Pi}$ at 5.5PN in the CM acceleration. To do so, one writes the relative acceleration ${\bm A} \equiv {\bm A}_1-{\bm A}_2$, and consistently impose at next-to-leading order the CM condition $\bm{G}= -\bm{\Gamma}$, or equivalently $\bm{P} = - \bm{\Pi}$ (using the notations of~\cite{BFT25}). In practice, we iteratively replace the individual positions and velocities in the acceleration using relations of the form $\bm{y}_1 = \bm{x} \,m_2/m -\delta \bm{G}/(m c^2)-\bm{\Gamma}/m$ and $\bm{v}_1 = \bm{v} \,m_2/m  - \delta \bm{P}/(m c^2)-\bm{\Pi}/m$. The result, extending Eq.~\eqref{eq:45PNnonloc} to the next order (see also Eq.~(6.10) of~\cite{BFST26}), can be expressed solely in terms of the relative position, velocity and $\bm{\Pi}$ as
\begin{subequations}\label{eq:ARRinonloc55PN}
	\begin{align}
		A^{i\,{\rm RR}}_\text{4.5PN\,$\Pi$} &= \frac{G \delta}{r^2c^2}\Big[2 n^i (v\Pi)+v^i (n\Pi)\Big]\,,\\
		A^{i\,{\rm RR}}_\text{5.5PN\,$\Pi$} &= \frac{G \delta}{r^2c^4}\Bigg[ n^{i} \left( \frac{3 \nu}{2} (nv) (n\Pi)\big(3 v^2-5 (nv)^2 \big)  + (v\Pi) \left(3\nu (nv)^2 -4\nu v^2 -\frac{G m}{r}(8-\nu) \right)\right) \nn\\
		& \qquad + v^i \left((n\Pi) \left( \frac{9\nu}{2}(nv)^2 +(2-\nu)\left(\frac{v^2}{2} - \frac{2G m}{r}\right)\right)-(4 + \nu) (n v) (v\Pi)\right)\Bigg]\, ,
	\end{align}
\end{subequations}
where $(n\Pi)=\bm{n}\cdot\bm{\Pi}$ and $(v\Pi)=\bm{v}\cdot\bm{\Pi}$. Note that after inserting the explicit solution~\eqref{eq:PiEllipticRes} we have found for $\bm{\Pi}$, the acceleration $A^{i\,{\rm RR}}_\text{5.5PN\,$\Pi$}$ has exactly the same structure as its local-in-time counterpart and can be treated in the same way as those of Sec.~\ref{sec:reslocal}. The full results are given in the Supplemental Material~\cite{ancillary}.

Finally, we return to the fact that the 3.5PN term in the RR acceleration (thus induced by the 2.5PN contribution in $\bm{\Pi}$) is actually a gauge term that can be removed by performing a suitable gauge transformation. We already found in Eqs.~\eqref{eq:45PNnonloc0}--\eqref{eq:xi} that the contribution of the constant part of $\bm{\Pi}$ is pure gauge at the lowest, 3.5PN, order in the RR acceleration. Here we extend the argument by considering the gauge transformation at the next 4.5PN order and computing it with the \textit{exact} expression~\eqref{eq:PiEllipticRes}, without any PN re-expansion of the phase variable. The expression of the gauge vector generalizing~\eqref{eq:xi} to the next order, thus composed of 3.5PN and 4.5PN terms, reads 
\begin{align}\label{eq:xiNLO}
	\xi^{i} = - \frac{r \delta}{c^2 m}\left[\left(1 + \frac{\nu}{c^2}\left(-\frac{3}{2}v^2+2\frac{G m}{r}\right)\right)v^i + \frac{G m}{2r c^2}\dot{r} \nu n^i\right] (n\Pi) \,.
\end{align}
This expression has been obtained by recasting the RR acceleration~\eqref{eq:ARRinonloc55PN} into a pure gauge transformation in the case of the constant $\bm{\Pi}_0$, and then by defining it in~\eqref{eq:xiNLO} with the full expression of $\bm{\Pi}$ given by~\eqref{eq:PiEllipticRes}. The associated gauge transformation of the RR acceleration at the requested order is given by
\begin{align}\label{eq:45PNnonloc0gauge}
	\delta_\xi A^{i} = \frac{\dd^2 \xi^{i}}{\dd t^2} - \xi^j \frac{\partial A^i}{\partial x^j} - \frac{\dd \xi^{j}}{\dd t} \frac{\partial A^i}{\partial v^j}  \,,
\end{align}
where the conservative acceleration~\eqref{eq:Acons} on the right-hand side also contains the 1PN correction, $\bm{A}\equiv \bm{A}^\text{cons}_\text{N}+\bm{A}^\text{cons}_\text{1PN}$. We find that this gauge transformation exactly cancels the 3.5PN term in the RR acceleration, so that after the gauge transformation ($A'^{i} = A^{i} - \delta_\xi A^{i}$) the 3.5PN terms arising from the 4.5PN acceleration through the 1PN order reduction of $\bm{\Pi}$ disappear, which again shows that they are merely an artifact of the harmonic coordinates we use. 

We conclude this section with a non-trivial consistency check: we compare the equations
describing the secular evolution of the orbital elements we have obtained with
those derived from the flux-balance equations through the chain-rule relation
\begin{equation}\label{eq:fluxbalance}
	\frac{\dd \bar{c}_\alpha}{\dd t} = - \frac{\partial \bar{c}_\alpha}{\partial E}\langle \calF_E \rangle - \frac{\partial \bar{c}_\alpha}{\partial J}\langle \calF_J \rangle\,,
\end{equation}
where $\calF_E$ is the total energy flux and $\calF_J$ the norm of the angular momentum flux. Inserting the expressions of the fluxes to 2PN relative order in the same consistent gauge as the QK parametrization (here in harmonic coordinates), from, \textit{e.g.},~\cite{ABIQ08,ABIS09} or the Supplemental Material of~\cite{H26}, one must recover the same results as the Lagrange method of variation of constants up to some possible Schott terms. We combine the local and non-local contributions to the secular orbital elements derived with the Lagrange method and compare them with those obtained from Eq.~\eqref{eq:fluxbalance}. We recover exactly the same expressions without needing to introduce Schott terms. Notice that if the 5.5PN term of the non-local acceleration in~\eqref{eq:ARRinonloc55PN} is omitted, the test fails. This means that the reduced-PN-order term of $\bm{\Pi}$, inserted into the non-local 5.5PN term of the acceleration, is crucial for a consistent calculation of the orbital evolution in harmonic gauge.

\subsection{The dissipative part of the tail term at 4PN}

The 4PN non-local contribution to the RR acceleration has been given by Eq.~\eqref{eq:tail} which we repeat here:
\begin{subequations}\label{eq:Integral4PNtail}
\begin{align}
	A^{i\,{\rm RR}}_{\text{4PN\,tail}} &= -\frac{4G^2M}{5 c^8} \,r n^j
	\,\mathcal{T}_{ij}(t)\,,\\
	\mathcal{T}_{ij}(t) &\equiv \int_0^{+\infty} \!\dd\tau
	\ln\left(\frac{\tau}{\tau_0}\right)\left[I_{ij}^{(7)}(t-\tau) +
	I_{ij}^{(7)}(t+\tau)\right]\,,\label{eq:Tijdef}
\end{align}
\end{subequations}
where $M$ is the ADM mass of the system and $I_{ij}$ is the source quadrupole moment of the binary. Contrary to the non-local terms of Section~\ref{sec:resnonlocal}, no PN order reduction can occur in the elliptic case. This is due to the fact that the integrand (besides the logarithmic kernel) is a total time derivative, which is not the case of Eq.~\eqref{Piintegral}. Thus, it is sufficient to consider the Newtonian expression of $I_{ij}$, and set $M=m$, $K = 1$ in both elliptic and hyperbolic cases.

Because of the time derivatives of the quadrupole there is no constant mode and the tail integral is confined to the orbital plane, hence the Cartesian components are 
\begin{align}\label{eq:tailcomponent}
	A^{i\,{\rm RR}}_{\text{4PN\,tail}} = -\frac{4G^2M}{5 c^8}
	\!\left(\begin{array}{l} \displaystyle x\mathcal{T}_{xx} + y\mathcal{T}_{xy}\\[0.5cm] \displaystyle x\mathcal{T}_{xy} + y\mathcal{T}_{yy}\end{array}\right)\,.
\end{align}

\subsubsection{Quasi-elliptic case}

We introduce a decomposition similar to the one in the non-local part at the 4.5PN order in the elliptic case (see Sec.~\ref{sec:secular}) to express the quadrupole in the complex basis $\{\bm{m}, \bm{m}^*, \bm{\ell}\}$ where $\bm{m}=\frac{1}{\sqrt{2}}(\bm{n}+\di\bm{\lambda})$ and $\bm{\ell}$ is the unit vector orthogonal to the orbital plane. The Newtonian quadrupole reads in this basis
\begin{equation}\label{eq:quadmm*}
	I_{ij} = m\nu x^{\langle i}x^{j\rangle} = \frac{1}{2} m \nu r^2 \Bigl[ m^{ij} + m_*^{ij} - \hat{\ell}^{ij} \Bigr]\,,
\end{equation}
where we pose $m^{ij}\equiv m^{i}m^{j}$ and ${m^*}^{ij}\equiv {m^*}^{i} {m^*}^{j}$ which are automatically STF tensors, and $\hat{\ell}^{ij}\equiv \ell^{\langle i}\ell^{j\rangle}=-2m^{\langle i} {m^*}^{j\rangle}$.

To perform the integration in~\eqref{eq:Tijdef} we express the vectors in the retarded or advanced basis at times $t \pm \tau$ in terms of the fixed basis at current time $t$ with the help of the formula~\eqref{eq:mpast}. Thus we can rewrite the quadrupole at retarded or advanced time (posing $s\equiv t\pm\tau$) as
\begin{equation}\label{eq:quadmm*bis}
	I_{ij}(s) = \frac{1}{2} m \nu \,r^2(s) \left( \de^{- 2\di \psi(s)} m_0^{ij} + \de^{ 2\di \psi(s)} {m_0^{*}}^{ij} - \hat{\ell}^{ij} \right)\,,
\end{equation}
where we use for convenience the phase angle with respect to the periastron or true anomaly $\psi=\phi-\omega$.

It is not possible to find a closed-form expression for the integral given by Eq.~\eqref{eq:Integral4PNtail} but we can compute its Fourier expansion. To do so, we shall use the following relation between the true anomaly $\psi$ and the mean anomaly~$\ell$ (well-known in celestial mechanics~\cite{duriez} and valid at Newtonian order)
\begin{align}
\label{eq:MainFourierExpansion4PN}
\Big(\frac{r}{a}\Big)^n \de^{\di m \psi} = \sum_{p = - \infty}^{+\infty} X_{p}^{n, m}(e) \,\de^{\di p \ell}\,,
\end{align}
where $X_{p}^{n, m}(e)$ are the Hansen coefficients depending on the eccentricity $e$ and given by
\begin{subequations}\label{eq:hansen}
\begin{align}
X_{p}^{n, m}(e) &= \frac{1}{(1+\beta^2)^{n+1}} \sum_{k=0}^{+\infty} (-\beta)^k \sum_{j=0}^{k} C^{n,m}_{k-j,j} \,J_{p-m+k-2j}(p e)\,,\\
C^{n,m}_{k-j,j} &\equiv \binom{n+m+1}{k-j} \binom{n-m+1}{j}\,,
\end{align}
\end{subequations}
where $\beta=(1-\sqrt{1-e^2})/e$ and $J_{q}(x)$ is the Bessel function of the first kind, valid for all integers $n$, $m$ and $p$. The sum over $k$ in~\eqref{eq:hansen} is actually finite for $|m|\leqslant n+1$ since the coefficients $C^{n,m}_{k-j,j}$ are zero whenever $k\geqslant 2n+3$. With this decomposition, we are able to write the Fourier transform of the quadrupole moment at early or late times $s$ as
\begin{align}\label{eq:FourierX}
	I_{ij}(s) = \frac{1}{2} m \nu a^2 \sum_{p=-\infty}^{+\infty} \biggl[ X_{p}^{2,-2}(e)\, m^{ij}_0 + X_{p}^{2,2}(e)\, m_{0*}^{ij}  - X_{p}^{2,0}(e)\, \hat{\ell}^{ij} \biggr]\de^{\di p \ell(s)}\,.
\end{align}
An explicit (well known) formulation is recalled in App.~\ref{app:IijFourier}. Next, we time differentiate the Fourier decomposition~\eqref{eq:FourierX}, which kills the non-oscillatory term, and insert it into the tail integral~\eqref{eq:Integral4PNtail}. The integration of the oscillating terms can be carried out using the formula (valid for $n>0$ and any $p\in\mathbb{Z}^*$)
\begin{subequations}
\begin{align}\label{eq:formule1}
	\int_{0}^{+\infty} \dd\tau \, \mathrm{ln}\left( \frac{\tau}{\tau_0}\right) \de^{-\di p n \tau} &=  \frac{\di}{p n} \,\biggl[\ln\left( |p|n \tau_0\right) + \gamma_{\rm E} + \frac{\di \pi}{2} \mathrm{sign}(p)\biggr]\,.
\end{align}
More specifically, each Fourier mode of the two terms in~\eqref{eq:Tijdef} combine, so that we only require the integral
\begin{align}\label{eq:formule2}
	\int_{0}^{+\infty} \dd\tau \, \mathrm{ln}\left( \frac{\tau}{\tau_0}\right) \Bigl(\de^{-\di p n \tau} + \de^{\di p n \tau} \Bigr) &= - \frac{\pi}{p n} \mathrm{sign}(p)\,,
\end{align}
\end{subequations}
in which we observe that the arbitrary scale $\tau_0$ disappears. After reintroducing the $\bm{m}(t) = \bm{m}_0 \exp[-\di \psi(t)]$, we obtain
\begin{align}\label{eq:Tijval}
	\mathcal{T}_{ij}(t) = \frac{\di\pi}{2} m \nu a^2 \sum_{p=-\infty}^{+\infty} (p n)^6 \mathrm{sign}(p) \biggl[ m^{ij}(t) \,\de^{2 \di \psi(t)} X_{p}^{2,-2}(e)+ {m^*}^{ij}(t) \,\de^{- 2 \di \psi(t)} X_{p}^{2,2}(e) - \hat{\ell}^{ij} X_{p}^{2,0}(e) \biggr]\de^{\di p \ell(t)}\, .
\end{align}
Finally, the Fourier expansion of the 4PN tail contribution to the acceleration~\eqref{eq:Integral4PNtail} reads 
\begin{align}\label{eq:ARRtailfinal}
A^{i\,{\rm RR}}_{\text{4PN\,tail}} = -\frac{\sqrt{2} \di \pi}{5 c^8}
	\,\frac{G^5 m^5 \nu}{a^7} r m^i \sum_{p=-\infty}^{+\infty} p^6 \,\mathrm{sign}(p) \biggl[ \de^{2 \di \psi} X_{p}^{2,-2}(e) + \frac{1}{3} X_{p}^{2,0}(e)\biggr] \de^{\di p \ell} + \text{c.c.}\, .
\end{align}
The small eccentricity expansion of the RR tail acceleration is straightforward from the Fourier coefficients~\eqref{eq:ARRtailfinal}. Since $J_q(p e) = \calO(e^{|q|})$, only a finite number of Fourier harmonics can contribute to a given order in eccentricity.

The derivation of the set of evolution equations for the parameters can be achieved by performing the Fourier expansion of all the terms multiplied by the 4PN acceleration. Their Fourier expansion can be obtained using~\eqref{eq:MainFourierExpansion4PN} and the equations for the evolution of the parameters are provided in the Supplemental Material~\cite{ancillary}. Here we give the evolution equations for the parameters in the adiabatic approximation [see Eq.~\eqref{eq:adiabapprox}]:
\begin{subequations}
\begin{align}
	\frac{\dd  \overline{a}^{\text{tail}}}{\dd t} &= \frac{32\pi}{15 c^8}\Bigl(\frac{G m}{a}\Bigr)^{9/2}\frac{\nu}{e^4} \sum_{q=-\infty}^{+\infty} q^3 \,\mathrm{sign}(q) \biggl [3 \bigl(1-e^2\bigr) e^2 J_{q+1}\bigl(e q\bigr)^2 \bigl(-1
	-  q^2
	+ e^2 q^2\bigr)\nn\\
	&-  3 \bigl(1-e^2\bigr) e J_{q}\bigl(e q\bigr) J_{q+1}\bigl(e q\bigr) \bigl(-2
	+ 4 q
	- 3 e^2 q
	- 2 q^2
	+ 2 e^2 q^2\bigr)\nn\\
	&+ J_{q}\bigl(e q\bigr)^2 \bigl(-6
	+ 6 e^2
	-  e^4
	+ 12 q
	- 21 e^2 q
	+ 9 e^4 q
	- 6 q^2
	+ 15 e^2 q^2
	- 12 e^4 q^2
	+ 3 e^6 q^2\bigr)\biggr]\, , \\
	\frac{\dd  \overline{e}^{\text{tail}}}{\dd t} &= 
	\frac{16\pi}{15 c^8}\Bigl(\frac{G m}{a}\Bigr)^{9/2}\frac{\nu (1-e^2)}{a e^5} \sum_{q=-\infty}^{+\infty} q^2 \,\mathrm{sign}(q) \biggl [3 \bigl(1-e^2\bigr) e^2 J_{q+1}\bigl(e q\bigr)^2 q \bigl(-5
	-  q^2
	+ e^2 q^2\bigr) \\
	&+ 3 e J_{q}\bigl(e q\bigr) J_{q+1}\bigl(e q\bigr) \bigl(-4
	+ 2 e^2
	+ 10 q
	- 10 e^2 q
	- 8 q^2
	+ 15 e^2 q^2
	- 7 e^4 q^2
	+ 2 q^3
	- 4 e^2 q^3
	+ 2 e^4 q^3\bigr)\nn\\
	&+  J_{q}\bigl(e q\bigr)^2 \bigl(12
	- 6 e^2
	- 30 q
	+ 36 e^2 q
	- 7 e^4 q
	+ 24 q^2
	- 45 e^2 q^2
	+ 21 e^4 q^2
	- 6 q^3
	+ 15 e^2 q^3
	- 12 e^4 q^3
	+ 3 e^6 q^3\bigr)\biggr]\, ,\nn\\
	\frac{\mathrm{d}  \overline{\iota}^{\text{tail}}}{\mathrm{d}t} &= \frac{\mathrm{d}  \overline{\bar{\varepsilon}}^{\text{tail}}}{\mathrm{d}t} = \frac{\mathrm{d}  \overline{\varpi}^{\text{tail}}}{\mathrm{d}t} = \frac{\mathrm{d}  \overline{\Omega}^{\text{tail}}}{\mathrm{d}t} = 0\,.
\end{align}
\end{subequations}
The general structure of the evolution equations for the RR corrections due to the tail [see Eq.~\eqref{eq:deltarrc}] is of the form
\begin{align}\label{eq:deltaRRtailstruct}
	\frac{\dd \delta^\text{RR} c_\alpha^{\text{tail}}}{\dd t} = \sum_{\substack{p=-\infty\\ p\neq 0}}^{+\infty} \de^{\di p n t} \sum_{q=-\infty}^{+\infty} \mathcal{C}_{p, q}\bigl[\overline{c}_\beta\bigr]\,.
\end{align}
Similarly, the oscillating contributions of the tail parts can be decomposed such that
\begin{align}\label{eq:IdeltaRRtailstruct}
	\delta^\text{RR} c_\alpha^{\text{tail}} = \sum_{\substack{p=-\infty\\ p\neq 0}}^{+\infty} \de^{\di p n t} \sum_{q=-\infty}^{+\infty} \mathcal{D}_{p,q}\bigl[\overline{c}_\beta\bigr]\,.
\end{align}
The above coefficients $\mathcal{C}_{p,q}$ and $\mathcal{D}_{p,q}$ are given in the Supplemental Material~\cite{ancillary}, together with the coefficients $\mathcal{D}_{p,q}$ of the phase parameters $\delta^{\text{RR}}\ell^{\text{tail}}$ and $\delta^{\text{RR}}\Lambda^{\text{tail}}$ defined by Eqs.~\eqref{eq:ell} and~\eqref{eq:Lambda}. They turn out to be combinations of products of Bessel functions $J_{p-q+\eta_1}((p-q)e) J_{q+\eta_2}(q e)$, with $\eta_{1,2}\in\{0,1\}$. The respective prefactors are simple rational functions of $e$ or $\sqrt{1-e^2}$.

For small eccentricities, only a finite number of the coefficients $\mathcal{C}_{p,q}$ and $\mathcal{D}_{p,q}$ contribute to the sums~\eqref{eq:deltaRRtailstruct}--\eqref{eq:IdeltaRRtailstruct} at a given order $N$ in the expansion parameter $e$. Indeed, due to the asymptotic behaviour $J_{n}(x)\sim x^{|n|}$ when $x\to0$, the products $J_{p-q+\eta_1}((p-q)e) J_{q+\eta_2}(q e)$ entering those coefficients are of order $e^{|p-q+\eta_1|+|q+\eta_2|}$. Moreover, their prefactors are functions of $e$, with leading orders $e^{N_0}$, $N_0\geqslant -3$, in the small eccentricity expansion. Thus, the terms that contribute to order $N$ in eccentricity all satisfy $|p-q+\eta_1|+|q+\eta_2|\leqslant N+3$, which implies $|p|\leqslant N+4$ and $|q|\leqslant N+4$. Truncating the sums~\eqref{eq:deltaRRtailstruct}--\eqref{eq:IdeltaRRtailstruct} accordingly, the dropped terms are all of order $\mathcal{O}(e^{N+1})$. For definiteness, we provide below the explicit expressions for the fast RR oscillations of the orbital elements associated with the tail effect, up to the third order in eccentricity:
\begin{subequations}
\begin{align}
\delta^\text{RR} a^\text{tail} &= \frac{G^4 m^4 \pi \nu}{a^3 c^8} \Big(- \frac{2904}{5} e \sin \ell
 -  \frac{15547}{5} e^3 \sin \ell
 -  \frac{18754}{15} e^2 \sin 2 \ell
 -  \frac{123562}{45} e^3 \sin 3 \ell\Big)\,,\\
\delta^\text{RR} e^\text{tail} &= \frac{G^4 m^4 \pi \nu}{a^4 c^8} \Big(- \frac{256}{5} \sin \ell
 - 766 e^2 \sin \ell
 -  \frac{2987}{15} e \sin 2 \ell
 -  \frac{33129}{40} e^3 \sin 2 \ell
 -  \frac{8882}{15} e^2 \sin 3 \ell\nn\\
& -  \frac{70115}{48} e^3 \sin 4 \ell\Big)\,,\\
\delta^\text{RR} \bar{\varepsilon}^\text{tail} &= \frac{G^4 m^4 \pi \nu}{a^4 c^8} \Big(\frac{6088}{15} e \cos \ell
 + \frac{1552}{3} e^3 \cos \ell
 + \frac{26963}{30} e^2 \cos 2 \ell
 + \frac{9051}{5} e^3 \cos 3 \ell\Big)\,,\\
\delta^\text{RR} \varpi^\text{tail} &= \frac{G^4 m^4 \pi \nu}{a^4 c^8} \Big(\frac{256 \cos \ell}{5 e}
 -  \frac{1522}{15} e \cos \ell
 -  \frac{2017}{10} e^3 \cos \ell
 + \frac{2987}{15} \cos 2 \ell
 -  \frac{13721}{40} e^2 \cos 2 \ell
 + \frac{8882}{15} e \cos 3 \ell\nn\\
& -  \frac{113863}{108} e^3 \cos 3 \ell
 + \frac{70115}{48} e^2 \cos 4 \ell
 + \frac{575911}{180} e^3 \cos 5 \ell\Big)\,,\\
 \delta^\text{RR} \Omega^\text{tail}={}& \delta^\text{RR} \iota^\text{tail}= 0\,.
\end{align}
\end{subequations}
As for the tail corrections to the RR oscillatory parts of the mean anomaly $\ell$ and the angle $\Lambda$, they read
\begin{subequations}
\begin{align}
\delta^\text{RR} \ell^\text{tail} &= \frac{G^4 m^4 \pi \nu}{a^4 c^8} \Big(- \frac{256 \cos \ell}{5 e}
 -  \frac{5458}{15} e \cos \ell
 -  \frac{59176}{15} e^3 \cos \ell
 -  \frac{2987}{15} \cos 2 \ell
 + \frac{36491}{120} e^2 \cos 2 \ell
 -  \frac{8882}{15} e \cos 3 \ell\nn\\
& + \frac{805451}{540} e^3 \cos 3 \ell
 -  \frac{70115}{48} e^2 \cos 4 \ell
 -  \frac{575911}{180} e^3 \cos 5 \ell\Big)\,,\\
\delta^\text{RR} \Lambda^\text{tail} &= \delta^\text{RR} \ell^\text{tail}\,.
\end{align}
\end{subequations}

\subsubsection{Quasi-hyperbolic case}

The non-local RR acceleration at 4PN order has been given by Eq.~\eqref{eq:Integral4PNtail}. In the hyperbolic case one must consider a continuous Fourier transform of the quadrupole moment instead of the discrete Fourier series considered in the previous subsection for the elliptic case. We thus pose (with $s=t\pm\tau$)
\begin{equation}\label{eq:FourierIij} 
	I_{ij}(s) = \int_{-\infty}^{+\infty} \frac{\dd\omega}{2\pi}\,\widetilde{I}_{ij}(\omega) \,\de^{- \di \omega s}\,,
\end{equation}
where $\widetilde{I}_{ij}(\omega)$ is the Fourier transform of the quadrupole moment, such that $\widetilde{I}_{ij}(-\omega)=\widetilde{I}_{ij}^*(\omega)$. It is then straightforward to obtain the continuous Fourier transform of the RR tail term~\eqref{eq:Tijdef} by means of the same integration formula~\eqref{eq:formule2} as in the elliptic case and we find
\begin{align}\label{eq:tailHyperexpr}
	\mathcal{T}_{ij}(t) = - \di \pi \int_0^{+\infty} \frac{\dd\omega}{2\pi}\,\widetilde{I}_{ij}(\omega) \,\omega^6\,\de^{- \di \omega t} + \text{c.c.}\,.
\end{align}
The latter result can be expressed in terms of the eccentric anomaly $v$ instead of time $t$ through the Newtonian relation $t=\sqrt{\hata_r^3/Gm}\,(\hat{e}_r \sinh v-v)$. The Fourier transform $\widetilde{I}_{ij}(\omega)$ of the quadrupole moment can be done exactly in terms of modified Bessel functions of the second kind $K_p(u)$ and $K_{p+1}(u)$, with order $p\equiv \di u/\hat{e}_r$ and argument $u\equiv\omega \,\hat{e}_r \sqrt{\hata_r^3/Gm}$ (see, \textit{e.g.}, Ref.~\cite{BG21}). The integral~\eqref{eq:tailHyperexpr} can then be computed by taking the large-eccentricity expansion of the integrand, leading to Bessel functions $K_\nu(u)$ of order $0$ and $1$, and their derivatives with respect to the order $\nu$ (evaluated at $\nu=0,1$). The components of the RR acceleration~\eqref{eq:Integral4PNtail} are then readily obtained following~\eqref{eq:tailcomponent}. The final result for the components of the acceleration in a large eccentricity expansion $\hat{e}_r\to\infty$ reads
\begin{subequations}
\begin{align}
	\hatA^{x\,{\rm RR}}_{\text{4PN\,tail}} &=
    \nu\left(\frac{Gm}{\hata_r^2}\right)\left(\frac{Gm}{c^2\hata_r}\right)^4 \Biggl\{\frac{1}{\hat{e}_r^4}\left[\left(\frac{1456}{\cosh(v)^9}-\frac{1520}{\cosh(v)^7}+\frac{1088}{5 \cosh(v)^5}\right) v\right.\nn\\
    &+\left. \left(\frac{1456}{\cosh(v)^8}-\frac{1648}{3
		\cosh(v)^6}-\frac{96}{5 \cosh(v)^4}\right) \sinh (v)\right]\nn\\
	&
	+\frac{1}{\hat{e}_r^5}\left[\left(\frac{4032 \left(4 v^2+\pi ^2\right)}{\cosh(v)^{11}}-\frac{112 \left(116 v^2+29 \pi ^2+27\right)}{\cosh(v)^9}+\frac{16 \left(80 v^2+20
		\pi ^2+119\right)}{\cosh(v)^7}-\frac{352}{15 \cosh(v)^5}\right) \sinh
	(v)\right.\nn\\
	&\left.
	+\left(-\frac{19152}{\cosh(v)^{10}}+\frac{22288}{\cosh(v)^8}-\frac{67264}{15 \cosh(v)^6}-\frac{512}{5 \cosh(v)^4}\right) v
	\right]\nn\\
	&
	+\frac{1}{\hat{e}_r^6}\left[\left(\frac{1848 \left(5 \pi ^2-132 v^2\right)}{\cosh(v)^{12}}-\frac{112 \left(-2004 v^2+62 \pi
		^2-255\right)}{\cosh(v)^{10}}+\frac{56 \left(-2932 v^2+31 \pi ^2-1630\right)}{5 \cosh(v)^8}\right.\right.\nn\\
	&\left.
	-\frac{32 \left(80 v^2-127\right)}{5
		\cosh(v)^6}+\frac{96}{5 \cosh(v)^4}\right) \sinh (v)
	+\left(-\frac{35112 \left(4 v^2+\pi ^2\right) }{\cosh(v)^{13}}+\frac{504 \left(436 v^2+109 \pi ^2+262\right) }{\cosh(v)^{11}}
	\right.\nn\\
	&\left.\left. -\frac{112 \left(792
		v^2+198 \pi ^2+1519\right) }{\cosh(v)^9}+\frac{64 \left(1380 v^2+345 \pi ^2+10582\right) }{15 \cosh(v)^7}+\frac{1984 }{15
		\cosh(v)^5}\right)v
	\right]
	+\calO\left(\frac1{\hat{e}_r^7}\right)
	\Biggr\}
	\,,\\
	\hat A^{y\,{\rm RR}}_{\text{4PN\,tail}} &=
    \left(\frac{Gm}{\hata_r^2}\right)\left(\frac{Gm}{c^2\hata_r}\right)^4 \nu
    \Biggl\{\frac{1}{\hat{e}_r^4}\left[-\frac{1456}{\cosh(v)^8}+\frac{3568}{3 \cosh(v)^6}+\frac{32}{3
		\cosh(v)^4}\right.\nn \\
      &+\left.\left(\frac{1456}{\cosh(v)^9}-\frac{704}{\cosh(v)^7}-\frac{256}{5 \cosh(v)^5}\right) v \sinh (v)\right]\nn\\
	&
	+\frac{1}{\hat{e}_r^5}\left[-\frac{4032 \left(4 v^2+\pi ^2\right)}{\cosh(v)^{11}}+\frac{56 \left(364 v^2+91 \pi ^2+54\right)}{\cosh(v)^9}-\frac{32 \left(152 v^2+38 \pi
		^2+105\right)}{\cosh(v)^7}-\frac{16 \left(80 v^2+20 \pi ^2-179\right)}{5 \cosh(v)^5}\right.\nn\\
	&\left.
	+\frac{96}{5	\cosh(v)^3}+\left(-\frac{19152}{\cosh(v)^{10}}+\frac{11984}{\cosh(v)^8}-\frac{1024}{15 \cosh(v)^6}\right) v \sinh (v)
	\right]\nn\\
	&
	+\frac{1}{\hat{e}_r^6}\left[-\frac{1848 \left(5 \pi ^2-132 v^2\right)}{\cosh(v)^{12}}+\frac{56 \left(-6060 v^2+199 \pi ^2-510\right)}{\cosh(v)^{10}}-\frac{56
		\left(-9372 v^2+181 \pi ^2-2835\right)}{5 \cosh(v)^8}\right.\nn\\
	&\left.
	-\frac{8 \left(144 v^2+708 \pi ^2+10619\right)}{15 \cosh(v)^6}-\frac{1328}{15
		\cosh(v)^4}+\left(-\frac{35112 \left(4 v^2+\pi ^2\right)}{\cosh(v)^{13}}+\frac{1008 \left(144 v^2+36 \pi ^2+131\right)
	}{\cosh(v)^{11}}\right.\right.\nn\\
	&\left.\left.
	-\frac{168 \left(160 v^2+40 \pi ^2+599\right) }{\cosh(v)^9}-\frac{32 \left(480 v^2+120 \pi ^2-3479\right) }{15
		\cosh(v)^7}+\frac{896 }{5 \cosh(v)^5}\right) v \sinh (v)
	\right]
	+\calO\left(\frac1{\hat{e}_r^7}\right)
	\Biggr\}
	\,.
\end{align}
\end{subequations}

The solutions for the RR corrections $\delta^{\rm RR}\hat{c}_A^{\,\rm tail}(t)$ of the parameters~\eqref{eq:ctildeA} are given in the Supplemental Material~\cite{ancillary}. In these tail solutions, the polylogarithms appear only through the combination
\begin{subequations}
\begin{align}
\mathrm{Ti}_k(x) \equiv -\frac{\di}{2}\Bigl[\mathrm{Li}_k(\di x)-\mathrm{Li}_k(-\di x)\Bigr]\,,
\end{align}
where $\mathrm{Ti}_k$ is the inverse tangent integral of order $k$ and the polylogarithm of
order $k\geqslant 2$ is given by
\begin{align}
    \mathrm{Li}_k(z) = \frac{(-1)^{k-1}}{(k-2)!}\int_0^z \frac{\dd s}{s}\,
    \ln^{k-2}\!\left(\frac{s}{z}\right)\ln(1-s)\,.
\end{align}
\end{subequations}
In practice, only $\mathrm{Ti}_2$ and $\mathrm{Ti}_3$ enter these expressions.

The corresponding perturbed orbit is given by 
\begin{subequations}
	\begin{align}
	\delta^{\rm RR} r^{\rm tail}(t) &= \frac1{\hatchi_r}\hat{a}_r\hat{e}_r\sinh v\,\delta^{\rm RR} \hat\ell^{\rm tail}(t)
	+\hatchi_r\delta^{\rm RR} \hat{a}_r^{\rm tail}(t) -
	\frac{\hat{a}_r}{\hat{e}_r}\left(1-\frac{\hat{e}_r^2-1}{\hatchi_r}\right)\delta^{\rm RR} \hat{e}_r^{\rm tail}(t)
	\,,\\
	\delta^{\rm RR} \phi^{\rm tail}(t) &=  \frac{\sqrt{\hat{e}_r^2-1}}{\hatchi_r^2}\delta^{\rm RR} \hat\ell^{\rm tail}(t)
	+\delta^{\rm RR} \hat c_\phi^{\rm tail}(t) -
	\frac{\sinh v}{\hatchi_r\sqrt{\hat{e}_r^2-1}}\left(1+\frac{\hat{e}_r^2-1}{\hatchi_r}\right)\delta^{\rm RR} \hat{e}_r^{\rm tail}(t)
	\,,
\end{align}
\end{subequations}
where, for example, 
\begin{align}
\delta^{\rm RR} \hat{e}_r^{\rm tail} &= \left(\frac{Gm}{c^2{\hat a}_r}\right)^4\nu\biggl\{
\frac{1}{\hat e_r^2}\left[\left(-\frac{208}{\cosh(v)^7}-\frac{112}{5 \cosh(v)^5}+\frac{320}{3 \cosh(v)^3}\right) v\right.\\
&+\left.\left(-\frac{208}{\cosh(v)^6}-\frac{2416}{15 \cosh(v)^4}-\frac{96}{5 \cosh(v)^2}\right) \sinh (v)\right]\nn\\
&+\frac{1}{\hat e_r^3}\biggl[
\left(\frac{2128}{\cosh(v)^8}-\frac{848}{3 \cosh(v)^6}-\frac{15776}{15 \cosh(v)^4}-\frac{384}{5 \cosh(v)^2}\right) v-\frac{3584}{45}\nn\\
&+
\biggl(-\frac{448 \left(4 v^2+\pi ^2\right)}{\cosh(v)^9}-\frac{48 \left(8 v^2+2 \pi ^2-7\right)}{\cosh(v)^7}+\frac{16 \left(84 v^2+21 \pi^2+29\right)}{3 \cosh(v)^5}-\frac{4672}{45 \cosh(v)^3}-\frac{3584}{45 \cosh(v)}\biggr) \sinh (v)\biggr]\nn\\
&+\frac{1}{\hat e_r^4}\biggl[
\biggl(\frac{3192 \left(4 v^2+\pi ^2\right)}{\cosh(v)^{11}}-\frac{56 \left(140 v^2+35 \pi ^2+212\right)}{\cosh(v)^9}-\frac{16 \left(312
   v^2+78 \pi ^2-233\right)}{\cosh(v)^7}\nn\\
   &+\frac{8 \left(2976 v^2+744 \pi ^2+11083\right)}{15 \cosh(v)^5}-\frac{5758}{15
   \cosh(v)^3}-\frac{573}{5 \cosh(v)}\biggr)v+\frac{1146}{5}\left( v\,{\rm Ti}_2(e^v)-{\rm Ti}_3(e^v)\right)\nn\\
&+\biggl(-\frac{168 \left(5 \pi ^2-132 v^2\right)}{\cosh(v)^{10}}-\frac{14 \left(-76 v^2+5 \pi
   ^2+176\right)}{\cosh(v)^8}+\frac{-101124 v^2+287 \pi ^2-9880}{15 \cosh(v)^6}\nn\\
   &-\frac{17652 v^2+1853 \pi ^2-32112}{60 \cosh(v)^4}-\frac{3
   \left(764 v^2+191 \pi ^2+760\right)}{40 \cosh(v)^2}\biggr) \sinh (v)
-\frac{573}{20} \left(4 v^2+\pi ^2\right) {\rm arctan}\left(e^v\right)
\biggr]
+\calO\biggl(\frac{1}{\hat e_r^5}\biggr)\biggr\}\,.\nn
\end{align}
The tail-related contribution to the (relative) RR scattering angle $\chi^{\rm RR}_{\rm tail} \equiv [\delta^{\rm RR}\phi^{\rm tail}]$ turns out to be given by Eq.~\eqref{chi_rr_tail} above, in agreement with Eq.~(C11) of Ref.~\cite{BDG23}, where the parameters $(p_\infty,j)$ are used instead of $(\hat{a}_r,\hat{e}_r)$. Integrating the kinematical part of the fluxes gives $ E_{\rm sys}^{\rm tail}=- E_{\rm rad}^{\rm tail}$ and $  J_{\rm sys}^{\rm tail}=-  J_{\rm rad}^{\rm tail}$, with 
\begin{subequations}
\begin{align}
	  E_{\rm rad}^{\rm tail} &= mc^2 \nu^2 \left( \frac{Gm}{c^2 \hata_r}\right)^5\Biggl[
	\frac{3136}{45 \hat{e}_r^4}+\frac{297 \pi ^3}{20 \hat{e}_r^5}+\frac{1}{\hat{e}_r^6}\left(\frac{15616}{45}+\frac{88576}{675}\pi ^2\right) + \calO\left(\frac{1}{\hat{e}_r^7}\right)
	\Biggr]\,,\\
	  J_{\rm rad}^{\rm tail} &= \frac{Gm^2}{c}\nu^2 \left( \frac{Gm}{c^2 \hata_r}\right)^{7/2} 
    \Biggl[
	\frac{448}{5 \hat{e}_r^3}+\frac{69 \pi ^3}{5 \hat{e}_r^4}+\frac{1}{\hat{e}_r^5}\left(\frac{2144}{15}+\frac{4352}{45}\pi^2\right)
	+ \calO\left(\frac{1}{\hat{e}_r^6}\right)
	\Biggr]\,.
\end{align}
\end{subequations}
The tail contribution to the radiated energy and angular momentum can also be found in Ref.~\cite{BDG21}, see Eqs.~(D26) and~(F2) respectively, as functions of the parameters $(p_\infty,j)$.

\section{Conclusions} 
\label{sec:conclusion} 

In this paper, we have investigated radiation-reaction (RR) effects on the gravitational-wave (GW) phasing of compact binaries on elliptic-like and hyperbolic-like orbits, extending previous results to include the complete RR contributions through 4.5PN order in harmonic coordinates~\cite{BFST26}. Our analysis relies on two ingredients: (i) the quasi-Keplerian (QK) parametrization of the 2PN conservative dynamics (for both elliptic-like and hyperbolic-like orbits)~\cite{DD85, DS88, SW93, MGS04, CGHL18,CTG22}, and (ii) the Lagrange method of variation of constants, which provides a systematic framework for describing the evolution of the orbital elements, and hence of the GW phasing, beyond the standard secular (adiabatic) approximation~\cite{DGI04, KG06, BDG23, BGR25}. A key feature of this approach is that it captures the distinct time scales on which RR acts in eccentric systems: secular evolution on the RR time scale, oscillations on the orbital time scale, and modulations associated with periastron precession. These effects are relevant both to the construction of accurate GW templates for eccentric binaries and to comparisons with effective-field-theory (EFT) results for hyperbolic encounters. 

Particular attention has been devoted to the non-local contributions entering the RR acceleration at the 4PN and the 4.5PN order. These arise, respectively, from the RR (dissipative) part of the classic GW tail effect and from the GW recoil associated with the emission of linear momentum. The contribution of the recoil is mandatory to consistently take into account the GW emission at the high 4.5PN order, as shown in Refs.~\cite{BFT25, BFST26}. 

In the elliptic case, we have presented two computations of the recoil: one is the standard PN integration of the linear-momentum flux which is obtained modulo an unspecified integration constant. We showed that the contribution of this constant in the RR 4.5PN acceleration can be gauged away and therefore has no intrinsic physical meaning. Next we presented a more sophisticated derivation of the recoil based on a multi-scale approach of the dynamics, taking into account the orbital time scale and the precession time scale. As a result we obtained a purely oscillating expression of the recoil (with no secular growing term), which is expected to be accurate over a time scale long with respect to the precession period (and of course, \textit{a fortiori}, the orbital period). We showed, however, that the two expressions of the recoil are consistent with each other in the appropriate limit. In this limit we found a contribution arising formally at 3.5PN order in the RR acceleration in harmonic coordinates, but which, again, is pure gauge and has no intrinsic meaning.

To summarize the recoil calculation in the elliptic case:
\begin{itemize}
\item There exist two possible PN-type expansions: the systematic PN expansion (\textit{\`a la} Fitchett~\cite{Fit83}) performed in Sec.~\ref{sec:4p5PNnonlocalelliptic} which retains only the leading 3.5PN order in the flux but leads to the presence of a secular-growing term. The multi-scale PN expansion in Sec.~\ref{sec:secular} which necessitates the inclusion of the next order 4.5PN in the flux and avoids the appearance of the secular term. 
\item The resummed approach (or ``multi-scale'' as it takes into account the orbital precession scale) is physically expected to stay accurate over a longer period of time and leads to a PN order reduction, implying a 2.5PN contribution in the recoil integral.
\end{itemize}

We have also shown that the elliptic-like and hyperbolic-like descriptions of the orbital dynamics due to the local RR contributions are related by analytic continuation. This provides a valuable consistency check on our results and highlights the common analytical structure underlying bound and unbound orbits (see also, for instance,~\cite{DD85, BS89, KP1, KP2}). Non-local effects, by contrast, cannot be obtained in this way and must be treated separately in each orbital regime, taking into account their specific characteristics. 

Overall, our results provide a unified description of RR effects through 4.5PN order for bound and unbound eccentric motion in harmonic coordinates. They lay the groundwork for further studies of GW phasing beyond the adiabatic approximation, with potential applications to high-accuracy waveform modeling and to the comparison between PN and EFT post-Minkowskian (PM) approaches to the two-body problem. Reaching this very high PN accuracy allows us to capture several PM orders (powers of $G$) in various observable quantities (\textit{e.g.}, radiative losses, scattering angle, periastron advance, \textit{etc.}) which can be directly compared against genuine PM-technology results, continuing a synergistic effort between these two approaches which has proven to be extremely useful in recent years. 

Future developments in this field will aim at reaching the next PN approximation level, \textit{i.e.}, the 5PN order, where for the first time ``quadratic effects'' in RR will appear [since we have $\text{(2.5PN)}^2=$ 5PN], and consequently modifications to any type of linear-response formula~\cite{BD12} should be considered too. Computing all building blocks necessary to accomplish this important task will be the object of forthcoming works.

\acknowledgments

Q.H. was supported by the Universitat de les Illes Balears (UIB) with funds from the Programa de Foment de la Recerca i la Innovació de la UIB 2024-2026 (supported by the yearly plan of the Tourist Stay Tax ITS2023-086); the Spanish Agencia Estatal de Investigación grants PID2022-138626NB-I00, RED2024-153978-E, RED2024-153735-E, funded by MICIU/AEI/10.13039/501100011033 and the ERDF/EU; and the Comunitat Autònoma de les Illes Balears through
the Conselleria d'Educació i Universitats with funds from the European Union - European Regional Development Fund (ERDF) (SINCO2022/18146 - Plataforma HiTech-IAC3-BIO). G.F. and E.S. acknowledge financial support from the Centre national d'études spatiales (CNES), France (ROR:https://ror.org/04h1h0y33) within the framework of the LISA mission. D.B. and S.R.A. acknowledge membership to the Italian Gruppo Nazionale per
la Fisica Matematica (GNFM) of the Istituto Nazionale
di Alta Matematica (INDAM).
A.G. is grateful to the Istituto
per le Applicazioni del Calcolo \lq\lq M. Picone," CNR, Rome (IT) for
past support and hospitality during the development of
the present project. S.R.A. is grateful to Luc Blanchet and Guillaume Faye for their hospitality at Institut d'Astrophysique de Paris, where part of the research presented in this article was carried out.
Most computations as well as the formatting of lengthy equations were performed using the \textit{xAct} library~\cite{xtensor} for \emph{Mathematica}.

\appendix

\section{Quasi-Keplerian representation at 3PN order} 
\label{app:QK}

In this Appendix we give exhaustive formulas for the quasi-Keplerian (QK) parametrization of the solution for both elliptic-like and hyperbolic-like orbits up to the 3PN level~\cite{DD85, DS88, SW93, MGS04, CGHL18}. In the present paper we need in fact this solution up to the 2PN order only. However there has been some confusion in the literature about the QK representation at 3PN order for hyperbolic-like orbits~\cite{CGHL18}.\footnote{See in particular the Appendix~D of~\cite{BDG23} for a list of typos in the 3PN-accurate hyperbolic QK representation in Ref.~\cite{CGHL18}.} In addition, to our knowledge the equivalence between the QK quasi-elliptic and quasi-hyperbolic representations through analytic continuation (AC) has not been clearly established at the 3PN order. Our aim is thus to present a comprehensive treatment at 3PN order and to prove the equivalence through AC.

As is well known, spurious (gauge-dependent) logarithms appear in standard harmonic coordinates at 3PN order, so we adopt here the modified harmonic coordinates which are free of such logarithms at 3PN, as defined in Refs.~\cite{MW03, MGS04, ABIQ08}. The 3PN QK representation for the quasi-elliptic orbit is given by~\cite{DD85, DS88, SW93, MGS04} 
\begin{subequations}\label{QK2PNelliptic3PN}
	\begin{align}
		r &= a_r\bigl( 1 -e_r \cos u\bigr)\,,\\
		\ell &= u - e_t \sin u + f_t \sin V_\phi +g_t\bigl( V_\phi - u\bigr)+h_t \sin 2 V_\phi + i_t \sin 3 V_\phi\,,\\
		\phi &= \omega + K\Bigl[ V_\phi + f_\phi \sin 2 V_\phi + g_\phi \sin 3 V_\phi + h_\phi \sin 4 V_\phi + i_\phi \sin 5 V_\phi \Bigr]\,, 
	\end{align}
\end{subequations}
where $\ell=n(t-t_0)+c_\ell$ and $V_\phi = u + 2 \arctan\left[\frac{\beta_\phi\sin u}{1-\beta_\phi\cos u}\right]$ with $\beta_\phi\equiv [1-(1-e_\phi^2)^{1/2}]/e_\phi$. Instead of the parameters $\omega$ and $c_\ell$ one may use alternative parameters $\varpi=\omega+\Omega$ (where $\Omega$ is the longitude of the ascending node) and $\bar{\varepsilon}=\varpi+c_\ell$. For the 3PN quasi-hyperbolic case we have~\cite{CGHL18}
\begin{subequations}\label{QK2PNhyperbolic3PN}
	\begin{align}
		r &= \hata_r\bigl( \hate_r \cosh v - 1\bigr)\,,\\
		\hat{\ell} &= \hate_t \sinh v - v + \hat{f}_t \sin \hatV_\phi +\hat{g}_t \hatV_\phi + \hat{h}_t \sin 2 \hatV_\phi+ \hat{i}_t \sin 3 \hatV_\phi \,,\\
		\phi &=\hatc_\phi + \hat{K}\Bigl[ \hatV_\phi + \hat{f}_\phi \sin 2 \hatV_\phi + \hat{g}_\phi \sin 3 \hatV_\phi+ \hat{h}_\phi \sin 4 \hatV_\phi + \hat{i}_\phi \sin 5 \hatV_\phi \Bigr]\,,
	\end{align}
\end{subequations}
where $\hatl=\hatn(t-t_0)+\hatc_\ell$ and $\hatV_\phi = 2 \arctan\left[\sqrt{\frac{\hate_\phi+1}{\hate_\phi-1}}\tanh \left(\frac{v}{2}\right) \right]$.

Apart from the parameters $c_\ell$ and $\omega$ in the elliptic case, and $\hatc_\ell$ and $\hatc_\phi$ in the hyperbolic case, which are free integration constants in the conservative dynamics, all the other parameters must be given as functions of the orbit's energy and angular momentum as defined by~\eqref{eq:EJ}. For the quasi-elliptic case we use the Newtonian-like parameters $a$ and $e$ defined by~\eqref{eq:Newtonianae}, \textit{i.e.}
\begin{align}\label{eq:appNewtonianae}
	a \equiv - \frac{G m}{2 \tilde{E}}\,,\qquad e \equiv \sqrt{ 1 + \frac{2\tilde{E} \tilde{J}^2}{G^2m^2}}\,,\qquad\text{($\tilde{E}<0)$}\,.
\end{align}
Up to 3PN order we have
\begin{subequations}\label{eq:parameterselliptic}
\begin{align}
	n &= \sqrt{\frac{G m}{a^3}} \Biggl\{1
	+ \frac{1}{8} \frac{G m}{a c^2} \bigl(-15
	+ \nu \bigr)
	+ \frac{G^2 m^2}{a^2 c^4} \biggl [\frac{3 \bigl(-5 + 2 \nu \bigr)}{2 \sqrt{1-e^2}}
	+ \frac{1}{128} \bigl(555
	+ 30 \nu
	+ 11 \nu^2\bigr)\biggr]\\
	& + \frac{G^3 m^3}{a^3 c^6} \biggl [- \frac{10080 - 13952 \nu + 123 \pi^2 \nu + 1440 \nu^2}{192 \bigl(1 -  e^2\bigr)^{3/2}}
	+ \frac{15 \bigl(17 - 9 \nu + 2 \nu^2\bigr)}{8 \sqrt{1-e^2}}
	+ \frac{15}{1024} \bigl(-653
	- 111 \nu
	- 7 \nu^2 + 3 \nu^3\bigr)\biggr]\Biggr\}\,,\nn\\
	K &= 1
	+ 3 \frac{G m}{a c^2 {\bigl(1-e^2\bigr)}}
	+ \frac{G^2 m^2}{a^2 c^4 {\bigl(1-e^2\bigr)}^2} \biggl [- \frac{3}{4} e^2 \bigl(-5
	+ 2 \nu \bigr)
	-  \frac{3}{2} \bigl(-15
	+ 4 \nu \bigr)\biggr]
	+ \frac{G^3 m^3}{a^3 c^6 {\bigl(1-e^2\bigr)}^3} \biggl [\frac{3}{16} e^4 \bigl(5
	- 5 \nu
	+ 4 \nu^2\bigr)\nn\\
	& + \frac{1}{32} \bigl(6750
	- 6542 \nu
	+ 123 \pi^2 \nu
	+ 84 \nu^2\bigr)
	+ \frac{1}{128} e^2 \bigl(9840
	- 13712 \nu
	+ 123 \pi^2 \nu
	+ 1248 \nu^2\bigr)\biggr]\,,\\
	a_r &= a
	+ \frac{1}{4} \frac{G m}{c^2} \bigl(-7
	+ \nu \bigr)
	+ \frac{G^2 m^2}{a c^4 {\bigl(1-e^2\bigr)}} \biggl [\frac{1}{16} e^2 \bigl(-1
	-  \nu^2\bigr)
	+ \frac{1}{16} \bigl(-63
	+ 112 \nu
	+ \nu^2\bigr)\biggr]\nn\\
	& + \frac{G^3 m^3}{a^2 c^6 {\bigl(1-e^2\bigr)}^2} \biggl [\frac{1}{6720} e^2 \bigl(-27090
	+ 215618 \nu
	- 4305 \pi^2 \nu
	- 47040 \nu^2
	- 210 \nu^3\bigr)
	+ \frac{1}{64} e^4 \bigl(1
	-  \nu
	+ \nu^3\bigr)\nn\\
	& + \frac{1}{2240} \bigl(-62685
	+ 162861 \nu
	- 4305 \pi^2 \nu
	- 4480 \nu^2
	+ 35 \nu^3\bigr)\biggr]\,,\\
	e_r &= e
	+ \frac{G m}{a c^2} \biggl [- \frac{5}{8} e \bigl(-3
	+ \nu \bigr)
	+ \frac{9 + \nu}{8 e}\biggr]
	+ \frac{G^2 m^2}{a^2 c^4 {\bigl(1-e^2\bigr)}} \biggl [\frac{1}{128} e^3 \bigl(-415
	+ 210 \nu
	- 7 \nu^2\bigr)
	-  \frac{5 \bigl(-135 + 34 \nu + \nu^2\bigr)}{128 e}\nn\\
	& + \frac{1}{128} e \bigl(845
	- 1814 \nu
	+ 13 \nu^2\bigr)
	-  \frac{\bigl(9 + \nu \bigr)^2}{128 e^3}\biggr]
	+ \frac{G^3 m^3}{a^3 c^6 {\bigl(1-e^2\bigr)}^2} \Biggl\{\frac{1}{107520} e^3 \biggl [-2184315
	+ \bigl(578087\nn\\
	& + 68880 \pi^2\bigr) \nu
	+ 80535 \nu^2
	- 1155 \nu^3\biggr]
	+ \frac{51543 + \bigl(-63499 + 984 \pi^2\bigr) \nu - 1935 \nu^2 - 9 \nu^3}{1536 e}\nn\\
	& -  \frac{5589 - 477 \nu - 113 \nu^2 + \nu^3}{1024 e^3}
	+ \frac{3}{1024} e^5 \bigl(1893
	- 1245 \nu
	+ 135 \nu^2
	+ \nu^3\bigr)\nn\\
	& + \frac{1}{53760} e \biggl [2685795 + \bigl(-9173827
	+ 206640 \pi^2\bigr) \nu
	+ 966525 \nu^2
	+ 735 \nu^3\biggr]
	+ \frac{\bigl(9 + \nu \bigr)^3}{1024 e^5}\Biggr\}\,,\\
	e_t &= e
	+ \frac{G m}{a c^2} \biggl [\frac{9 + \nu}{8 e}
	+ \frac{1}{8} e \bigl(-17
	+ 7 \nu \bigr)\biggr]
	+ \frac{G^2 m^2}{a^2 c^4} \Biggl\{\frac{e}{\sqrt{1-e^2}} \bigl(- \frac{15}{2}
	+ 3 \nu \bigr)
	+ \frac{1}{{\bigl(1-e^2\bigr)}} \biggl [\frac{1}{128} e^3 \bigl(-607
	+ 138 \nu\nn\\
	& - 79 \nu^2\bigr)
	+ \frac{99 - 18 \nu + 19 \nu^2}{128 e}
	+ \frac{1}{128} e \bigl(1101
	- 998 \nu
	+ 61 \nu^2\bigr)
	-  \frac{\bigl(9 + \nu \bigr)^2}{128 e^3}\biggr]\Biggr\}\nn\\
	& + \frac{G^3 m^3}{a^3 c^6} \Biggl[\frac{1}{{\bigl(1-e^2\bigr)}^{3/2}} \Biggl\{\frac{1}{192} e \biggl [-1980
	- 41 \bigl(-220
	+ 3 \pi^2\bigr) \nu
	- 360 \nu^2\biggr]
	+ \frac{3 \bigl(-45 + 13 \nu + 2 \nu^2\bigr)}{16 e}\nn\\
	&-  \frac{3}{4} e^3 \bigl(45 - 31 \nu
	+ 8 \nu^2\bigr)\Biggr\}
	+ \frac{1}{{\bigl(1-e^2\bigr)}^2} \Biggl\{\frac{1}{3072} e^3 \bigl(91959
	- 82531 \nu
	+ 36093 \nu^2
	- 2097 \nu^3\bigr)\nn\\
	& -  \frac{2997 - 81 \nu + 71 \nu^2 + 13 \nu^3}{1024 e^3}
	+ \frac{23895 + \bigl(-37747 + 984 \pi^2\bigr) \nu - 1455 \nu^2 + 207 \nu^3}{1536 e}\nn\\
	&+ \frac{1}{1024} e^5 \bigl(-10801 + 1405 \nu
	- 771 \nu^2
	+ 407 \nu^3\bigr)\nn\\
	&+ \frac{1}{53760} e \biggl [-39165
	+ \bigl(-2958103
	+ 103320 \pi^2\bigr) \nu
	- 54075 \nu^2
	+ 8715 \nu^3\biggr] + \frac{\bigl(9 + \nu \bigr)^3}{1024 e^5}\Biggr\}\Biggr]\,,\\
	e_\phi &= e
	+ \frac{G m}{a c^2} \biggl [\frac{1}{8} e \bigl(15
	-  \nu \bigr)
	+ \frac{9 + \nu}{8 e}\biggr]
	+ \frac{G^2 m^2}{a^2 c^4 {\bigl(1-e^2\bigr)}} \biggl [\frac{1}{128} e^3 \bigl(-415
	+ 94 \nu
	- 11 \nu^2\bigr)
	+ \frac{675 - 98 \nu + 3 \nu^2}{128 e}\nn\\
	& -  \frac{3}{128} e \bigl(-495
	+ 114 \nu
	+ 17 \nu^2\bigr)
	-  \frac{\bigl(9 + \nu \bigr)^2}{128 e^3}\biggr]
	+ \frac{G^3 m^3}{a^3 c^6 {\bigl(1-e^2\bigr)}^2} \Biggl\{\frac{1}{53760} e \biggl [5508195
	+ \bigl(-9974953\nn\\
	& + 137760 \pi^2\bigr) \nu
	+ 89145 \nu^2
	- 25935 \nu^3\biggr]
	+ \frac{60183 + \bigl(-38779 + 984 \pi^2\bigr) \nu - 927 \nu^2 - 57 \nu^3}{1536 e}\nn\\
	& -  \frac{5589 - 153 \nu - 41 \nu^2 + 5 \nu^3}{1024 e^3}
	+ \frac{1}{1024} e^5 \bigl(5679
	- 2143 \nu
	+ 237 \nu^2
	+ 59 \nu^3\bigr)
	+ \frac{1}{107520} e^3 \biggl [-3487995\nn\\
	& + \bigl(339047
	- 17220 \pi^2\bigr) \nu
	- 203385 \nu^2
	+ 20685 \nu^3\biggr]
	+ \frac{\bigl(9 + \nu \bigr)^3}{1024 e^5}\Biggr\}\,,\\
	f_t &= \frac{G^2 m^2}{a^2 c^4 {\bigl(1-e^2\bigr)}^{3/2}} \Biggl\{- \frac{1}{8} e \bigl(-15
	+ \nu \bigr) \nu
	+ \frac{1}{8} e^3 \bigl(-15
	+ \nu \bigr) \nu\nn\\
	&+ \frac{G m}{a c^2} \biggl [- \frac{\bigl(-15 + \nu \bigr) \nu \bigl(9 + \nu \bigr)}{64 e}
	+ \frac{1}{64} e^3 \nu \bigl(297
	- 175 \nu + 23 \nu^2\bigr)\nn\\
	&+ \frac{1}{2240} e \bigl(22400
	+ 28531 \nu
	- 1435 \pi^2 \nu
	- 15050 \nu^2
	- 385 \nu^3\bigr)\biggr]\Biggr\}\,,\\
	g_t &= \frac{G^2 m^2}{a^2 c^4 {\bigl(1-e^2\bigr)}^{3/2}} \Biggl\{- \frac{3}{2} \bigl(-5
	+ 2 \nu \bigr)
	+ \frac{3}{2} e^2 \bigl(-5
	+ 2 \nu \bigr)\nn\\
	&+ \frac{G m}{a c^2} \biggl [\frac{3}{16} e^2 \bigl(95
	- 55 \nu
	+ 18 \nu^2\bigr)
	+ \frac{1}{192} \bigl(6660
	- 11972 \nu + 123 \pi^2 \nu
	+ 792 \nu^2\bigr)\biggr]\Biggr\}\,,\\
	h_t &= \frac{e^2 G^3 m^3 \nu}{16 a^3 c^6 {\bigl(1-e^2\bigr)}^{3/2}} \bigl(116
	- 49 \nu
	+ 3 \nu^2\bigr)\,,\\
	i_t &= \frac{e^3 G^3 m^3 \nu}{192 a^3 c^6 {\bigl(1-e^2\bigr)}^{3/2}} \bigl(23
	- 73 \nu
	+ 13 \nu^2\bigr)\,,\\
	f_\phi &= \frac{G^2 m^2}{a^2 c^4 {\bigl(1-e^2\bigr)}^3} \Biggl\{\frac{1}{8} e^2 \bigl(1
	+ 19 \nu
	- 3 \nu^2\bigr)
	+ \frac{1}{8} e^4 \bigl(-1
	- 19 \nu
	+ 3 \nu^2\bigr)
	+ \frac{G m}{a c^2} \biggl [- \frac{1}{32} \bigl(9
	+ \nu \bigr) \bigl(-1
	- 19 \nu
	+ 3 \nu^2\bigr)\nn\\
	& + \frac{1}{26880} e^2 \bigl(58800
	+ 708544 \nu
	- 30135 \pi^2 \nu
	- 228480 \nu^2
	- 11760 \nu^3\bigr)
	+ \frac{1}{96} e^4 \bigl(3
	+ 506 \nu
	- 357 \nu^2 + 36 \nu^3\bigr)\biggr]\Biggr\}\,,\\
	g_\phi &= \frac{G^2 m^2}{a^2 c^4 {\bigl(1-e^2\bigr)}^3} \Biggl\{- \frac{1}{32} e^3 \nu \bigl(-1
	+ 3 \nu \bigr)
	+ \frac{1}{32} e^5 \nu \bigl(-1
	+ 3 \nu \bigr)
	+ \frac{G m}{a c^2} \biggl [- \frac{3}{256} e \nu \bigl(9
	+ \nu \bigr) \bigl(-1
	+ 3 \nu \bigr)\nn\\
	&
	+ \frac{1}{256} e^5 \nu \bigl(14 - 49 \nu
	+ 26 \nu^2\bigr)
	-  \frac{1}{8960} e^3 \nu \bigl(-34726
	+ 1435 \pi^2
	+ 24080 \nu
	+ 70 \nu^2\bigr)\biggr]\Biggr\}\,, \\
	h_\phi &= \frac{e^4  G^3 m^3 \nu}{192 a^3 c^6 {\bigl(1-e^2\bigr)}^3} \bigl(82
	- 57 \nu
	+ 15 \nu^2\bigr)\,, \\
	i_\phi &= \frac{e^5 G^3 m^3 \nu}{256 a^3 c^6 {\bigl(1-e^2\bigr)}^3} \bigl(1
	- 5 \nu
	+ 5 \nu^2\bigr)\,.
\end{align}
\end{subequations}

In the quasi-hyperbolic case, employing the parameters $\hat{a}$ and $\hat{e}$ defined by
\begin{align}\label{eq:appNewtonianaetilde}
	\hata \equiv \frac{G m}{2 \tilde{E}}\,,\qquad \hate \equiv \sqrt{ 1 + \frac{2\tilde{E} \tilde{J}^2}{G^2m^2}}\,,\qquad\text{($\tilde{E}>0)$}\,.
\end{align}
we have
\begin{subequations}\label{eq:parametershyperbolic}
\begin{align}
	\hat{n} &= \sqrt{\frac{G m}{\hat{a}^3}} \biggl [1
	+ \frac{1}{8} \frac{G m}{\hat{a} c^2} \bigl(15
	-  \nu \bigr)
	+ \frac{1}{128} \frac{G^2 m^2}{\hat{a}^2 c^4} \bigl(555
	+ 30 \nu
	+ 11 \nu^2\bigr)
	-  \frac{15}{1024} \frac{G^3 m^3}{\hat{a}^3 c^6} \bigl(-653
	- 111 \nu
	- 7 \nu^2 + 3 \nu^3\bigr)\biggr]\,,\\
	\hat{K} &= 1
	+ 3 \frac{G m}{\hat{a} c^2 {\bigl(\hat{e}^2-1\bigr)}}
	+ \frac{G^2 m^2}{\hat{a}^2 c^4 {\bigl(\hat{e}^2-1\bigr)}^2} \biggl[ - \frac{3}{4} \hat{e}^2 \bigl(-5
	+ 2 \nu \bigr)
	-  \frac{3}{2} \bigl(-15
	+ 4 \nu \bigr)\biggr]
	+ \frac{G^3 m^3}{\hat{a}^3 c^6 {\bigl(\hat{e}^2-1\bigr)}^3} \biggl [\frac{3}{16} \hat{e}^4 \bigl(5
	- 5 \nu\nn\\
	& + 4 \nu^2\bigr)
	+ \frac{1}{32} \bigl(6750
	- 6542 \nu
	+ 123 \pi^2 \nu
	+ 84 \nu^2\bigr)
	+ \frac{1}{128} \hat{e}^2 \bigl(9840
	- 13712 \nu
	+ 123 \pi^2 \nu
	+ 1248 \nu^2\bigr)\biggr]\,,\\
	\hat{a}_r &= \hat{a}
	+ \frac{1}{4} \frac{G m}{c^2} \bigl(7
	-  \nu \bigr)
	+ \frac{G^2 m^2}{\hat{a} c^4} \biggl [\frac{1}{{\bigl(\hat{e}^2-1\bigr)}} \bigl(4
	- 7 \nu \bigr)
	+ \frac{1}{16} \bigl(1
	+ \nu^2\bigr)\biggr]\nn\\
	& 
	+ \frac{G^3 m^3}{\hat{a}^2 c^6} \biggl [\frac{1}{1680} \frac{1}{{\bigl(\hat{e}^2-1\bigr)}^2} \bigl(53760 - 176024 \nu
	+ 4305 \pi^2 \nu
	+ 15120 \nu^2\bigr)\nn\\
	& 
	+ \frac{1}{6720} \frac{1}{{\bigl(\hat{e}^2-1\bigr)}} \bigl(26880
	- 215408 \nu
	+ 4305 \pi^2 \nu
	+ 47040 \nu^2\bigr)
	+ \frac{1}{64} \bigl(-1 + \nu
	-  \nu^3\bigr)\biggr]\,,\\
	\hat{e}_r &= \hat{e}
	+ \frac{G m}{\hat{a} c^2} \biggl [- \frac{9 + \nu}{8 \hat{e}}
	+ \frac{5}{8} \hat{e} \bigl(-3
	+ \nu \bigr)\biggr]
	+ \frac{G^2 m^2}{\hat{a}^2 c^4 {\bigl(\hat{e}^2-1\bigr)}} \biggl [\frac{1}{128} \hat{e} \bigl(-845
	+ 1814 \nu
	- 13 \nu^2\bigr)\nn\\
	& + \frac{5 \bigl(-135 + 34 \nu + \nu^2\bigr)}{128 \hat{e}}
	+ \frac{1}{128} \hat{e}^3 \bigl(415
	- 210 \nu
	+ 7 \nu^2\bigr)
	+ \frac{\bigl(9 + \nu \bigr)^2}{128 \hat{e}^3}\biggr]\nn\\
	& + \frac{G^3 m^3}{\hat{a}^3 c^6 {\bigl(\hat{e}^2-1\bigr)}^2} \Biggl\{\frac{1}{53760} \hat{e} \biggl [-2685795
	+ \bigl(9173827
	- 206640 \pi^2\bigr) \nu
	- 966525 \nu^2
	- 735 \nu^3\biggr]\nn\\
	& + \frac{5589 - 477 \nu - 113 \nu^2 + \nu^3}{1024 \hat{e}^3}
	-  \frac{3}{1024} \hat{e}^5 \bigl(1893
	- 1245 \nu
	+ 135 \nu^2
	+ \nu^3\bigr)\nn\\
	& -  \frac{51543 -  \bigl(63499 - 984 \pi^2\bigr) \nu - 1935 \nu^2 - 9 \nu^3}{1536 \hat{e}}\nn\\
	& 
	+ \frac{1}{107520} \hat{e}^3 \biggl [2184315
	-  \bigl(578087
	+ 68880 \pi^2\bigr) \nu - 80535 \nu^2
	+ 1155 \nu^3\biggr]
	-  \frac{\bigl(9 + \nu \bigr)^3}{1024 \hat{e}^5}\Biggr\}\,, \\
	\hat{e}_t &= \hat{e}
	+ \frac{G m}{\hat{a} c^2} \biggl [\frac{1}{8} \hat{e} \bigl(17
	- 7 \nu \bigr)
	-  \frac{9 + \nu}{8 \hat{e}}\biggr]
	+ \frac{G^2 m^2}{\hat{a}^2 c^4 {\bigl(\hat{e}^2-1\bigr)}} \biggl [\frac{1}{128} \hat{e} \bigl(-1101
	+ 998 \nu
	- 61 \nu^2\bigr)
	-  \frac{99 - 18 \nu + 19 \nu^2}{128 \hat{e}}\nn\\
	& + \frac{1}{128} \hat{e}^3 \bigl(607
	- 138 \nu
	+ 79 \nu^2\bigr)
	+ \frac{\bigl(9 + \nu \bigr)^2}{128 \hat{e}^3}\biggr]
	+ \frac{G^3 m^3}{\hat{a}^3 c^6 {\bigl(\hat{e}^2-1\bigr)}^2} \Biggl\{\frac{1}{53760} \hat{e} \biggl [39165
	+ \bigl(2958103\nn\\
	& - 103320 \pi^2\bigr) \nu
	+ 54075 \nu^2
	- 8715 \nu^3\biggr]
	+ \frac{1}{1024} \hat{e}^5 \bigl(10801
	- 1405 \nu
	+ 771 \nu^2
	- 407 \nu^3\bigr)\nn\\
	& -  \frac{23895 -  \bigl(37747 - 984 \pi^2\bigr) \nu - 1455 \nu^2 + 207 \nu^3}{1536 \hat{e}}
	+ \frac{2997 - 81 \nu + 71 \nu^2 + 13 \nu^3}{1024 \hat{e}^3}\nn\\
	& 
	+ \frac{1}{3072} \hat{e}^3 \bigl(-91959 + 82531 \nu
	- 36093 \nu^2
	+ 2097 \nu^3\bigr)
	-  \frac{\bigl(9 + \nu \bigr)^3}{1024 \hat{e}^5}\Biggr\}\,,\\
	\hat{e}_\phi &= \hat{e}
	+ \frac{G m}{\hat{a} c^2} \biggl [- \frac{9 + \nu}{8 \hat{e}}
	+ \frac{1}{8} \hat{e} \bigl(-15
	+ \nu \bigr)\biggr]
	+ \frac{G^2 m^2}{\hat{a}^2 c^4 {\bigl(\hat{e}^2-1\bigr)}} \biggl [- \frac{675 - 98 \nu + 3 \nu^2}{128 \hat{e}}
	+ \frac{1}{128} \hat{e}^3 \bigl(415
	- 94 \nu
	+ 11 \nu^2\bigr)\nn\\
	& + \frac{3}{128} \hat{e} \bigl(-495
	+ 114 \nu
	+ 17 \nu^2\bigr)
	+ \frac{\bigl(9 + \nu \bigr)^2}{128 \hat{e}^3}\biggr]
	+ \frac{G^3 m^3}{\hat{a}^3 c^6 {\bigl(\hat{e}^2-1\bigr)}^2} \Biggl\{\frac{1}{107520} \hat{e}^3 \biggl [3487995
	+ \bigl(-339047\nn\\
	& + 17220 \pi^2\bigr) \nu
	+ 203385 \nu^2
	- 20685 \nu^3\biggr]
	+ \frac{1}{1024} \hat{e}^5 \bigl(-5679
	+ 2143 \nu
	- 237 \nu^2
	- 59 \nu^3\bigr)\nn\\
	& + \frac{5589 - 153 \nu - 41 \nu^2 + 5 \nu^3}{1024 \hat{e}^3}
	-  \frac{60183 -  \bigl(38779 - 984 \pi^2\bigr) \nu - 927 \nu^2 - 57 \nu^3}{1536 \hat{e}}\nn\\
	& 
	+ \frac{1}{53760} \hat{e} \biggl [-5508195 + \bigl(9974953
	- 137760 \pi^2\bigr) \nu
	- 89145 \nu^2
	+ 25935 \nu^3\biggr]
	-  \frac{\bigl(9 + \nu \bigr)^3}{1024 \hat{e}^5}\Biggr\}\,,\\
	\hat{f}_t &= \frac{G^2 m^2}{\hat{a}^2 c^4 {\bigl(\hat{e}^2-1\bigr)}^{3/2}} \Biggl\{\frac{1}{8} \hat{e} \bigl(-15
	+ \nu \bigr) \nu
	-  \frac{1}{8} \hat{e}^3 \bigl(-15
	+ \nu \bigr) \nu
	+ \frac{G m}{\hat{a} c^2} \biggl [- \frac{\bigl(-15 + \nu \bigr) \nu \bigl(9 + \nu \bigr)}{64 \hat{e}}\nn\\
	& 
	+ \frac{1}{64} \hat{e}^3 \nu \bigl(297
	- 175 \nu + 23 \nu^2\bigr)
	+ \frac{1}{2240} \hat{e} \bigl(22400
	+ 28531 \nu
	- 1435 \pi^2 \nu
	- 15050 \nu^2
	- 385 \nu^3\bigr)\biggr]\Biggr\}\,,\\
	\hat{g}_t &= \frac{G^2 m^2}{\hat{a}^2 c^4 {\bigl(\hat{e}^2-1\bigr)}^{3/2}} \Biggl\{\frac{3}{2} \bigl(-5
	+ 2 \nu \bigr)
	-  \frac{3}{2} \hat{e}^2 \bigl(-5
	+ 2 \nu \bigr)\nn\\
	& 
	+ \frac{G m}{\hat{a} c^2} \biggl [\frac{3}{16} \hat{e}^2 \bigl(95
	- 55 \nu
	+ 18 \nu^2\bigr)
	+ \frac{1}{192} \bigl(6660
	- 11972 \nu + 123 \pi^2 \nu
	+ 792 \nu^2\bigr)\biggr]\Biggr\}\,,\\
	\hat{h}_t &= \frac{\hat{e}^2 G^3 m^3 \nu}{16 \hat{a}^3 c^6 {\bigl(\hat{e}^2-1\bigr)}^{3/2}} \bigl(116
	- 49 \nu
	+ 3 \nu^2\bigr)\,,\\
	\hat{i}_t &= \frac{\hat{e}^3 G^3 m^3 \nu}{192 \hat{a}^3 c^6 {\bigl(\hat{e}^2-1\bigr)}^{3/2}} \bigl(23
	- 73 \nu
	+ 13 \nu^2\bigr)\,,\\
	\hat{f}_\phi &= \frac{G^2 m^2}{\hat{a}^2 c^4 {\bigl(\hat{e}^2-1\bigr)}^3} \Biggl\{\frac{1}{8} \hat{e}^4 \bigl(1
	+ 19 \nu
	- 3 \nu^2\bigr)
	+ \frac{1}{8} \hat{e}^2 \bigl(-1
	- 19 \nu
	+ 3 \nu^2\bigr)
	+ \frac{G m}{\hat{a} c^2} \biggl [- \frac{1}{32} \bigl(9
	+ \nu \bigr) \bigl(-1
	- 19 \nu
	+ 3 \nu^2\bigr)\nn\\
	& + \frac{1}{26880} \hat{e}^2 \bigl(58800
	+ 708544 \nu
	- 30135 \pi^2 \nu
	- 228480 \nu^2
	- 11760 \nu^3\bigr)
	+ \frac{1}{96} \hat{e}^4 \bigl(3
	+ 506 \nu
	- 357 \nu^2 + 36 \nu^3\bigr)\biggr]\Biggr\}\,,\\
	\hat{g}_\phi &= \frac{G^2 m^2}{\hat{a}^2 c^4 {\bigl(\hat{e}^2-1\bigr)}^3} \Biggl\{\frac{1}{32} \hat{e}^3 \nu \bigl(-1
	+ 3 \nu \bigr)
	-  \frac{1}{32} \hat{e}^5 \nu \bigl(-1
	+ 3 \nu \bigr)
	+ \frac{G m}{\hat{a} c^2} \biggl [- \frac{3}{256} \hat{e} \nu \bigl(9
	+ \nu \bigr) \bigl(-1
	+ 3 \nu \bigr)\nn\\
	& 
	+ \frac{1}{256} \hat{e}^5 \nu \bigl(14 - 49 \nu
	+ 26 \nu^2\bigr)
	-  \frac{1}{8960} \hat{e}^3 \nu \bigl(-34726
	+ 1435 \pi^2
	+ 24080 \nu
	+ 70 \nu^2\bigr)\biggr]\Biggr\}\,,\\
	\hat{h}_\phi &= \frac{\hat{e}^4  G^3 m^3 \nu}{192 \hat{a}^3 c^6 {\bigl(\hat{e}^2-1\bigr)}^3} \bigl(82
	- 57 \nu
	+ 15 \nu^2\bigr)\,,\\
	\hat{i}_\phi &= \frac{\hat{e}^5 G^3 m^3 \nu}{256 \hat{a}^3 c^6 {\bigl(\hat{e}^2-1\bigr)}^3} \bigl(1
	- 5 \nu
	+ 5 \nu^2\bigr)\,.
\end{align}
\end{subequations}
All the parameters in the quasi-elliptic case~\eqref{eq:parameterselliptic} and in the quasi-hyperbolic case~\eqref{eq:parametershyperbolic} are equivalent \textit{via} the analytic continuation. The two eccentric anomalies are related by
\begin{subequations}\label{eq:AnalyticContinuationQKparameters}
\begin{align}
	v = - \di \,u\,.
\end{align}
The correspondence reads
\begin{align}
	\qquad\hatn &= \di \,n \bigl(1+g_t\bigr) \,, & \hat{K} &= K \,, & \hata_r &= -a_r \,, &&&&\\
	\hate_r &= e_r \,, & \hate_t &= e_t\bigl(1+g_t\bigr) \,, & \hate_\phi &= e_\phi \,, &&&&\\
	\hat{f}_t &=  \di \,f_t \,, & \hat{g}_t &=  \di \,g_t \,, & \hat{f}_\phi &= f_\phi \,,
	& \hat{g}_\phi &= g_\phi \,, &&\\
	\hat{h}_t &=  \di \,h_t \,, & \hat{i}_t &=  \di \,i_t \,, & \hat{h}_\phi &= h_\phi \,,
	& \hat{i}_\phi &= i_\phi \,.&&
\end{align}
\end{subequations}
Note the interesting role played by the parameter $g_t=-\di\,\hat{g}_t$ (which enters at the 2PN order in the QK representation) in ensuring the equivalence up to 3PN order. We remind that the mean motion $n$ and periastron precession $K$ (or their equivalent $\hat{n}$ and $\hat{K}$) are gauge-invariant --- taking the same expression in modified harmonic coordinates and, for instance, in ADM coordinates. Since $g_t$ appears in the relation between $n$ and $\hat{n}$ we infer that it must also be gauge-invariant, which we have checked by comparing its expression in modified harmonic and ADM coordinates.

In perturbations, all the parameters~\eqref{eq:AnalyticContinuationQKparameters} vary. In addition, note that the variations of the ``integration constants'' $\hat{c}_\ell$ and $\hat{c}_\phi$ for local contributions in the quasi-hyperbolic case are related through the analytic continuation to the corresponding variations of $c_\ell$ and $\omega$ in the quasi-elliptic case. This is given by Eqs.~\eqref{eq:relationcellcphi} of the main text.

\section{Formulas for integrals in the elliptic case} 
\label{app:formulas}

\subsection{Indefinite integrals}

We present a series of formulas (without proofs) for relevant indefinite integrals involving powers of $\chi(u) \equiv 1-e \cos u$ where $u$ is the eccentric anomaly. For definiteness we use the eccentricity defined in Eq.~\eqref{eq:Newtonianae}. The analogue of the function~\eqref{eq:defV} for the eccentricity $e$ reads
\begin{subequations}\label{eq:defV2}
	\begin{align}
		V &= u + 2 \arctan\biggl[\frac{\beta\sin u}{1-\beta\cos u}\biggr]\,, \\
		\text{with}\quad\beta &\equiv \frac{1-\sqrt{1-e^2}}{e}\,.
	\end{align}
\end{subequations}
Posing $w\equiv\frac{1}{2}(V-u)=\arctan[\frac{\beta\sin u}{1-\beta\cos u}]$ (which is a periodic function of $u$) we can write the alternative form\footnote{The fact that $w$ under this form is real can be checked with ($\epsilon=\pm 1$)
\begin{align*}
	&\ln\bigl(1-\beta \de^{\epsilon \di u}\bigr) = \di \arg \bigl(1-\beta \de^{\epsilon \di u}\bigr)+\frac{1}{2}\ln\bigl(1-2\beta \cos u+\beta^2\bigr)\,,\\
	\text{hence}\quad & w = -\frac{1}{2}\Bigl[\arg \bigl(1-\beta \de^{\di u}\bigr) - \arg \bigl(1-\beta \de^{-\di u}\bigr)\Bigr]\,.
\end{align*}
}
\begin{align}\label{eq:w}
	w = \frac{\di}{2}\Bigl[\ln\bigl(1-\beta \de^{\di u}\bigr) - \ln\bigl(1-\beta \de^{-\di u}\bigr)\Bigr]\,.
\end{align}
The indefinite integral with respect to $u$ (valid up to a constant) is given in terms of the dilogarithm function
\begin{subequations}
\begin{align}\label{eq:T}
	&T \equiv \int\dd u \,w = - \frac{1}{2}\Bigl[\text{Li}_2(\beta \de^{\di u}) + \text{Li}_2(\beta \de^{-\di u})\Bigr]\,,\\
	\text{with}\quad &\mathrm{Li}_2(z) \equiv - \int_0^z \frac{\dd s}{s}\, \ln(1-s)\,.\label{eq:Li2}
\end{align}
\end{subequations}
We need to compute the following indefinite integrals, for $n\in\mathbb{N}$:
\begin{subequations}
	\begin{align}
		&H_n(u) = \int \frac{\dd u \sin(u)}{\chi^n} \,,\qquad
		I_n(u) = \int \frac{\dd u}{\chi^n} \,,\qquad
		J_n(u) = \int \dd u \,\frac{V}{\chi^n} \,,\qquad\\
		&K_n(u) = \,\int \dd u \,\frac{u}{\chi^n}\,,\qquad\quad~
		L_n(u) = \int \dd u \,\bigl(V-u\bigr)\chi^n \,.
	\end{align}
\end{subequations}
These integrals are obtained using some recursive relations, with generic terms:
\begin{subequations}
	\begin{align}
		H_n &= -\frac{1}{e (n-1)\chi^{n-1}}\,,\\
		I_n &= \frac{1}{(n-1)(1-e^2)}\left[(2n-3)I_{n-1}-(n-2)I_{n-2}+e\frac{\sin u}{\chi^{n-1}} \right]\,,\\
		J_n &= \frac{1}{(n-1)(1-e^2)}\biggl[(2n-3)J_{n-1}-(n-2)J_{n-2} +e\frac{V \sin u}{\chi^{n-1}}  + \frac{\sqrt{1-e^2}}{n-1}\frac{1}{\chi^{n-1}}\biggr]\,,\\
		K_n &= \frac{1}{(n-1)(1-e^2)}\left[(2n-3)K_{n-1}-(n-2)K_{n-2}+e\frac{u \sin u}{\chi^{n-1}} + \frac{1}{(n-2)\chi^{n-2}}\right]\,,\\
		L_n &= \frac{1}{n}\biggl[(2n-1)L_{n-1}-(n-1)(1-e^2)L_{n-2}
		- e (V-u)\chi^{n-1}\sin u + \sqrt{1-e^2}\frac{\chi^{n-1}}{n-1} - \frac{\chi^n}{n} \biggr]\,.
\end{align}
\end{subequations}
The recursions start with the following initial expressions 
\begin{subequations}
	\begin{align}
		H_1 &= \frac{\ln\chi}{e}\,,\\
		I_1 &= \frac{V}{\sqrt{1-e^2}}\,,\\
		J_1 &= \frac{V^2}{2\sqrt{1-e^2}}\,,\\
		K_1 &= \frac{1}{\sqrt{1-e^2}}\left[u V-\frac{u^2}{2} -2 T\right]\,,\\
		K_2 &= \frac{1}{1-e^2}\left[K_1+e \frac{u \sin u}{\chi} - \ln\chi\right]\,,\\
		L_0 &= 2T\,,\\
		L_1 &= 2T - e (V-u)\sin u + e \cos u + \sqrt{1-e^2} \ln\chi\,.
	\end{align}
\end{subequations}
Note the presence of the function $T$ defined by~\eqref{eq:T} as well as of the logarithm of $\chi$. Similar expressions hold also in the quasi-hyperbolic case. In the elliptic case, the angles $V$ and $u$ always appear through the combination $V-u = 2 w$, which means that we always deal with combinations of $J_n - K_n$.

\subsection{Definite integrals}

We present a list of definite integrals for computing the averages of functions of the eccentric anomaly $u$,
\begin{align}
	\langle f\rangle \equiv \int_0^{2\pi} \frac{\dd u}{2\pi} \,f(u) \,.
\end{align}
With $\chi=1-e\cos{u}$, $\beta = (1-\sqrt{1-e^2})/e$, $w$ defined by~\eqref{eq:defV2}--\eqref{eq:w}, we have 
\begin{subequations}
\begin{align}
	\langle \frac{1}{\chi^n}\rangle &= \frac{1}{(1-e^2)^{n-1/2}} \sum_{i=0}^{\lfloor \frac{n-1}{2} \rfloor}
	\frac{(n-1)!}{(n-1-2i)! \, (i!)^{2}} \left(\frac{e}{2}\right)^{2i}\qquad \text{for $n\geqslant 1$} \,,\\
	\langle \chi^n\rangle &= \sum_{i=0}^{\lfloor \frac{n}{2} \rfloor}
		\frac{n!}{(n-2i)! \, (i!)^{2}} \left(\frac{e}{2}\right)^{2 i}\qquad \text{for $n\geqslant 0$}\,,\\
	\langle w^2 \chi\rangle&= \frac{1}{2} \,\left[\Li_{2}(\beta^{2})-1+\sqrt{1-e^2}\left(1-2 \ln{(1 - \beta^2)}\right)\right]\,, \\
	\langle w^2\rangle &= \,\frac{\Li_{2}(\beta^{2})}{2}\,, \\
	\langle w \sin u \rangle &= \frac{\beta}{2}\,, \\
	\langle \frac{w \sin u}{\chi} \rangle &= - \frac{\ln(1-\beta^{2})}{e}\,, \\
	\langle  \frac{w \sin u}{\chi^n} \rangle &= \frac{1}{2(n-1)e} \left[\sqrt{1-e^2} \,\langle\frac{1}{\chi^n}\rangle - \langle\frac{1}{\chi^{n-1}}\rangle\right] \qquad \text{for $n\geqslant 2$}\,,\\
	\langle \ln\chi \rangle &= - \ln (1+\beta^{2})\,,\\
	\langle \chi\ln\chi \rangle &= - \ln (1+\beta^{2}) + 1 - \sqrt{1-e^2}\,,\\
	\langle \Li_{2}(\beta e^{\di u}) \rangle &= 0\,,\\
	\langle  \Li_{2}(\beta e^{\di u}) \chi \rangle &= \frac{\sqrt{1-e^2}-1}{2}\,,\\
	\langle  \frac{\Li_{2}(\beta e^{\di u})}{\chi} \rangle &= \frac{\Li_{2}(\beta^2)}{\sqrt{1-e^2}}\,,\\
	\langle  \frac{\Li_{2}(\beta e^{\di u})}{\chi^2} \rangle &= \frac{\Li_{2}(\beta^2)}{(1-e^2)^{3/2}} - \frac{\ln{(1-\beta^2)}}{1-e^2}\,,\\
	\langle  \frac{\Li_{2}(\beta e^{\di u})}{\chi^n} \rangle &= \frac{1}{(n-1)(1-e^2)} \left[ (2n-3) \langle  \frac{\Li_{2}(\beta e^{\di u})}{\chi^{n-1}} \rangle - (n-2) \langle \frac{\Li_{2}(\beta e^{\di u})}{\chi^{n-2}} \rangle \right.\nn\\ & \left. \qquad\qquad\qquad\qquad\qquad\qquad + \frac{1}{2(n-2)} \left( \sqrt{1-e^2} \langle\frac{1}{\chi^{n-1}}\rangle - \langle\frac{1}{\chi^{n-2}}\rangle \right) \right] \qquad \text{for $n\geqslant 3$}\,.
\end{align}
\end{subequations}

\section{Fourier decomposition of the Newtonian source mass quadrupole}\label{app:IijFourier}

The Fourier coefficients of the quadrupole moment are customarily denoted (see \textit{e.g.} the Appendix~B of~\cite{BLL25}),
\begin{align}
	I_{ij}(t) = \sum_{p=-\infty}^{+\infty} \,\mathop{\mathcal{I}}_{p}{}_{\!\!ij} \,\de^{\di p n t}\,,\qquad\quad
	\mathop{\mathcal{I}}_{p}{}_{\!\!ij} = \mathop{\mathcal{A}}_{p}
	m_0^{ij} + \mathop{\mathcal{B}}_{p}
	\,{m_0^{*}}^{ij} + \mathop{\mathcal{C}}_{p}
	\hat{\ell}^{ij}\,,
\end{align}
where $\bm{m}_0=\frac{1}{\sqrt{2}}(\bm{i}+\di\,\bm{j})$, with $\bm{i}$, $\bm{j}$ being the orthonormal basis in the orbital plane, such that $\bm{i}$ in the direction of the periastron. From Eq.~\eqref{eq:FourierX} we have for all $p$
\begin{equation}
\mathop{\mathcal{A}}_{p} = \frac{1}{2}m \nu a^2 X_{p}^{2, -2}(e)\,,\qquad
\mathop{\mathcal{B}}_{p} = \frac{1}{2}m \nu a^2 X_{p}^{2, 2}(e)\,,\qquad
\mathop{\mathcal{C}}_{p} = - \frac{1}{2}m \nu a^2 X_{p}^{2, 0}(e)\,.
\end{equation}
We note also the closed-form expressions of the Fourier coefficients (see the Appendix~B of~\cite{BLL25})
\begin{align}
	\mathop{\mathcal{A}}_{p} &= \frac{m\nu\,a^2}{p^2 e^2}\biggl\{ 2 e
	J_{p-1}\left(p e\right)\left( p
	(1-e^2)+\sqrt{1-e^2}\right)\nn\\& \qquad\qquad + J_{p}\left(p
	e\right)\left(-2(p+1)\left(1+\sqrt{1-e^2}\right) +
	e^2\left[1+2p\left(1+\sqrt{1-e^2}\right)\right]\right)\biggr\}\,,\\
	\mathop{\mathcal{B}}_{p} &= \frac{m\nu\,a^2}{p^2 e^2}\biggl\{ 2 e
	J_{p-1}\left(p e\right)\left( p
	(1-e^2)-\sqrt{1-e^2}\right)\nn\\& \qquad\qquad + J_{p}\left(p
	e\right)\left(-2(p+1)\left(1-\sqrt{1-e^2}\right) +
	e^2\left[1+2p\left(1-\sqrt{1-e^2}\right)\right]\right)\biggr\}\,,\\
	\mathop{\mathcal{C}}_{p} &= \frac{m\nu\,a^2}{p^2} J_{p}\left(p e\right)\,,
\end{align}
which are valid whenever $p\not= 0$; in the case $p=0$ we have
\begin{align}
	\mathop{\mathcal{A}}_{0} = \mathop{\mathcal{B}}_{0} =
	\frac{5}{4}m\nu\,a^2\,e^2\,,\qquad\mathop{\mathcal{C}}_{0} =
	-\frac{1}{2}m\nu\,a^2\left(1+\frac{3}{2}e^2\right)\,.
\end{align}

\bibliography{ListeRef_QK4p5PN.bib}

@misc{ancillary,
  howpublished = "The ancillary file
  \texttt{SupplementalMaterial.wl} contains the results in machine-readable format.  Note that this file is optimally read using \emph{Mathematica}, but it can be straightforwardly parsed by any text editor."
}

@Article{ABIQ08,
  Title                    = {Inspiralling compact binaries in quasi-elliptical
 orbits: The complete 3{PN} energy flux},
  Author                   = {Arun, K.G. and Blanchet, L. and Iyer, B.R. and
 Qusailah, M.S.},
  Journal                  = {Phys. Rev. D},
  Year                     = {2008},
  Pages                    = {064035},
  Volume                   = {77},

  Eprint                   = {arXiv:0711.0302 [gr-qc]}
}

@Article{ABIS09,
  Title                    = {Third post-{N}ewtonian angular momentum flux and the
 secular evolution of orbital elements for
 inspiralling compact binaries in quasi-elliptical
 orbits},
  Author                   = {Arun, K.G. and Blanchet, L. and Iyer, B.R. and
 Siddhartha, S.},
  Journal                  = {Phys. Rev. D},
  Year                     = {2009},
  Pages                    = {124018},
  Volume                   = {80},

  Eprint                   = {arXiv:0908.3854 [gr-qc]}
}

@Article{ABIQ04,
  Title                    = {The 2.5{PN} gravitational wave polarisations from 
 inspiralling compact binaries in circular orbits},
  Author                   = {Arun, K.G. and Blanchet, L. and Iyer, B. R. and
 Qusailah, M. S.},
  Journal                  = {Class. Quant. Grav.},
  Year                     = {2004},
  Note                     = {Erratum \textit{Class. Quant. Grav.}, 22:3115, 2005},
  Pages                    = {3771},
  Volume                   = {21},

  Eprint                   = {gr-qc/0404185}
}

@Article{BBFM17,
  Title                    = {Center-of-Mass Equations of Motion and Conserved Integrals of Compact Binary Systems at the Fourth Post-{N}ewtonian Order},
  Author                   = {Bernard, Laura and Blanchet, Luc and Faye, Guillaume and Marchand, Tanguy},
  Journal                  = {Phys. Rev. D},
  Year                     = {2018},
  Pages                    = {044037},
  Volume                   = {97},
  
  Eprint                   = {arXiv:1711.00283},
  primaryclass = {gr-qc},
}

@Article{BlanchetLR_2024,
  Title                    = {Post-{N}ewtonian theory for gravitational waves},
  Author                   = {Blanchet, L.},
  Journal                  = {Living Rev. Relativ.},
  Year                     = {2024},
  Pages                    = {4},
  Volume                   = {27},

  Eprint                   = {arXiv:1310.1528},
  primaryclass = {gr-qc},
}

@Article{B97,
  Title                    = {Gravitational radiation reaction and balance
 equations to post-{N}ewtonian order},
  Author                   = {Blanchet, L.},
  Journal                  = {Phys. Rev. D},
  Year                     = {1997},
  Pages                    = {714-732},
  Volume                   = {55},

  Eprint                   = {gr-qc/9609049},
  Slaccitation             = {%%CITATION = GR-QC 9609049;%%}
}

@Article{Blanchet,
  Title                    = {A class of non-metric couplings to gravity},
  Author                   = {Blanchet, L.},
  Journal                  = {Phys. Rev. Lett.},
  Year                     = {1992},
  Pages                    = {559},
  Volume                   = {69}
}

@Article{BD92,
  Title                    = {Hereditary effects in gravitational radiation},
  Author                   = {Blanchet, L. and Damour, T.},
  Journal                  = {Phys. Rev. D},
  Year                     = {1992},
  Pages                    = {4304-4319},
  Volume                   = {46},

  Slaccitation             = {%%CITATION = PHRVA,D46,4304;%%}
}

@Article{BD88,
  Title                    = {Tail-transported temporal correlations in the
 dynamics of a gravitating system},
  Author                   = {Blanchet, L. and Damour, T.},
  Journal                  = {Phys. Rev. D},
  Year                     = {1988},
  Pages                    = {1410},
  Volume                   = {37},

  Slaccitation             = {%%CITATION = PHRVA,D37,1410;%%}
}

@Article{BD84,
  Title                    = {Multipolar radiation reaction in general relativity},
  Author                   = {Blanchet, L. and Damour, T.},
  Journal                  = {Phys. Lett. A},
  Year                     = {1984},
  Pages                    = {82-86},
  Volume                   = {104},

  Slaccitation             = {%%CITATION = PHLTA,A104,82;%%}
}

@Article{BQW05,
  Title                    = {Gravitational recoil of inspiraling black-hole binaries 
 to second post-{N}ewtonian order},
  Author                   = {Blanchet, L. and Qusailah, M. S. and Will, Clifford M.},
  Journal                  = {Astrophys. J.},
  Year                     = {2005},
  Pages                    = {508},
  Volume                   = {635},

  Eprint                   = {astro-ph/0507692}
}

@Article{BS89,
  Title                    = {Higher order gravitational radiation losses in binary
 systems},
  Author                   = {Blanchet, L. and Sch{\"a}fer, G.},
  Journal                  = {Mon. Not. Roy. Astron. Soc.},
  Year                     = {1989},
  Pages                    = {845-867},
  Volume                   = {239},

  Slaccitation             = {%%CITATION = MNRAA,239,845;%%}
}

@Article{BoR61,
  Title                    = {Transport of momentum by gravitational waves -
 Linear approximation},
  Author                   = {Bonnor, W.B. and Rotenberg, M.A.},
  Journal                  = {Proc. R. Soc. London, Ser. A},
  Year                     = {1961},
  Pages                    = {109},
  Volume                   = {265},
  Referencedin             = {2002-3blanchet}
}

@Book{BrouwerClemence,
  Title                    = {Methods of celestial mechanics},
  Author                   = {Brouwer, D. and Clemence, G.M.},
  Publisher                = {Academic Press},
  Year                     = {1961},

  Address                  = {New York}
}

@Article{Bu71,
  Title                    = {Gravitational radiation damping of slowly moving
 systems calculated using matched asymptotic
 expansions},
  Author                   = {Burke, W.L.},
  Journal                  = {J. Math. Phys.},
  Year                     = {1971},
  Pages                    = {401},
  Volume                   = {12},
  Referencedin             = {2002-3blanchet}
}

@InProceedings{BuTh70,
  Title                    = {Gravitational Radiation Damping},
  Author                   = {Burke, W.L. and Thorne, K.S.},
  Booktitle                = {Relativity},
  Year                     = {1970},

  Address                  = {New York and London},
  Editor                   = {Carmeli, M. and Fickler, S.I. and Witten, L.},
  Pages                    = {209--228},
  Publisher                = {Plenum Press},
  Referencedin             = {2002-3blanchet}
}

@Article{CE70,
  Title                    = {The 5/2-post-{N}ewtonian equations of
 hydrodynamics and radiation reaction in general
 relativity},
  Author                   = {Chandrasekhar, S. and Esposito, F.P.},
  Journal                  = {Astrophys. J.},
  Year                     = {1970},
  Pages                    = {153--179},
  Volume                   = {160},
  Referencedin             = {2002-3blanchet}
}

@Article{3mn,
  Title                    = {The last three minutes: Issues in
 gravitational-wave measurements of coalescing
 compact binaries},
  Author                   = {Cutler, C. and Apostolatos, T.A. and Bildsten,
 L. and Finn, L.S. and Flanagan, E.E. and Kennefick,
 D. and Markovic, D.M. and Ori, A. and Poisson,
 E. and Sussman, G.J. and Thorne, K.S.},
  Journal                  = {Phys. Rev. Lett.},
  Year                     = {1993},
  Pages                    = {2984--2987},
  Volume                   = {70},
  Referencedin             = {2002-3blanchet}
}

@Article{D83,
  Title                    = {Gravitational radiation reaction in the binary pulsar 
 and the quadrupole formula controvercy},
  Author                   = {Damour, T.},
  Journal                  = {Phys. Rev. Lett.},
  Year                     = {1983},
  Pages                    = {1019-1021},
  Volume                   = {51},

  Slaccitation             = {%%CITATION = PRLTA,51,1019;%%}
}

@Article{D82,
  Title                    = {Probl{\`e}me des deux corps et freinage de rayonnement
 en relativit{\'e} g{\'e}n{\'e}rale},
  Author                   = {Damour, T.},
  Journal                  = {C. R. Acad. Sc. Paris},
  Year                     = {1982},
  Pages                    = {1355},
  Volume                   = {294}
}

@Article{DD85,
  Title                    = {General relativistic celestial mechanics of binary systems 
 I. The post-{N}ewtonian motion},
  Author                   = {Damour, T. and Deruelle, N.},
  Journal                  = {Annales Inst. H. Poincar{\'e} Phys. Th\'eor.},
  Year                     = {1985},
  Pages                    = {107-132},
  Volume                   = {43}
}

@Article{DD81a,
  Title                    = {Radiation reaction and angular momentum loss in small angle
 gravitational scattering},
  Author                   = {Damour, T. and Deruelle, N.},
  Journal                  = {Phys. Lett. A},
  Year                     = {1981},
  Pages                    = {81},
  Volume                   = {87},

  Slaccitation             = {%%CITATION = PHLTA,A87,81;%%}
}

@Article{DJS14,
  Title                    = {Non-local-in-time action for the fourth
 post-{N}ewtonian conservative dynamics of two-body
 systems},
  Author                   = {T.~Damour and P.~Jaranowski and G.~Sch{\"a}fer},
  Journal                  = {Phys. Rev. D},
  Year                     = {2014},
  Pages                    = {064058},
  Volume                   = {89},

  Eprint                   = {arXiv:1401.4548},
  primaryclass = {gr-qc},
}

@Article{DS88,
  Title                    = {Higher order relativistic periastron advances 
 in binary pulsars},
  Author                   = {Damour, T. and Sch{\"a}fer, G.},
  Journal                  = {Nuovo Cim.},
  Year                     = {1988},
  Pages                    = {127},
  Volume                   = {B101},

  Slaccitation             = {%%CITATION = NUCIA,B101,127;%%}
}

@Misc{duriez,
  Title                    = {Cours de m\'ecanique c\'eleste classique},

  Author                   = {Duriez, L.},
  Year                     = {2002},

  Address                  = {Universit\'e de Lille}
}

@Article{Ehl80,
  Title                    = {Isolated systems in general relativity},
  Author                   = {Ehlers, J.},
  Journal                  = {Ann. N.Y. Acad. Sci.},
  Year                     = {1980},
  Pages                    = {279},
  Volume                   = {336}
}

@Article{Fit83,
  Title                    = {The influence of gravitational wave momentum
 losses on the centre of mass motion of a {N}ewtonian
 binary system},
  Author                   = {Fitchett, M.~J.},
  Journal                  = {Mon. Not. Roy. Astron. Soc.},
  Year                     = {1983},
  Pages                    = {1049},
  Volume                   = {203}
}

@Article{FStail,
  Title                    = {Tail terms in gravitational radiation reaction via
 effective field theory},
  Author                   = {S.~Foffa and R.~Sturani},
  Journal                  = {Phys. Rev. D},
  Year                     = {2012},
  Pages                    = {044056},
  Volume                   = {87},

  Eprint                   = {arXiv:1111.5488},
  primaryclass = {gr-qc},
}

@Article{GLPR16,
  Title                    = {Tail effect in gravitational radiation reaction:
 Time nonlocality and renormalization group
 evolution},
  Author                   = {C.~R.~Galley and A.~K.~Leibovich and R.~A.~Porto
 and A.~Ross},
  Journal                  = {Phys. Rev. D},
  Year                     = {2016},
  Pages                    = {124010},
  Volume                   = {93},
  Eprint                   = {arXiv:1511.07379 [gr-qc]}
}

@Article{GII97,
  Title                    = {Second post-{N}ewtonian gravitational radiation reaction for
 two-body systems: Nonspinning bodies},
  Author                   = {Gopakumar, A. and Iyer, B. R. and Iyer, Sai},
  Journal                  = {Phys. Rev. D},
  Year                     = {1997},
  Pages                    = {6030-6053},
  Volume                   = {55},

  Eprint                   = {gr-qc/9703075}
}

@Article{HulseTaylor,
  Title                    = {Discovery of a pulsar in a binary system},
  Author                   = {Hulse, R.A. and Taylor, J.H.},
  Journal                  = {Astrophys. J.},
  Year                     = {1975},
  Pages                    = {L51},
  Volume                   = {195}
}

@Article{itoh3,
  Title                    = {Third-and-a-half order post-{N}ewtonian equations of
 motion for relativistic compact binaries using the
 strong field point particle limit},
  Author                   = {Itoh, Y.},
  Journal                  = {Phys. Rev. D},
  Year                     = {2009},
  Pages                    = {024003},
  Volume                   = {80},

  Eprint                   = {arXiv:0911.4232},
  primaryclass = {gr-qc},
}

@Article{IW95,
  Title                    = {Post-{N}ewtonian gravitational radiation reaction
 for two-body systems: Nonspinning bodies},
  Author                   = {Iyer, B.R. and Will, C.M.},
  Journal                  = {Phys. Rev. D},
  Year                     = {1995},
  Pages                    = {6882--6893},
  Volume                   = {52},
  Referencedin             = {2002-3blanchet}
}

@Article{IW93,
  Title                    = {Post-{N}ewtonian gravitational-radiation reaction
 for two-body systems},
  Author                   = {Iyer, B.R. and Will, C.M.},
  Journal                  = {Phys. Rev. Lett.},
  Year                     = {1993},
  Pages                    = {113--116},
  Volume                   = {70},
  Referencedin             = {2002-3blanchet}
}

@Article{JaraS97,
  Title                    = {Radiative 3.5 post-{N}ewtonian {ADM} {H}amiltonian for
 many body point - mass systems},
  Author                   = {Jaranowski, P. and Sch{\"a}fer, G.},
  Journal                  = {Phys. Rev. D},
  Year                     = {1997},
  Pages                    = {4712-4722},
  Volume                   = {55}
}

@Article{KFS03,
  Title                    = {The binary black-hole dynamics at the
 third-and-a-half post-{N}ewtonian order in the
 {ADM}-formalism},
  Author                   = {K{\"o}nigsd{\"o}rffer, C. and Faye, G. and
 Sch{\"a}fer, G.},
  Journal                  = {Phys. Rev. D},
  Year                     = {2003},
  Pages                    = {044004},
  Volume                   = {68},

  Eprint                   = {astro-ph/0305048},
  Slaccitation             = {%%CITATION = GR-QC 9902030;%%}
}

@Article{KG06,
  Title                    = {Phasing of gravitational waves from inspiralling eccentric binaries at the third-and-a-half post-{N}ewtonian order},
  Author                   = {K\"onigsd\"orffer, Christian and Gopakumar, Achamveedu},
  Journal                  = {Phys. Rev. D},
  Year                     = {2006},

  Month                    = {Jun},
  Pages                    = {124012},
  Volume                   = {73},

  Doi                      = {10.1103/PhysRevD.73.124012},
  Eprint                   = {gr-qc/0603056},
  Issue                    = {12},
  Numpages                 = {24},
  Publisher                = {American Physical Society},
  Url                      = {https://link.aps.org/doi/10.1103/PhysRevD.73.124012}
}

@Article{K80a,
  Title                    = {Finite reduced hydrodynamic equations in the slow-motion
 approximation to general relativity. I. First
 post-{N}ewtonian equations},
  Author                   = {Kerlick, G.D.},
  Journal                  = {Gen. Relativ. Gravit.},
  Year                     = {1980},
  Pages                    = {467--482},
  Volume                   = {12}
}

@Article{K80b,
  Title                    = {Finite reduced hydrodynamic equations in the slow-motion
 approximation to general relativity. II. Radiation
 reaction and higher-order divergent terms},
  Author                   = {Kerlick, G.D.},
  Journal                  = {Gen. Relativ. Gravit.},
  Year                     = {1980},
  Pages                    = {521--543},
  Volume                   = {12}
}

@Misc{xtensor,
  Title                    = {{xAct}: Efficient tensor computer algebra for
 {Mathematica}},

  Author                   = {J.~M.~Mart\'in-Garc\'ia and A. Garc\'ia-Parrado and
 A. Stecchina and B. Wardell and C. Pitrou and
 D. Brizuela and D. Yllanes and G. Faye and L. Stein
 and R. Portugal and T. B\"ackdahl},
  Note                     = {http://www.xact.es/},
  Year                     = {GPL 2002--2012}
}

@Article{MGS04,
  Title                    = {Third post-{N}ewtonian accurate generalized
 quasi-Keplerian parametrization for compact binaries
 in eccentric orbits},
  Author                   = {Memmesheimer, R. and Gopakumar, A. and Sch{\"a}fer, G.},
  Journal                  = {Phys. Rev. D},
  Year                     = {2004},
  Pages                    = {104011},
  Volume                   = {70},

  Eprint                   = {gr-qc/0407049}
}

@Article{Miller74,
  Title                    = {The Effect of Gravitational Radiation-Reaction on the Evolution of the Riemann S-Type Ellipsoids},
  Author                   = {Miller, B.D.},
  Journal                  = {Astrophys. J.},
  Year                     = {1974},
  Pages                    = {609--620},
  Volume                   = {187},
  Referencedin             = {2002-3blanchet}
}

@Article{MW03,
  Title                    = {A post-{N}ewtonian diagnostic of quasi-equilibrium binary 
 configurations of compact objects},
  Author                   = {Mora, Thierry and Will, Clifford M.},
  Journal                  = {Phys. Rev. D},
  Year                     = {2004},
  Pages                    = {104021},
  Volume                   = {69},

  Eprint                   = {gr-qc/0312082},
  Slaccitation             = {%%CITATION = GR-QC 0311052;%%}
}

@Article{NB05,
  Title                    = {Gravitational radiation reaction in the equations of motion of 
 compact binaries to 3.5 post-{N}ewtonian order},
  Author                   = {Nissanke, S. and Blanchet, L.},
  Journal                  = {Class. Quant. Grav.},
  Year                     = {2005},
  Pages                    = {1007},
  Volume                   = {22},

  Eprint                   = {gr-qc/0412018}
}

@Article{Papa71,
  Title                    = {\'Etude syst\'ematique du rayonnement
 gravitationnel 4-polaire. \'Energie-impulsion et
 moment cin\'etique du rayonnement},
  Author                   = {Papapetrou, A.},
  Journal                  = {Ann. Inst. Henri Poincar{\'e}},
  Year                     = {1971},
  Pages                    = {79},
  Volume                   = {XIV},
  Referencedin             = {2002-3blanchet}
}

@Article{PapaL81,
  Title                    = {Equation of motion including the reaction of 
 gravitational radiation},
  Author                   = {Papapetrou, A. and Linet, B.},
  Journal                  = {Gen. Relativ. Gravit.},
  Year                     = {1981},
  Pages                    = {335},
  Volume                   = {13}
}

@Article{PW02,
  Title                    = {Post-{N}ewtonian gravitational radiation and
 equations of motion via direct integration of the
 relaxed Einstein equations. II. Two-body equations
 of motion to second post-{N}ewtonian order, and
 radiation-reaction to 3.5 post-{N}ewtonian order},
  Author                   = {Pati, M.E. and Will, C.M.},
  Journal                  = {Phys. Rev. D},
  Year                     = {2002},
  Pages                    = {104008},
  Volume                   = {65},

  Eprint                   = {gr-qc/0201001}
}

@Article{SW93,
  Title                    = {Second post-{N}ewtonian motion of compact binaries},
  Author                   = {Sch{\"a}fer, G. and Wex, N.},
  Journal                  = {Phys. Lett. A},
  Year                     = {1993},
  Note                     = {Erratum Phys. Lett. A {\bf 177}, 461 (1993)},
  Pages                    = {196},
  Volume                   = {174}
}

@Article{TW82,
  Title                    = {A New Test of General Relativity: Gravitational
 Radiation and the Binary Pulsar PSR 1913+16},
  Author                   = {Taylor, J.H. and Weisberg, J.M.},
  Journal                  = {Astrophys. J.},
  Year                     = {1982},
  Pages                    = {908--920},
  Volume                   = {253},
  Referencedin             = {2002-3blanchet}
}

@Article{Th80,
  Title                    = {Multipole expansions of gravitational radiation},
  Author                   = {Thorne, K.S.},
  Journal                  = {Rev. Mod. Phys.},
  Year                     = {1980},
  Pages                    = {299--339},
  Volume                   = {52},
  Referencedin             = {2002-3blanchet}
}

@Article{WalkW80,
  Title                    = {The approximation of radiative effects in
 relativistic gravity --- Gravitational radiation
 reaction and energy loss in nearly {N}ewtonian
 systems},
  Author                   = {Walker, M. and Will, C.M.},
  Journal                  = {Astrophys. J.},
  Year                     = {1980},
  Pages                    = {L129},
  Volume                   = {242}
}

@Article{Wi92,
  Title                    = {Coalescing binary systems of compact objects to
 (post)$^{5/2}$-{N}ewtonian order. II. Higher-order
 wave forms and radiation recoil},
  Author                   = {Wiseman, A.~G.},
  Journal                  = {Phys. Rev. D},
  Year                     = {1992},
  Pages                    = {1517},
  Volume                   = {46}
}

@article{KP1,
    author = "Kälin, Gregor and Porto, Rafael A.",
    title = "{From Boundary Data to Bound States}",
    eprint = "1910.03008",
    archivePrefix = "arXiv",
    primaryClass = "hep-th",
    reportNumber = "DESY 19-167, UUITP-40/19, DESY-19-167",
    doi = "10.1007/JHEP01(2020)072",
    journal = "JHEP",
    volume = "01",
    pages = "072",
    year = "2020"
}

@article{KP2,
    author = "Kälin, Gregor and Porto, Rafael A.",
    title = "{From boundary data to bound states. Part II. Scattering angle to dynamical invariants (with twist)}",
    eprint = "1911.09130",
    archivePrefix = "arXiv",
    primaryClass = "hep-th",
    reportNumber = "DESY 19-201, SLAC-PUB-17487, UUITP-46/19, DESY-19-201",
    doi = "10.1007/JHEP02(2020)120",
    journal = "JHEP",
    volume = "02",
    pages = "120",
    year = "2020"
}

@article{H26,
    author = "Henry, Quentin",
    title = "{Adiabatic tides in compact binaries on quasielliptic orbits: Radiation at the second-and-a-half relative post-Newtonian order}",
    eprint = "2601.01794",
    archivePrefix = "arXiv",
    primaryClass = "gr-qc",
    doi = "10.1103/bszd-247c",
    journal = "Phys. Rev. D",
    volume = "114",
    number = "6",
    pages = "064062",
    year = "2026"
}

@article{BFHLT23b,
    author = "Blanchet, Luc and Faye, Guillaume and Henry, Quentin and Larrouturou, Fran\c{c}ois and Trestini, David",
    title = "{Gravitational-wave flux and quadrupole modes from quasicircular nonspinning compact binaries to the fourth post-Newtonian order}",
    eprint = "2304.11186",
    archivePrefix = "arXiv",
    primaryClass = "gr-qc",
    reportNumber = "DESY-23-044",
    doi = "10.1103/PhysRevD.108.064041",
    journal = "Phys. Rev. D",
    volume = "108",
    number = "6",
    pages = "064041",
    year = "2023"
}

@article{BF19,
  title={Flux-balance equations for linear momentum and center-of-mass position of self-gravitating post-{N}ewtonian systems},
  author={Blanchet, Luc and Faye, Guillaume},
  journal={Classical and Quantum Gravity},
  volume={36},
  number={8},
  pages={085003},
  year={2019},
  publisher={IOP Publishing}
}

@article{LPY23,
  title={Radiation Reaction for Non-Spinning Bodies at 4.5 PN in the Effective Field Theory Approach},
  author={Leibovich, Adam K and Pardo, Brian A and Yang, Zixin},
  journal={arXiv preprint arXiv:2302.11016},
  year={2023}
}

@article{HK23,
    author = "Henry, Quentin and Khalil, Mohammed",
    title = "{Spin effects in gravitational waveforms and fluxes for binaries on eccentric orbits to the third post-Newtonian order}",
    eprint = "2308.13606",
    archivePrefix = "arXiv",
    primaryClass = "gr-qc",
    journal = "Phys. Rev. D",
    volume = "108",
    number = "10",
    pages = "104016",
    year = "2023"
}

@article{CTG22,
  title={Generalized quasi-Keplerian solution for eccentric, nonspinning compact binaries at 4PN order and the associated inspiral-merger-ringdown waveform},
  author={Cho, Gihyuk and Tanay, Sashwat and Gopakumar, Achamveedu and Lee, Hyung Mok},
  journal={Physical Review D},
  volume={105},
  number={6},
  pages={064010},
  year={2022},
  publisher={APS},
    eprint = "2110.09608",
    archivePrefix = "arXiv",
    primaryClass = "gr-qc"
}

@article{BFT25,
   title={Gravitational radiation reaction for compact binary systems at the fourth-and-a-half post-Newtonian order},
   volume={42},
   ISSN={1361-6382},
   url={http://dx.doi.org/10.1088/1361-6382/adac9d},
   DOI={10.1088/1361-6382/adac9d},
   number={6},
   journal={Classical and Quantum Gravity},
   publisher={IOP Publishing},
   author={Blanchet, Luc and Faye, Guillaume and Trestini, David},
   year={2025},
   month=mar, pages={065015} 
}

@article{T25,
    author = "Trestini, David",
    title = "{Schott term in the binding energy for compact binaries on circular orbits at fourth post-Newtonian order}",
    eprint = "2504.13245",
    archivePrefix = "arXiv",
    primaryClass = "gr-qc",
    doi = "10.1103/lsbb-sv45",
    journal = "Phys. Rev. D",
    volume = "112",
    number = "2",
    pages = "024076",
    year = "2025"
}

@misc{BDG24,
      title={Explicit solution of the gravitational two-body problem at the second post-Minkowskian order}, 
      author={Donato Bini and Thibault Damour and Andrea Geralico},
      year={2024},
      eprint={2408.17193},
      archivePrefix={arXiv},
      primaryClass={gr-qc},
      url={https://arxiv.org/abs/2408.17193}, 
}

@article{DGI04,
    author = "Damour, Thibault and Gopakumar, Achamveedu and Iyer, Bala R.",
    title = "{Phasing of gravitational waves from inspiralling eccentric binaries}",
    eprint = "gr-qc/0404128",
    archivePrefix = "arXiv",
    doi = "10.1103/PhysRevD.70.064028",
    journal = "Phys. Rev. D",
    volume = "70",
    pages = "064028",
    year = "2004"
}

@article{BDG23,
    author = "Bini, Donato and Damour, Thibault and Geralico, Andrea",
    title = "{Radiated momentum and radiation reaction in gravitational two-body scattering including time-asymmetric effects}",
    eprint = "2210.07165",
    archivePrefix = "arXiv",
    primaryClass = "gr-qc",
    doi = "10.1103/PhysRevD.107.024012",
    journal = "Phys. Rev. D",
    volume = "107",
    number = "2",
    pages = "024012",
    year = "2023"
}

@article{BGR25,
    author = "Bini, Donato and Geralico, Andrea and Rufrano Aliberti, Sara",
    title = "{Radiation-reaction correction to scattering binary dynamics at the next-to-leading post-Newtonian order}",
    eprint = "2509.17853",
    archivePrefix = "arXiv",
    primaryClass = "gr-qc",
    doi = "10.1103/ggbn-fw7j",
    journal = "Phys. Rev. D",
    volume = "112",
    number = "10",
    pages = "104005",
    year = "2025"
}

@article{BFST26,
    author = "Blanchet, Luc and Faye, Guillaume and Seraille, Emeric and Trestini, David",
    title = "{Gravitational radiation reaction for compact binary systems at the fourth-and-a-half post-Newtonian order in harmonic coordinates}",
    doi = {10.1088/1361-6382/ae6411},
    url = {https://doi.org/10.1088/1361-6382/ae6411},
    year = {2026},
    month = {may},
    publisher = {IOP Publishing},
    volume = {43},
    number = {10},
    pages = {105009},
    journal = {Classical and Quantum Gravity},
    eprint = "2601.06743",
    archivePrefix = "arXiv",
    primaryClass = "gr-qc"
}

@article{BD12,
    author = "Bini, Donato and Damour, Thibault",
    title = "{Gravitational radiation reaction along general orbits in the effective one-body formalism}",
    eprint = "1210.2834",
    archivePrefix = "arXiv",
    primaryClass = "gr-qc",
    doi = "10.1103/PhysRevD.86.124012",
    journal = "Phys. Rev. D",
    volume = "86",
    pages = "124012",
    year = "2012"
}

@article{BG21,
    author = "Bini, Donato and Geralico, Andrea",
    title = "{Frequency domain analysis of the gravitational wave energy loss in hyperbolic encounters}",
    eprint = "2108.02472",
    archivePrefix = "arXiv",
    primaryClass = "gr-qc",
    doi = "10.1103/PhysRevD.104.104019",
    journal = "Phys. Rev. D",
    volume = "104",
    number = "10",
    pages = "104019",
    year = "2021"
}

@article{BDG21,
    author = "Bini, Donato and Damour, Thibault and Geralico, Andrea",
    title = "{Radiative contributions to gravitational scattering}",
    eprint = "2107.08896",
    archivePrefix = "arXiv",
    primaryClass = "gr-qc",
    doi = "10.1103/PhysRevD.104.084031",
    journal = "Phys. Rev. D",
    volume = "104",
    number = "8",
    pages = "084031",
    year = "2021"
}

@article{CGHL18,
   title={Gravitational waves from compact binaries in post-Newtonian accurate hyperbolic orbits},
   volume={98},
   ISSN={2470-0029},
   url={http://dx.doi.org/10.1103/PhysRevD.98.024039},
   DOI={10.1103/physrevd.98.024039},
   number={2},
   journal={Physical Review D},
   publisher={American Physical Society (APS)},
   author={Cho, Gihyuk and Gopakumar, Achamveedu and Haney, Maria and Lee, Hyung Mok},
   year={2018},
   month={July} 
}

@article{BLL25,
  title={Innermost stable circular orbit of arbitrary-mass compact binaries at fourth post-Newtonian order},
  author={Blanchet, Luc and Langlois, David and Ligout, Etienne},
  journal={Physical Review D},
  volume={112},
  number={6},
  pages={064025},
  year={2025},
  publisher={APS}
}

@article{HHLL23b,
    author = "Henry, Quentin and Larrouturou, Fran{\c{c}}ois and Le Poncin-Lafitte, Christophe",
    title = "{Electromagnetic fields in compact binaries: Post-Newtonian wave generation and application to double white dwarfs systems}",
    eprint = "2310.03785",
    archivePrefix = "arXiv",
    primaryClass = "gr-qc",
    doi = "10.1103/PhysRevD.109.084048",
    journal = "Phys. Rev. D",
    volume = "109",
    number = "8",
    pages = "084048",
    year = "2024"
}

@article{HH25,
    author = "Henry, Quentin and Heffernan, Anna",
    title = "{Adiabatic tides in compact binaries on quasielliptic orbits: Dynamics at the second-and-a-half relative post-Newtonian order}",
    eprint = "2512.06489",
    archivePrefix = "arXiv",
    primaryClass = "gr-qc",
    doi = "10.1103/z6pb-hx1w",
    journal = "Phys. Rev. D",
    volume = "114",
    number = "6",
    pages = "064061",
    year = "2026"
}

@article{THS13,
    author = "Tessmer, Manuel and Hartung, Johannes and Schafer, Gerhard",
    title = "{Aligned Spins: Orbital Elements, Decaying Orbits, and Last Stable Circular Orbit to high post-Newtonian Orders}",
    eprint = "1207.6961",
    archivePrefix = "arXiv",
    primaryClass = "gr-qc",
    doi = "10.1088/0264-9381/30/1/015007",
    journal = "Class. Quant. Grav.",
    volume = "30",
    pages = "015007",
    year = "2013"
}

\end{document}